\documentclass[a4paper, 11pt]{article}
\usepackage{jheppub}

\usepackage[dvipsnames]{xcolor}
\usepackage{amsthm}

\usepackage[utf8]{inputenc}
\usepackage{graphicx}
\usepackage{systeme}
\usepackage{amssymb, mathrsfs}
\usepackage{amsmath}

\usepackage{etoolbox}
\usepackage{enumitem}
\usepackage{mathtools}
\usepackage{tikz-cd}
\usepackage{tikz}
\usepackage{pgfplots}
\usetikzlibrary{positioning}
\usetikzlibrary{decorations.markings}
\usetikzlibrary{snakes}
\usepgfplotslibrary{fillbetween}
\usetikzlibrary{intersections}
\usepackage{float}
\usepackage{simpler-wick}
\usepackage{amsbsy}
\usepackage{hyperref}

\usepackage[toc]{appendix}

\newtheoremstyle{theoremstyle}
    {12pt} 
    {12pt} 
    {\itshape} 
    {} 
    {\bfseries} 
    {.} 
    {.5em} 
    {} 

\newtheoremstyle{definitionstyle}
    {12pt} 
    {12pt} 
    {} 
    {} 
    {\bfseries} 
    {.} 
    {.5em} 
    {} 

\newtheoremstyle{remarkstyle}
    {12pt} 
    {12pt} 
    {} 
    {} 
    {\itshape} 
    {.} 
    {.5em} 
    {} 

\swapnumbers{}

\theoremstyle{theoremstyle}
\newtheorem{theorem}{Theorem}[section]

\newcommand{\thistheoremname}{}
\newtheorem{genericthm}[theorem]{\thistheoremname}

\theoremstyle{theoremstyle}
\newtheorem{lemma}[theorem]{Lemma}
\newtheorem{proposition}[theorem]{Proposition}
\newtheorem{corollary}[theorem]{Corollary}

\theoremstyle{definitionstyle}

\theoremstyle{remarkstyle}

\newcommand{\K}{\mathcal{K}}
\newcommand{\B}{\mathcal{B}}
\newcommand{\R}{\mathbb{R}}

\newcommand{\Z}{\mathbb{Z}}
\newcommand{\C}{\mathbb{C}}
\newcommand{\T}{\mathbb{T}}
\newcommand{\M}{\mathcal{M}}
\renewcommand{\H}{\mathcal{H}}
\renewcommand{\P}{\mathcal{P}}
\newcommand{\F}{\mathcal{F}}

\newcommand{\D}{\mathcal{D}}
\newcommand{\A}{\mathcal{A}}
\newcommand{\U}{\mathcal{U}}
\renewcommand{\O}{\mathcal{O}}

\newcommand{\id}{\mathrm{id}}

\newcommand{\sign}[1]{\mathrm{sgn}(#1)}

\newcommand{\nobarfrac}{\genfrac{}{}{0pt}{}}

\let\temp\epsilon
\let\epsilon\varepsilon
\let\varepsilon\temp

\let\temp\phi
\let\phi\varphi
\let\varphi\temp

\author[a,b]{Koen Schouten}
\author[a]{and Mikhail Isachenkov}
\affiliation[a]{Institute of Physics, University of Amsterdam, Amsterdam, the Netherlands}
\affiliation{Korteweg-de Vries Institute for Mathematics, University of Amsterdam, Amsterdam, the Netherlands}
\affiliation[b]{School of Mathematics, Trinity College Dublin, Dublin, Ireland}

\emailAdd{kschoute@tcd.ie}
\emailAdd{m.isachenkov@uva.nl}

\title{\boldmath The von Neumann algebraic quantum group $\mathrm{SU}_q(1,1)\rtimes \mathbb{Z}_2$ and the DSSYK model}

\abstract{The double-scaling limit of the SYK (DSSYK) model is known to possess an underlying $\mathcal{U}_q(\mathfrak{su}(1,1))$ quantum group symmetry. In this paper, we provide, for the first time, a von Neumann algebraic quantum group-theoretical description of the degrees of freedom and the dynamics of the DSSYK model. In particular, we construct the operator-algebraic quantum Gauss decomposition for the von Neumann algebraic quantum group $\mathrm{SU}_q(1,1)\rtimes \mathbb{Z}_2$, i.e. the $q$-deformation of the normaliser of $\mathrm{SU}(1,1)$ in $\mathrm{SL}(2,\mathbb{C})$, and derive the Casimir action on its quantum homogeneous spaces. We then show that the dynamics on quantum $\mathrm{AdS}_{2,q}$ space reduces to that of the DSSYK model. Furthermore, we argue that the extension of the global symmetry group to its normaliser is not only necessary for a consistent definition of the locally compact quantum group, but that, moreover, the reduction to the DSSYK model works exclusively at the level of the normaliser. The von Neumann algebraic description is shown to give a natural restriction on the allowed quantised coordinates, elegantly ensuring length positivity and non-negative integer chord numbers. Lastly, we make remarks on the correlation function related to the strange series representation, which is argued to interpolate between the AdS and dS regions of our $q$-homogeneous space.}

\begin{document}
\hspace{0cm}

\maketitle

\newpage

\section{Introduction}
In recent years, the study and application of operator algebras, in particular von Neumann algebras, have gained considerable attention in the context of quantum gravity \cite{LL2023, LL2023_2}. The essence is that the norm topology on the operator algebra is too fine for many physical purposes. Namely, even if one would progressively, more precisely, measure matrix elements of a given operator, with the result converging to a definite value, the corresponding sequence of operators might not converge in the norm topology. To accommodate for such a natural idealisation of experimentally possible situations, one instead requires the operator algebra to be closed under the coarser weak topology\footnote{An operator algebra $\A\subseteq \B(\H)$ is closed under the weak topology if and only if for any sequence $\{a_n\} \subseteq \A$ and $a\in \B(\H)$ such that $\langle v|a_n-a|w\rangle \xrightarrow{n\to \infty} 0$ for all $|v\rangle, |w\rangle\in \H$, the limit is also included in the algebra, i.e. $a\in \A$ \cite{M1990}.}, which by definition makes it into a von Neumann algebra. These von Neumann algebras are known to have a canonical decomposition (over the centre) into three distinct types of factors (algebras with trivial centre) \cite{TakesakiOperatorAlgebras}, based on the existence of pure states and density matrices, which play a crucial role in the understanding of black hole physics and the black hole information problem \cite{HV2025}. In particular, when going from a coarse-grained towards a more microscopic description of quantum gravity, resolving inconsistencies of its semiclassical description necessitates a change in the operator-algebraic picture. At the level of the von Neumann algebras of observables associated with local regions, this can be seen as the need to change the von Neumann algebra from Type III to Type II, or eventually to Type I \cite{LL2023, LL2023_2, W2022, HV2025, Chandrasekaran2023AnalgebraObservers, Chandrasekaran2023LargeN, Witten2022WhydoesQFTmakesense, PeningtonWitten2023algebrasstatesjtgravity} (see \cite{Sorce2024Notesontype} for a review). At the level of spacetime itself and its symmetries, this can be interpreted as a certain kind of discretisation. A well-studied method for discretising/deforming continuous symmetries is through specific deformations of groups, known as {\it quantum groups} \cite{QGAR, KasselQuantumGroups, CPAGtQG}. In any sensible quantum theory, these discretised/quantum group symmetries should be compatible with the von Neumann algebra of observables, and therefore have a von Neumann algebraic description themselves.\\

A specific toy model of quantum gravity that exhibits such a quantum group symmetry is the {\it double-scaled Sachdev-Ye-Kitaev (DSSYK) model} \cite{SY1993, K2015, Cotler_2017, BNS2018, BINT2019, BINN2023}, which is characterised by a parameter $q$. This model has received considerable attention recently, due to its solvability and proposed duality with gravity models \cite{NV2025, V2025, VZ2025, Tietto2025microscopicmodelsitterspacetime, O2025, Almheiri2025Holographyquantumdisk, BLMPP2025, JafferisKolchmeyerJTgravity, S2021, S2022, S2025, Susskind2023sitterspacechordsconfined, Sekino2025doublescaledsykqcdflat, Bhattacharjee2023Krylov, BM2024, BMT2025, HVX2025QuantumSymmetry, Blommaert2025qschwarzianliouvillegravity, Berkooz2024Parisi, OkuyamaSuzuki2023Correlators, LS2022, GoelNarovlansky2023semiclassicalgeo, BerkoozBrukner2024Chaosintegrabilitydoublescaled, Verlinde2024doublescaledsykchordssitter, AlmheiriGoel2024QuantumGravity, Bossi2025Sinedilatongravityvsdoublescaled, Xu2025chorddynamicscomplexitygrowth, Heller2025Krylov, GaiottoVerline2025SYKSchur, Susskind2023Paradoxresolutionillustrateprinciples, Aguilargutierrez2024complexitysitterspacedoublescaled, aguilargutierrez2024t2deformationsdoublescaledsyk,AguilarGutierrez2025BuildingHol, AguilarGutierrez2025Geometry, AguilarGutierrez2025Evolution, Milekhin2024revisitingbrowniansykpossible}. In particular, in the $q\to 1$ and low-energy limit, this model reduces to Liouville quantum mechanics \cite{MS2016}, which represents a mini-superspace limit of the Liouville CFT, and describes the same low-energy dynamics as JT-gravity \cite{J1985, T1983}, thus giving a tractable example of (nearly) AdS$_2$/CFT$_1$ holography \cite{M1999, Susskind1995WorldAsHologram, Witten199Aantidesitterspacethermal, SusskindWitten1998holographicboundantidesitter, HarlowOoguri2019, Gubser1998GaugeTheory, Mukhametzhanov2023largepsykchord}. For general values of $q$, the DSSYK model can be solved exactly at all energy scales using combinatorial objects called chord diagrams \cite{BNS2018, BINT2019} (see also \cite{BM2025} for a recent review), and it was found that this model possesses a quantum group symmetry with respect to the $\mathcal{U}_q\left(\mathfrak{su}(1,1)\right)$ and $\mathrm{SU}_q(1,1)$ quantum groups \cite{BINT2019, BINN2023, LinStanfordSymmetryAlgDSSYK}. For example, the correlation functions are constructed of the 3j- and 6j-symbols of this quantum group via Feynman-type rules. Moreover, the algebra of observables for this model was shown to be of Type II$_1$ \cite{SniadyFactoriality2004, RicardFactoriality2005, L2022, Xu2025vonneumannalgebrasdoublescaled}. This quantum group symmetry suggests that the corresponding gravitational bulk theory is defined on a quantum homogeneous space, hereby turning the spacetime into a non-commutative geometry (or, correspondingly, discretising it).\\

In the context of this paper, by a non-commutative geometry description, we mean having full control over the `measurable functions on the non-commutative space' and their symmetries, i.e. having the von Neumann algebra of functions on the space, and the corresponding von Neumann algebraic quantum group acting on it. As in the commutative case, there are further, potentially more refined, possible meanings of this term, corresponding to `geometrising' different analytical classes of functions on a deformed space in question, and thus probing different flavours of its `non-commutative points'. One of them is the $C^*$-algebraic version of the operator-algebraic quantum group construction, which we do not consider here, as it was shown to be equivalent to the von Neumann algebraic quantum group description in the theory developed by Kustermans and Vaes  \cite{KV2000, KV2000_1}.\footnote{Another approach that is expected to provide more subtle geometric information, is to construct a suitable Connes' spectral triple \cite{ConnesNoncommutativeGeometry1990} for the non-commutative spaces in question, which would then capture information on the smooth structure of the latter. This is a work in progress \cite{WPJort}. See \cite{SU11-Jort} for treating the classical SU$(1,1)$ Lie group case in this framework.}\\

Combining the above ingredients, we naturally conclude that the right representation-theoretical object to consider in the context of symmetries of the DSSYK model is a von Neumann algebraic version of the quantum group $\mathrm{SU}_q(1,1)$. Although this was already advocated in \cite[(Section~7)]{BINT2019}, followed by \cite{BINN2023}, and since then several other attempts at a quantum group theoretical description have been developed \cite{BM2024, BMT2025, HVX2025QuantumSymmetry}, the details of a von Neumann algebraic description for DSSYK had not yet been worked out.
Despite various technicalities in defining von Neumann algebraic quantum groups (perhaps, one of the reasons for the slow progress on that front), a considerable amount is fortunately known about the simplest rank-one groups. As is the case for classical Lie groups, most difficulties arise due to the non-compactness. Most famously, it was shown by Woronowicz \cite{W1991} in the early days of operator-algebraic quantum groups, that the coproduct for $\mathrm{SU}_q(1,1)$ ceases to be well-defined when one tries to upgrade the algebraic quantum group to an operator-algebraic one, i.e. when adding analytical structure. Korogodsky showed that this negative result is directly related to the non-existence of a global Poisson dressing action on $\mathrm{SU}(1,1)$ \cite{K1994}. In particular, it was shown that this no-go result can be bypassed by extending the group to the normaliser of $\mathrm{SU}(1,1)$ in $\mathrm{SL}(2,\C)$, which we denote by $\mathrm{SU}(1,1)\rtimes \Z_2$, as this normaliser does allow for a global dressing action. Finally, Koelink and Kustermans \cite{KK2003} explicitly constructed the von Neumann algebraic quantum group $\mathrm{SU}_q(1,1)\rtimes \Z_2$ in the framework developed by Kustermans and Vaes \cite{KV2000, KV2000_1}. In a later paper, the Pontryagin dual\footnote{By Pontryagin dual, we mean the full, non-abelian unitary dual of the operator-algebraic quantum group that turns out to also have a structure of an operator-algebraic quantum group. Existence of such a nice `quantum group Pontryagin duality' was one of the main motivations behind developing the theory of operator-algebraic quantum groups by Kustermans and Vaes \cite{KV2000, KV2000_1}. } for the quantum group was also constructed \cite{GKK2010}.\\

The {\it von Neumann algebraic quantum group} $\mathrm{SU}_q(1,1)\rtimes \Z_2$ is the main character of this paper. In the case of ordinary Lie groups, examples of spaces with a transitive $G$-action (homogeneous $G$-spaces) come from the coset construction. Similarly, in the $q$-deformed setting, any quantum group symmetry naturally acts on quantum homogeneous spaces, which can similarly be constructed as a coset of the quantum group. The main objective of this paper is to provide an explicit operator-algebraic construction of such a quantum homogeneous space that describes the degrees of freedom and dynamics of the DSSYK model. In particular, as argued before, this quantum homogeneous space corresponds to a quantisation of AdS$_2$ space. It is known that the Poincaré patch of an ordinary AdS$_2$ space can be constructed as a coset of $\mathrm{SU}(1,1)$ corresponding to the Gauss decomposition with respect to parabolic subgroups in $\mathrm{SU}(1,1)$. The primary goal of this paper is therefore to construct the quantum Gauss decomposition of the von Neumann algebraic quantum group $\mathrm{SU}_q(1,1)\rtimes \Z_2$ and the corresponding quantum homogeneous space.\\

\noindent The outline and main results of the paper are as follows:
\begin{itemize}[leftmargin=*]
    \item \textbf{In Section~\ref{sec:classicalSUandLiouville}} we review the classical Lie group $\mathrm{SU}(1,1)$ and Liouville quantum mechanics. This section is primarily intended as preparation for the quantum group description, as the classical Lie group should provide us with the necessary intuitions for the generalisation to the $q$-deformed case. In Section~\ref{sec:classicalNormaliserandLiouvilleTheory}, we provide arguments for why it is natural for Liouville quantum mechanics to consider the extension of the symmetry group to the normaliser. Then, in Section~\ref{sec:LiouvillePartitionFunction}, we recall the computation of the partition function and the $n$-point functions of Liouville quantum mechanics, emphasising the representation-theoretical aspects of $\mathrm{SU}(1,1)$.

    \item \textbf{In Section~\ref{sec:quantumsu11}} we construct the von Neumann algebraic quantum group $\mathrm{SU}_q(1,1)\rtimes \Z_2$ and compute its quantum Gauss decomposition. In particular, in Section~\ref{sec:quantumcoordinatealgebra}, we review the representation theory of the quantum coordinate algebra, and show why the extension of the normaliser is necessary for the coproduct to be well-defined. Then, in Section~\ref{sec:locallycompactquantumgroup}, based on \cite{KK2003, GKK2010}, we explicitly construct the von Neumann algebraic quantum group in the Cartan decomposition and give the associated regular representations in Section~\ref{sec:quantumregularrepresentations}. The main results of this paper are then presented in Sections~\ref{sec:quantumGaussDecomp} and \ref{sec:actionCasimiroperator}, in which, for the first time, we explicitly calculate the quantum Gauss decomposition for the operator-algebraic quantum group, and calculate the action of the Casimir element in this decomposition. This decomposition then explicitly shows how the coordinates in the associated quantum homogeneous spaces become quantised. The Casimir action will then be related to the Hamiltonian of a quantum-mechanical particle moving on the quantum homogeneous space, which describes the dynamics of the DSSYK model. We remark that this section is mathematically heavy. A more physics-oriented reader might therefore want to skip this section on their first read-through and proceed to Section~\ref{sec:reductiondsSYK} where its physical applications are discussed.

    \item \textbf{In Section~\ref{sec:reductiondsSYK}} we show how one can reduce to the DSSYK model by considering the dynamics on a von Neumann algebraic quantum homogeneous space. In particular, the eigenstates in the two-sided Hilbert space for the DSSYK model are shown to be given by matrix elements of the normaliser. In fact, it is argued that the reduction works exclusively on the level of the normaliser. Moreover, in the von Neumann algebraic description, the allowed eigenvalues of the twisted primitive elements, i.e. the quantised momenta, are restricted. In particular, it is shown in Section~\ref{sec:lengthpositivity} that this implies length positivity, i.e. only non-negative chord numbers are allowed. The operator-algebraic approach is therefore a significant improvement over the quantum group descriptions in \cite{BM2024, BMT2025}, where length positivity either had to be imposed by hand, or required the introduction of structures with unusual unitarity properties. In Section~\ref{sec:sykpartitionfunction}, we then derive the partition function and $n$-point functions of a quantum particle moving on this homogeneous space, which are indeed the same as for the DSSYK model. Additionally, it was shown in \cite{GKK2010} that the strange series appear in the regular representations of the quantum group. In Section~\ref{sec:strangematter}, we discuss the 2-point functions for these representations, and in particular argue that these interpolate between two-dimensional Anti-de Sitter and de Sitter space. Lastly, in Sections~\ref{sec:onesidedHilbertspace} and \ref{sec:quantumgroupactiononAdS}, we comment on the one-sided Hilbert space and the quantum group action on AdS$_{2,q}$.

    \item \textbf{In Appendix~\ref{sec:specialfunctions}} we recall some necessary facts on $q$-calculus and special functions, which are used in this paper to work out the operator-algebraic versions of the quantum Iwasawa and Gauss decomposition. In particular, we provide some new $q$-difference relations for the little $q$-Jacobi function, the Al-Salam-Carlitz polynomials, the Al-Salam-Carlitz II polynomials and the Ciccoli-Koelink-Koornwinder functions. \textbf{In Appendix~\ref{sec:calcsqgactions},} we give the explicit derivations of the (operator-algebraic) quantum group action of the quantum universal enveloping algebra $\U_q(\mathfrak{su}(1,1))$ on the quantum Iwasawa and Gauss decomposition.
\end{itemize}

\noindent \textbf{Conventions:} Throughout this paper, we will use the tensor product in several mathematically different contexts: the algebraic tensor product of elements in a (Hopf) algebra, the tensor product of Hilbert spaces and the spatial tensor product of von Neumann algebras. We will use the same notation $\otimes$ for each of these, as it should be clear from context which one is meant. For the $q$-deformed objects considered in this paper, we always assume $0<q<1$.

\section{Review on $\mathrm{SU}(1,1)$ and Liouville QM}
\label{sec:classicalSUandLiouville}
We begin our discussion on some of the representation-theoretical aspects of the classical Lie group $\mathrm{SU}(1,1)$ (see \cite{K1986, VK1991, SL2R} or \cite{K2018} for references), as this will form the basis for some of our intuition behind the quantum group structure that we will describe in Section~\ref{sec:quantumsu11}. Recall that $\mathrm{SU}(1,1)$ is given by the matrix Lie group
\begin{equation}
    \label{eq:definitionsu11}
     \mathrm{SU}(1,1) \coloneq \left\{\begin{bmatrix}\alpha & \bar \gamma \\ \gamma & \bar \alpha\end{bmatrix} : \alpha, \gamma \in \C \text{ and }|\alpha|^2 - |\gamma|^2 = 1\right\},
\end{equation}
and the corresponding (complex) Lie algebra $\mathfrak{sl}_2$ is spanned by $H = \left[\begin{smallmatrix}
    1 & 0\\ 0 &-1
\end{smallmatrix}\right]$, $E = \left[\begin{smallmatrix}
    0 & 1\\ 0 & 0
\end{smallmatrix}\right]$ and $F = \left[\begin{smallmatrix}
    0 & 0\\ 1 & 0
\end{smallmatrix}\right]$. Importantly, when working with quantum groups, one does not have access to actual group elements. Rather, quantum groups are described through the space of functions over the group. Therefore, in this chapter, we will create some intuition behind these function spaces and formulate results without referring to the group elements themselves. In Section~\ref{sec:classicalLocallyCompactGroup}, we discuss the Hopf algebra structure of the coordinate and universal enveloping algebra. Then, we show how the Haar measure upgrades these algebraic objects to analytic ones, giving rise to a von Neumann algebraic (quantum) group. In Section~\ref{sec:classicalGaussdecomp}, we describe the Gauss decomposition, and we discuss its applications to Liouville quantum mechanics in Sections~\ref{sec:classicalNormaliserandLiouvilleTheory} and \ref{sec:LiouvillePartitionFunction}.

\subsection{$\mathrm{SU}(1,1)$ as a locally compact quantum group}
\label{sec:classicalLocallyCompactGroup}
In the context of quantum groups, there are some well-known rigidity theorems (see \cite{KasselQuantumGroups}) that state that the multiplication in Lie groups, or correspondingly, the Lie bracket in Lie algebras, cannot be non-trivially deformed. To allow for more flexibility, one instead considers the dual description, consisting of some appropriate space of functions over the Lie group, which forms a Hopf algebra. In this chapter, we will discuss some of the properties of this algebra and how all the information about the Lie group is contained in it. Although we do introduce the basic concepts here, we refer to \cite{QGAR, CPAGtQG, KasselQuantumGroups} for complete discussions. Let $\A(\mathrm{SU}(1,1))$ be the associative, commutative and unital algebra of functions $\mathrm{SU}(1,1) \to \C$ with pointwise multiplication, generated by $\alpha, \beta, \gamma, \delta$ given by
\begin{equation}
    \label{eq:classicalAgenerators}
    \alpha\left(\begin{bmatrix}
        a & b\\c & d
    \end{bmatrix}\right) \coloneq a, \quad \beta\left(\begin{bmatrix}
        a & b\\c & d
    \end{bmatrix}\right) \coloneq b, \quad \gamma\left(\begin{bmatrix}
        a & b\\c & d
    \end{bmatrix}\right) \coloneq c, \quad \delta\left(\begin{bmatrix}
        a & b\\c & d
    \end{bmatrix}\right) \coloneq d.
\end{equation}
From the definition of $\mathrm{SU}(1,1)$ \eqref{eq:definitionsu11}, it follows that for every $g\in \mathrm{SU}(1,1)$, we have $\beta(g) = \overline{\gamma(g)}$ and $\delta(g) = \overline{\alpha(g)}$, and moreover $|\alpha(g)|^2 - |\gamma(g)|^2 = 1$. Therefore, $\A(\mathrm{SU}(1,1))$ naturally has the $*$-structure given by $\alpha^* = \delta$ and $\gamma^* = \beta$. In what follows we will write $\A$ for $\A(\mathrm{SU}(1,1))$. On this algebra, we define the \textit{coproduct} $\Delta \colon \A \to \A\otimes \A$ given by $\Delta(f)(g_1, g_2) = f(g_1g_2)$, the \textit{counit} $\epsilon \colon \A \to \C$ given by $\epsilon(f) \mapsto f(\mathbb{I}_{2\times 2})$ and the \textit{antipode} $S \colon \A\to \A$ given by $S(f)(g) = f(g^{-1})$. For the generators $\alpha$ and $\gamma$, these maps can be explicitly calculated to be given by
\begin{equation}
    \label{eq:classicalHopfstructure}
    \begin{gathered}
        \Delta(\alpha) = \alpha \otimes \alpha + \gamma^* \otimes \gamma, \quad \Delta(\gamma) = \gamma\otimes \alpha + \alpha^*\otimes \gamma,\\
        \epsilon(\alpha) = 1,\quad \epsilon(\gamma) = 0, \quad S(\alpha) = \alpha^*, \quad S(\gamma) = -\gamma.
    \end{gathered}
\end{equation}
One should note that the coproduct on the level of functions contains information about the multiplication on the group level, the counit contains information about the identity element, and the antipode contains information about the inverse. Therefore, the coordinate algebra $\A$ together with these maps fully describes the algebraic properties of the Lie group. Let us moreover define the maps $m\colon \A\otimes \A \to \A$ given by $m(a\otimes b) = ab$ and $\eta \colon \C\to \A$ given by $\eta(z)(g) = z$, then these maps satisfy the identities
\begin{equation}
    \label{eq:Hopfalgebraconditions}
    \begin{gathered}
        (\Delta\otimes \id)\circ \Delta = (\id\otimes \Delta)\circ \Delta,\\
        (\epsilon \otimes \id)\circ \Delta = \id = (\id \otimes \epsilon)\circ \Delta,\\
        m\circ (S\otimes \id)\circ \Delta = \eta \circ \epsilon = m\circ (\id \otimes S) \circ \Delta.
    \end{gathered}
\end{equation}
The above relations are exactly what define a \textit{Hopf algebra} (see \cite{QGAR, CPAGtQG}). Since $\A$ has a $*$-structure, it is in fact a Hopf $*$-algebra, and we call $\A(\mathrm{SU}(1,1))$ the \textit{coordinate algebra} of $\mathrm{SU}(1,1)$.\\

Similarly, the \textit{universal enveloping algebra} of the Lie algebra $\mathfrak{su}(1,1)$ is a Hopf $*$-algebra. That is, let $\U(\mathfrak{su}(1,1))$ be the associative unital algebra generated by $H,E,F$ that satisfy the algebraic relations 
\begin{equation}
    \label{eq:sl2algebra}
    [H,E] = 2E, \quad [H,F] = -2F,\quad [E,F] = H
\end{equation}
and with the $*$-structure given by $H^* = H$ and $E^* = -F$. We will often denote $\U(\mathfrak{su}(1,1))$ simply by $\U$. The coproduct, counit and antipode are given by $\Delta(X) = X\otimes 1 + 1\otimes X$, $\epsilon(X) = 0$ and $S(X) = -X$ for $X\in \{E,F,H\}$. The coproduct can be recognised as being the action of angular momentum/spin operators on tensor product spaces. The \textit{Casimir element}, i.e. the generator of the centre of $\U(\mathfrak{su}(1,1))$, is given by
\begin{equation}
    \label{eq:classicalCasimir}
    \Omega = \frac{1}{4}(H^2 - 2H + 4EF),
\end{equation}
which is defined up to normalisation. The above normalisation is chosen such that the Casimir operator is equal to the squared magnitude of the angular momentum operator. Since the Casimir element commutes with all the elements of $\mathfrak{su}(1,1)$, i.e. the symmetry generators, the Casimir element can correspond to the Hamiltonian of a physical system. Indeed, in our context, as we will see in Section~\ref{sec:classicalNormaliserandLiouvilleTheory}, it corresponds to the Liouville Hamiltonian in the Gauss decomposition.\\

The Lie algebra and Lie group are directly related to each other, with the former being the tangent space at the identity of the latter. Similarly, the coordinate algebra and the universal enveloping algebra constructed above are related, and both (at least, under favourable circumstances --- e.g. connectedness and simply-connectedness of the group) contain the same information about the underlying Lie group. In particular, there is a non-degenerate dual pairing between them. For $a\in \A$ and $u = X_1X_2\cdots X_k \in \U$ with $X_1,X_2,...,X_k \in \mathfrak{su}(1,1)$, we define the bilinear map $\langle \cdot, \cdot\rangle \colon \U\times \A \to \C$ given by
\begin{equation}
    \label{eq:classicalDualPairing}
    \langle u, a\rangle = \frac{\partial^k}{\partial t_1\cdots \partial t_k}\bigg|_{t_1 = \cdots = t_k = 0} a\left(\exp(t_1 X_1)\cdots \exp(t_k X_k)\right).
\end{equation}
It can be easily verified that this map is non-degenerate and satisfies the following properties:
\begin{equation}
    \begin{alignedat}{3}
        \label{eq:dualpairing}
        \langle \Delta_\U(u), a\otimes b\rangle &= \langle u, ab\rangle, \quad &\langle u, 1_\A\rangle &= \epsilon_\U(u), \quad &\langle u^*, a\rangle &= \overline{\langle u, S_\A(a)^*\rangle },\\
        \langle u\otimes v, \Delta_\A(a)\rangle &= \langle uv, a\rangle, \quad &\langle 1_\U, a\rangle& = \epsilon_\A(a), \quad &\langle u, a^*\rangle& = \overline{\langle S_\U(u)^*, a\rangle},
    \end{alignedat}
\end{equation}
where $\langle u\otimes v, a\otimes b\rangle \coloneq \langle u,a\rangle\langle v,b\rangle$. From \eqref{eq:classicalDualPairing}, it can be seen that this dual pairing exactly describes the link between the Lie group and its Lie algebra as the tangent space at the identity. Moreover, the dual pairing properties \eqref{eq:dualpairing} relate the coproduct on the coordinate algebra with multiplication on the universal enveloping algebra, and vice versa. Similarly, the counit on one side is related to the unit on the other side. Lastly, the antipode on one side is related to the adjoint on the other. Using this dual pairing, one can turn $\A$ into a left and right $\U$-module by defining the left and right actions of $\U$ on $\A$ by
\begin{equation}
    \label{eq:leftrightdualaction}
    u\triangleright a\coloneq ((\id \otimes \langle u,\cdot \rangle)\circ\Delta)(a)\quad \text{and}\quad a\triangleleft u \coloneq ((\langle u, \cdot \rangle \otimes \id)\circ\Delta)(a).
\end{equation}
Explicitly, these actions are given by
\begin{equation}
    \label{eq:algebraAction}
    (X\triangleright a)(g) = \frac{d}{dt}\bigg|_{t= 0} a\left(g\exp(t X)\right)\quad \text{and} \quad   (a\triangleleft X)(g) = \frac{d}{dt}\bigg|_{t= 0} a\left(\exp(t X)g\right)
\end{equation}
for $a\in \A(\mathrm{SU}(1,1))$, $g\in \mathrm{SU}(1,1)$ and $X\in \mathfrak{su}(1,1)$. This action, therefore, corresponds to taking a derivative of the function along the direction of the Lie algebra element, i.e. the action of the left- and right-invariant vector fields.\\

It can be seen that the information about the algebraic properties of $\mathrm{SU}(1,1)$ is fully contained within the coordinate algebra and the universal enveloping algebra. However, as $\mathrm{SU}(1,1)$ is a locally compact Lie group, it additionally has analytical properties that are crucial for describing its structure. Most prominently, it has a Haar measure, which becomes important when describing physical models as it defines the corresponding space of square-integrable functions. The above description should therefore be enhanced to include a Haar measure (see e.g. \cite{TakesakiOperatorAlgebras, CK2010, TItQGaD} for complete discussions). In particular, instead of the coordinate algebra, we consider the abelian von Neumann algebra of essentially bounded functions\footnote{That is, the set of functions $\mathrm{SU}(1,1)\to \C$ that are only allowed to be unbounded on zero-measure sets with respect to the Haar measure. Moreover, two functions that differ only on the zero-measure sets are considered equivalent.} $L^\infty(\mathrm{SU}(1,1))$ with coproduct $\Delta \colon L^\infty(\mathrm{SU}(1,1)) \to L^\infty(\mathrm{SU}(1,1))\otimes L^\infty(\mathrm{SU}(1,1))$\footnote{We identify $L^\infty(\mathrm{SU}(1,1))\otimes L^\infty(\mathrm{SU}(1,1))$ with $L^{\infty}(\mathrm{SU}(1,1)\times \mathrm{SU}(1,1))$, i.e. $(f\otimes g)(x,y) = f(x)g(y)$.} given by $\Delta(x)(g_1, g_2) = x(g_1g_2)$. Here, the tensor product denotes the spatial tensor product of von Neumann algebras. Let $L^\infty(\mathrm{SU}(1,1))^+$ be the set of positive elements, i.e. functions that map to the non-negative real line. The Haar measure $\mu$ on $\mathrm{SU}(1,1)$ now defines the (whenever finite) linear map $\phi \colon L^\infty(\mathrm{SU}(1,1))^+\to [0,\infty]$, called the \textit{Haar weight}, given by
\begin{equation}
    \phi(x) = \int_{\mathrm{SU}(1,1)}d\mu(g)x(g).
\end{equation}
One should note here that we allow $\infty$ to be part of the codomain. In particular, $\phi(1) =\infty$, which signifies the non-compactness and divergent volume of $\mathrm{SU}(1,1)$. The Haar measure and the Haar weight should be thought of as dual objects to each other, in the sense of the duality between spaces and algebras of functions on them. As we will see in Section~\ref{sec:locallycompactquantumgroup}, the Haar weight is the one that's better suited to use for quantum groups. Left- and right-invariance of the Haar measure is encoded into the Haar weight as follows. Let $L^\infty(\mathrm{SU}(1,1))_*$ denote the predual\footnote{\label{footnote:predualdefinition}The predual of a von Neumann algebra is the space of bounded linear functionals on the von Neumann algebra that are $\sigma$-weakly continuous. It can be shown that it is a Banach space, and that $L^{\infty}(\mathrm{SU}(1,1)) = (L^{\infty}(\mathrm{SU}(1,1))_*)^*$, hence the name predual \cite{CK2010}.} of $L^\infty(\mathrm{SU}(1,1))$, which is isomorphic to $L^1(\mathrm{SU}(1,1))$ via the map $\omega_{\bullet}: L^1(\mathrm{SU}(1,1)) \to L^\infty(\mathrm{SU}(1,1))_*$, $k \mapsto \omega_k$ given by $\omega_k(x) = \int d\mu(g) x(g)k(g)$ \cite{CK2010}. Left- and right-invariance of the Haar measure is now equivalent to
\begin{equation}
    \label{eq:classicalleftrightinvariance}
    \begin{split}
        \phi((\omega_k\otimes \id)\Delta(x)) &= \int d\mu(g)d\mu(h) x(gh)k(g) = \int d\mu(g)d\mu(h) x(h)k(g) = \phi(x)\omega_k(1),\\
        \phi((\id\otimes \omega_k)\Delta(x)) &= \int d\mu(g)d\mu(h) x(gh)k(h) = \int d\mu(g)d\mu(h)x(g)k(h) = \phi(x)\omega_k(1),
    \end{split} 
\end{equation}
for all $k\in L^1(\mathrm{SU}(1,1))$ and $x\in L^{\infty}(\mathrm{SU}(1,1))^+$. The above properties are the essential ingredients for defining the locally compact quantum group, as we will see in Section~\ref{sec:locallycompactquantumgroup}.\\

The (equivalence classes of) square-integrable functions $L^2(\mathrm{SU}(1,1))$ with respect to the Haar measure correspond to the Gelfand-Naimark-Segal (GNS) construction of $L^\infty(\mathrm{SU}(1,1))$ for the Haar weight $\phi$. In particular, it is known that the essentially bounded functions define an action on the square-integrable functions $f\in L^{2}(\mathrm{SU}(1,1))$ through point-wise multiplication, given by the map
\begin{equation}
    \label{eq:classicalLinftyaction}
    \pi \colon L^\infty(\mathrm{SU}(1,1))\to \B(L^2(\mathrm{SU}(1,1))), \quad \pi(x)f = xf.
\end{equation}
If we now consider the subspace $\mathcal{N}_\phi \coloneq \{x \in L^{\infty}(\mathrm{SU}(1,1)) : \phi(x^*x) < \infty\}$ and the map $\Lambda\colon \mathcal{N}_\phi \to L^2(\mathrm{SU}(1,1))$ given by $\Lambda(x) = x$, then $(L^2(\mathrm{SU}(1,1)), \pi, \Lambda)$ is a GNS construction for the Haar weight $\phi$. That is, we have the following properties:
\begin{enumerate}[label = (\alph*)]
    \item $\Lambda(\mathcal{N}_{\phi})$ is dense in $L^2(\mathrm{SU}(1,1))$;
    \item $\langle \Lambda(a),\Lambda(b)\rangle = \phi(a^*b)$ for all $a,b\in \mathcal{N}_\phi$;
    \item $\pi(x)\Lambda(a) = \Lambda(xa)$ for all $x\in L^\infty(\mathrm{SU}(1,1))$ and $a\in \mathcal{N}_{\phi}$.
\end{enumerate}
Physically, one can interpret the GNS-map $\Lambda$ as the state-operator correspondence between the square-integrable functions and the bounded functions. Also, one might have noticed that we have not made any remark on the antipode and counit in this von Neumann algebraic picture. However, it turns out that in the analytical setting, the left- and right-invariant Haar weights already contain the information about the antipode and the counit \cite{CK2010, KV2000}, and thus they do not need to be defined separately.\\

We will end this section with a discussion about the regular representations of $\mathrm{SU}(1,1)$ in the \textit{Cartan decomposition}. Importantly, the Cartan decomposition can be parametrised through coordinates $\theta\in [0,2\pi)$, $\psi \in [-2\pi, 2\pi)$ and $\rho \geq 0$. That is, these coordinates define a one-to-one parametrisation of the group elements in $\mathrm{SU}(1,1)$ \eqref{eq:definitionsu11} given by \cite{VK1991}
\begin{equation}
    \label{eq:classicalCartanDecomp}
    g(\theta, \rho , \psi) \coloneq e^{i\frac{\theta}{2} H}e^{\frac{\rho}{2}(E+F)}e^{i\frac{\psi}{2}H} = \begin{bmatrix}
        e^{i\frac{\theta+ \psi}{2}}\cosh \frac{\rho}{2} & e^{i\frac{\theta - \psi}{2}}\sinh \frac{\rho}{2}\\
        e^{i\frac{\psi - \theta}{2}}\sinh \frac{\rho}{2} & e^{-i\frac{\theta + \psi}{2}}\cosh \frac{\rho}{2}
    \end{bmatrix}.
\end{equation}
Through this parametrisation, any function $f \colon \mathrm{SU}(1,1)\to \C$ is equivalent to a function $f\colon [0,2\pi) \times [0,\infty) \times [-2\pi, 2\pi)\to \C$ given by $f(\theta, \rho, \psi) \coloneq f(g(\theta, \rho, \psi))$.
The Haar measure, and therefore the inner product on these functions, can then also be expressed in terms of this parametrisation by \cite{VK1991}
\begin{equation}
    \label{eq:classicalCartanHaarmeasure}
    \langle f , k\rangle \coloneq \int_{\mathrm{SU}(1,1)} d\mu(g)\overline{f(g)}k(g) = \int_0^{2\pi}\frac{d\theta}{2\pi}\int_{-2\pi}^{2\pi}\frac{d\psi}{4\pi} \int_0^{\infty}d\rho\sinh \rho~\overline{f(\theta, \rho, \psi)} k(\theta, \rho, \psi),
\end{equation}
which is defined up to a normalisation factor. Equation \eqref{eq:algebraAction} now defines an action of the universal enveloping algebra $\U$ on an appropriate dense subspace of the square-integrable functions. For example, for the Cartan element $H$, it can be readily verified that
\begin{equation}
    (H\triangleright f)(\theta, \rho, \psi) = 2i\frac{d}{dt}\bigg|_{t=0}f(\theta, \rho, \psi + t) = 2i \partial_\psi f(\theta, \rho, \psi).
\end{equation}
Of course, such an action is only well-defined if $f$ is a smooth function in $\psi$. Moreover, the action of the universal enveloping algebra is generally given by unbounded operators, such that one has to restrict this action to the dense subset of compactly supported and smooth functions. The left action of the universal enveloping algebra can then be derived to be given by \cite{K2018}
\begin{equation}
    \label{eq:classicalCartanactions}
    \begin{split}
        (H\triangleright f)(\theta, \rho, \psi) &= 2i\partial_\psi f (\theta, \rho, \psi),\\
        (E\triangleright f)(\theta, \rho, \psi) &= -e^{-i\psi}\left(\partial_\rho - \frac{\cosh \rho}{\sinh \rho} (i\partial_\psi) + \frac{1}{\sinh \rho}(i\partial_\theta)\right)f(\theta, \rho, \psi),\\
        (F\triangleright f)(\theta, \rho, \psi) &= -e^{i\psi}\left(\partial_\rho + \frac{\cosh \rho}{\sinh \rho}(i\partial _\psi) - \frac{1}{\sinh \rho}(i\partial_\theta)\right)f(\theta, \rho, \psi),
    \end{split}
\end{equation}
with a similar expression for the right action. In Section~\ref{sec:quantumregularrepresentations}, we will discuss the quantum analogues of these regular representations.

\subsection{The Gauss decomposition}
\label{sec:classicalGaussdecomp}
In this section, we will discuss the Gauss decomposition of the Lie group, as it is exactly in this decomposition that one can reduce to Liouville quantum mechanics. Indeed, as we will also see in Section~\ref{sec:classicalNormaliserandLiouvilleTheory}, Liouville quantum mechanics can be obtained from a particle-on-$\mathrm{SU}(1,1)$ model by constraining the states to have fixed left and right actions from the mixed \textit{parabolic elements}\footnote{In this paper, by `parabolic elements' we mean the generators of the one-parametric subgroups in a parabolic conjugacy class.} \cite{DVV1992}. We will discuss the Gauss decomposition in terms of the space of functions, rather than the usual group elements, such that the generalisation to the quantum group becomes manifest.\\ 

The parabolic elements are related to the generators of $\U(\mathfrak{su}(1,1))$, as introduced in the previous section, via the Cayley transformation. That is, let us consider a different set of generators of the universal enveloping algebra $\U(\mathfrak{su}(1,1))$, given by
\begin{equation}
    \label{eq:classicalParabolicElements}
    \tilde E \coloneq \frac{1}{2i}\left[E - F - H\right], \quad \tilde F \coloneq \frac{1}{2i}\left[E-F+H\right], \quad \tilde H \coloneq E+F.
\end{equation}
It is easily verified that the above generators also satisfy the commutation relations for $\mathfrak{sl}_2$ Lie algebra, as given in \eqref{eq:sl2algebra}. Moreover, their (formal) adjoints are given by $\tilde E^* = -\tilde E$, $\tilde F^* = -\tilde F$ and $\tilde H^* = - \tilde H$. The anti-self-adjoint elements $\tilde E$ and $\tilde F$ are therefore generators of one-parametric subgroups of $\mathrm{SU}(1,1)$ in parabolic conjugacy classes.\\

The group $\mathrm{SU}(1,1)$ now has the property that any element in the vicinity of the identity admits a Gauss decomposition \cite{BarutRaczka1977TheoryGroup, ForgacsWipfBalogFeherORaifeataigh}. We denote $\mathrm{SU}^\circ(1,1)$ for the subset of elements that can be written in the form
\begin{equation}
    \label{eq:classicalGaussDecomposition}
    g(\gamma_L, \varphi, \gamma_R) = e^{\gamma_L \tilde F} e^{\varphi \tilde H}e^{\gamma_R \tilde E}, \quad \text{for } \gamma_L, \varphi, \gamma_R \in \R,
\end{equation}
such that the group $\mathrm{SU}(1,1)$ can be covered by the four (overlapping) patches $\pm \mathrm{SU}^\circ(1,1)$, $\pm \mathrm{SU}^\circ(1,1)w$, where $w = \left[\begin{smallmatrix}
    i & 0\\ 0 & -i
\end{smallmatrix}\right]$ is the Weyl element\footnote{We will call this particular representative of the non-trivial element of the Weyl group $W \simeq \mathbb{Z}_2$ for $\mathrm{SU}(1,1)$ (with respect to the standard split real torus) the {\it Weyl element}, to shorten notations.}. Moreover, $\mathrm{SU}^\circ(1,1) \cup \left(-\mathrm{SU}^\circ(1,1)\right)$ is dense in $\mathrm{SU}(1,1)$ \cite{VK1991}.\footnote{This therefore also implies that $\mathrm{SU}^\circ(1,1)$ is isomorphic to a dense subset of the projective group $\mathrm{PSU}(1,1) \coloneq \mathrm{SU}(1,1)/\{\pm \mathbb{I}\}$.} Any function $f:\,\mathrm{SU}(1,1)\to \C$, when restricted to this dense subset, can therefore be written as a pair of functions $f^\pm(\gamma_L, \varphi, \gamma_R) \coloneq f(\pm g(\gamma_L, \varphi, \gamma_R))$ of the coordinates $\gamma_{L}, \varphi, \gamma_R\in \R$.\\

To have a well-defined action of the universal enveloping algebra, we again consider a dense subspace (of the $L^2$ space) of functions which are smooth and compactly supported. Note that for this function subspace, it is indeed enough to know a function on a dense subset (including the identity) of $\mathrm{SU}(1,1)$ to reconstruct it on the whole group. Equation~\eqref{eq:algebraAction} then defines a left and right action of the parabolic elements $\tilde E$ and $\tilde F$, given by
\begin{equation}\label{eq:EF_as_derivatives}
    \tilde E\triangleright f = \partial_{\gamma_R} f \quad \text{and}\quad f\triangleleft \tilde F = \partial_{\gamma_L} f.
\end{equation}
As noted before, Liouville quantum mechanics can be obtained from the representation theory of $\mathrm{SU}(1,1)$ by considering eigenfunctions with respect to these left and right actions of the parabolic elements. Therefore, by using the spectral decomposition of the two commuting operators in \eqref{eq:EF_as_derivatives}, we Fourier-transform the functions in the coordinates $\gamma_L$ and $\gamma_R$ to their conjugate coordinates $p_L$ and $p_R$, respectively, such that the functions $f \in L^2(\mathrm{SU}(1,1))$ can be decomposed into coefficient functions $f_{p_L, p_R}(\varphi)$, which satisfy $\tilde E\triangleright f_{p_L, p_R}(\varphi) = ip_R f_{p_L, p_R}(\varphi)$ and $f_{p_L, p_R}(\varphi)\triangleleft \tilde F = ip_L f_{p_L, p_R}(\varphi)$. The inner product for such functions, when restricted to $\mathrm{SU}^\circ(1,1)$, is given by \cite{VK1991}
\begin{equation}
    \label{eq:innerProductGauss}
        \langle f,g\rangle_{p_L, p_R} =\int_{-\infty}^{\infty}d\varphi e^{2\varphi} \overline{f_{p_L, p_R}(\varphi)} g_{p_L, p_R}(\varphi)
\end{equation}
where we denote by $\langle\cdot,\cdot\rangle_{p_L, p_R}$ the inner product descended to the subspace of functions with fixed mixed parabolic eigenvalues $p_L$ and $p_R$.\\

The biregular (i.e. left $\otimes$ right) representation of $\mathrm{SU}(1,1) \times \mathrm{SU}(1,1)$ acting on the space of square-integrable functions decomposes into irreducible representations as
\begin{equation}
    L^2(\mathrm{SU}(1,1)) \cong \int^{\oplus} d\mu(\pi) \pi \otimes \pi^*,
\end{equation}
where the direct integral is over all the (equivalence classes of) irreducible representations of $\mathrm{SU}(1,1)$ that appear in the regular square-integrable representation (i.e. {\it tempered)}. Moreover, we introduce the \textit{Plancherel measure} $d\mu(\pi)$ on this (tempered) unitary dual. For a given irreducible representation $\pi$, let $|p_L\rangle_\pi$ and $|p_R\rangle _\pi$ be the corresponding eigenstates of the parabolic elements $\tilde F$ and $\tilde E$, respectively, which form (generalised) bases of the representations. By the Plancherel theorem, any (coefficient) function $f_{p_L, p_R}(\varphi)$ can then be decomposed in terms of the \textit{mixed matrix elements}\footnote{We remark here that, regarded in the context of the $\mathrm{SU}(1,1)$ representation theory as here, these are not the commonly used matrix elements indexed by elements of a (generalised) ordered basis of a corresponding irreducible representation, since their definition also includes the overlap amplitude between two different bases. In this paper, we will still refer to them as (mixed) matrix elements. One well-known example of such mixed matrix elements are the so-called {\it Whittaker functions} \cite{Jacquet1967Whittaker, Hashizume1979WhittakerMF, Hasizume1982Whittaker}.} $\pi_{p_L,p_R}(\varphi) \coloneq \langle p_L | \pi(\exp{\varphi \tilde H} )|p_R\rangle_\pi$ as $f_{p_L, p_R}(\varphi) = \int d\mu(\pi) f^\pi\pi_{p_L, p_R}(\varphi)$. Moreover, by Schur orthogonality, the inner product between two matrix elements is given by
\begin{equation}
    \langle \pi^1_{a,b}, \pi^2_{c,d}\rangle = \delta(a-c)\delta(b-d)\delta_{\mu(\pi)}(\pi^1 - \pi^2),
\end{equation}
where the subscripts $a,b,c,d$ refer to different (generalised) basis elements of the representation spaces for $\pi^1$ and $\pi^2$, and $\delta_{\mu(\pi)}$ is the delta function with respect to the Plancherel measure. Explicitly, for a family of irreducible representations parametrised by a continuous parameter $s$, the delta function is given by 
\begin{equation}
    \label{eq:Plancherelmeasureformula}
    \delta_{\mu(\pi^s)}(\pi^s - \pi^{s'}) = \left(\frac{d\mu(\pi^s)}{ds}\right)^{-1}\delta(s - s').
\end{equation} 
For discrete families of representations, the Dirac delta function is naturally replaced by a Kronecker delta function. 

\subsection{The normaliser and Liouville QM}
\label{sec:classicalNormaliserandLiouvilleTheory}
It is known that the eigenstates in Liouville quantum mechanics correspond to matrix elements of the principal unitary series of $\mathrm{SU}(1,1)$ in the Gauss decomposition \cite{DVV1992, BMV2018, BMV2019}. However, as we will see in Section~\ref{sec:reductiondsSYK}, in the $q$-deformed case, the eigenstates of the double-scaled SYK model correspond to matrix elements of the normaliser $\mathrm{SU}_q(1,1)\rtimes \Z_2$. As the $q\to 1$ limit is then known to describe the Liouville quantum mechanics, one might wonder what the role of the normaliser is in this case. In this section, we will argue that the matrix elements in the normaliser have a natural interpretation of describing states in the two-sided wormhole. Recall that the normaliser of $\mathrm{SU}(1,1)$ in $\mathrm{SL}(2,\C)$ is given by
\begin{equation}
    \label{eq:definitionnormalisersu11}
    \mathrm{SU}(1,1)\rtimes \Z_2 \coloneq \mathcal{N}_{\mathrm{SL}(2,\C)}(\mathrm{SU}(1,1)) = \mathrm{SU}(1,1)\cup \mathrm{SU}(1,1)v,
\end{equation}
where $v = \left[\begin{smallmatrix}
    0 & 1\\ -1 & 0
\end{smallmatrix}\right]$ corresponds to the generator of $\Z_2$ in the semidirect product, which acts on $\mathrm{SU}(1,1)$ by conjugation. Note that this group is distinct from the often considered double cover of $\mathrm{SU}(1,1)$, the metaplectic group, which corresponds to a non-trivial group extension of $\mathrm{SU}(1,1)$, whereas the group we consider here is just a semidirect product with $\Z_2$.\\

Let us begin by considering the action of the normaliser on the \textit{AdS$_2$ hyperboloid}\footnote{We emphasise that we do not consider the `global AdS' (i.e. the universal cover of that hyperboloid) here. We still refer to the patch in global coordinates that covers the entire hyperboloid as the \textit{global patch}.}. Recall that AdS$_2$ can be realised as an embedding in $\R^{2,1}$ with the additional constraint $-T_1^2 - T_2^2 + X^2 = -1$, where we set the AdS radius to $1$ for simplicity. Equivalently, the AdS$_2$ coordinates correspond to the set of symmetric matrices $\mathbf{X} = \left[\begin{smallmatrix}
    X + T_2 & T_1\\ T_1 & X-T_2
\end{smallmatrix}\right]$ with $\det \mathbf{X} = -1$. The (transitive) action of $\mathrm{SL}(2,\R)$\footnote{Throughout this section, we will consider the equivalent (by Cayley transform) explicit realisation of $\mathrm{SL}(2,\R)$, instead of $\mathrm{SU}(1,1)$, as it is more natural for this context. It should be noted, however, that, unlike the classical case, the $q$-deformations $\mathrm{SL}_q(2,\R)$ and $\mathrm{SU}_q(1,1)$ are not isomorphic \cite{QGAR}.  In the $\mathrm{SL}(2,\R)$ realisation, the explicit matrix forms of the parabolic elements that we use are given by $\tilde E = \left[\begin{smallmatrix}
    0 & 1\\ 0 & 0
\end{smallmatrix}\right]$ and $\tilde F = \left[\begin{smallmatrix}
    0 & 0\\ 1 & 0
\end{smallmatrix}\right]$.} on AdS$_2$ is then given by $g\triangleright\mathbf{X} \coloneq g\mathbf{X}g^{T}$. It can be verified that the fixed points of the action of the parabolic elements $\tilde E$ and $\tilde F$ of $\mathrm{SL}(2,\R)$ are given by $(T_1, T_2, X) = (0,\infty,\infty)$ and $(T_1, T_2, X) = (0,\infty,-\infty)$, respectively, which correspond to points on the right and left boundaries of the AdS$_2$ hyperboloid. In particular, we can identify the left and right boundaries of AdS$_2$ with the cosets $\mathrm{SL}(2,\R)/\overline{B}$ and $\mathrm{SL}(2,\R)/B$, respectively, where $B$ is the Borel subgroup\footnote{The Borel subgroup $B$ of $\mathrm{SL}(2,\R)$ is the subgroup of upper triangular matrices. Moreover, we denote by $\overline B$ the opposite (by Weyl reflection) of this subgroup, i.e. the subgroup of lower triangular matrices.} of $\mathrm{SL}(2,\R)$. We then define the states on the left and right boundaries to have a fixed eigenvalue under the action of $\tilde F$ and $\tilde E$, respectively, where the eigenvalues correspond to the boundary conditions. The action of $\tilde v = \left[\begin{smallmatrix}
    i & 0\\ 0 & -i
\end{smallmatrix}\right]$, where $\tilde v$\footnote{Here, $\tilde{v}$ denotes the corresponding Cayley-transformed generator of $\Z_2$ in the semidirect product $\mathrm{SL}(2,\R)\rtimes \Z_2$, i.e. the one relevant for $\mathrm{SL}(2,\R)$ as opposed to $\mathrm{SU}(1,1)$.} is the generator of $\Z_2$ for the normaliser $\mathrm{SL}(2,\R)\rtimes \Z_2$, can be calculated to be given by $\tilde v\triangleright(T_1, T_2, X) = (T_1, -T_2, -X)$. Therefore, $\tilde v$ is also an isometry on AdS$_2$ and corresponds to a reflection. Importantly, $\tilde v$ maps between the left and right boundaries. We emphasise that the extension to the normaliser is necessary to get such a mapping between the two boundaries, as $\mathrm{SL}(2,\R)$ only acts transitively on each boundary independently, but does not map between them. For example, the action of the Weyl element $\tilde w = \left[\begin{smallmatrix}
    0  & 1\\
    -1 & 0
\end{smallmatrix}\right] \in \mathrm{SL}(2,\R)$ is given by $\tilde w \triangleright (T_1, T_2, X) = (-T_1, -T_2, X)$, i.e. a reflection in the temporal coordinates but not in the spatial coordinate.\\
\begin{figure}
    \centering
    \includegraphics[width=0.95\textwidth]{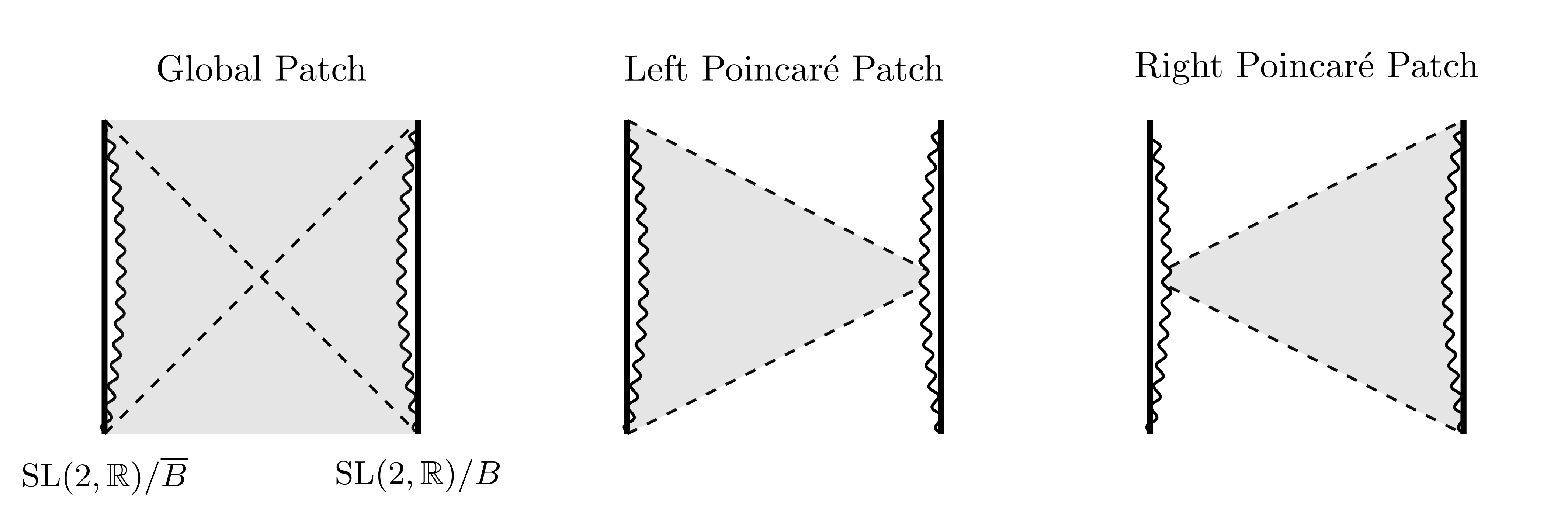}
    \caption{The global patch, the left Poincaré patch and the right Poincaré patch of nearly-AdS$_2$ space. The regions that are covered by the different patches are coloured in grey. Moreover, the left and right boundaries are identified with $\mathrm{SL}(2,\R)/\overline{B}$ and $\mathrm{SL}(2,\R)/B$, respectively.}
    \label{fig:NAdS2patches}
\end{figure}

To describe gravity on AdS$_2$, one needs to impose additional asymptotic boundary conditions \cite{AP2015, MSY2016, EMV2016, IliesiuPufu2019exactquantization} (See \cite{MT2023, Turiaci2025leshoucheslecturestwodimensional} for reviews). In particular, one introduces a UV cutoff at the two boundaries, turning the AdS$_2$ space into what is called \textit{nearly-AdS$_2$ space}. These boundary cutoffs are described by boundary trajectories near the left and right boundaries, given in terms of the coordinates on the Poincaré patch. It should be noted, however, that the Poincaré patch does not cover the entire AdS$_2$ hyperboloid and has only one boundary. Therefore, one needs a left Poincaré patch to describe the left boundary trajectory and a right Poincaré patch to describe the right boundary trajectory. The global and left/right Poincaré patches of nearly-AdS$_2$ space are shown in Figure~\ref{fig:NAdS2patches}. The Liouville Hamiltonian is then given in terms of the geodesic length between the two boundary trajectories \cite{HJ2020}. Let $\H_L$, $\H_R$ be the Hilbert spaces on the left and right boundaries and let $|p_L\rangle \in \H_L$ and $|p_R\rangle \in \H_R$ be the states that correspond to some left and right boundary conditions, i.e. are eigenstates of $\tilde F$ and $\tilde E$, respectively.\footnote{In the Schwarzian theory, the action of $\mathrm{SL}(2,\R)$ on the boundary trajectories is a gauge symmetry. In that context, picking eigenstates of the parabolic elements corresponds to choosing a gauge on the left and right boundaries.} Since $\mathrm{SL}(2,\R)$ acts as isometries on AdS$_2$, the state corresponding to the two boundary trajectories being separated by some geodesic distance must be given in terms of matrix elements of $\mathrm{SL}(2,\R)$.\footnote{Indeed, let $|h_L\rangle$ and $|h_R\rangle$ be highest weight states of $\H_L$ and $\H_R$, respectively, and suppose that $g_L|h_L\rangle$ and $g_R|h_R\rangle$ are two boundary states. Since $\mathrm{SL}(2,\R)$ acts as isometries on AdS$_2$, the geodesic distance between the two boundaries corresponds to a $\mathrm{SL}(2,\R)$-invariant mapping $\mathbf{d} \colon \H_L\otimes \H_R \to \R$. In particular, it satisfies $\mathbf{d}(g\cdot g_L|h_L\rangle, g\cdot g_R|h_R\rangle)=\mathbf{d}(g_L|h_L\rangle, g_R|h_R\rangle)$, and thus can only depend on $g_L^{-1}g_R$. Therefore, we can identify $\mathbf{d}$ with another function (depending on the initially chosen highest weight states) $\tilde{\mathbf{d}} \colon \mathrm{SL}(2,\R) \to \R$ such that $\tilde{\mathbf{d}}(g_L^{-1}g_R) = \mathbf{d}(g_L|h_L\rangle, g_R|h_R\rangle)$, which, when square-integrable, can be decomposed in terms of matrix elements as discussed in Section~\ref{sec:classicalGaussdecomp}.} Importantly, however, these can not simply be the matrix elements of the form $\langle p_L | \cdot | p_R\rangle$, as the states $|p_L\rangle$ and $|p_R\rangle$ are elements of the different Hilbert spaces $\H_L$ and $\H_R$. Indeed, the right boundary trajectory is defined only on the right Poincaré patch, which does not contain any information about the left boundary of the AdS$_2$ hyperboloid. Therefore, to take an inner product between the two boundary states, one needs to additionally map the state on the right boundary to the Hilbert space on the left boundary.  Note that such a mapping is exactly given by the action of $\tilde v$. In particular, we thus argue that the geodesic distance between the boundary trajectories is related to the function $\varphi \mapsto \langle p_L| \exp(\varphi \tilde H)\tilde v |p_R\rangle$, i.e. a matrix element of the normaliser $\mathrm{SL}(2,\R)\rtimes \Z_2$.\\

Additionally, let us discuss why it is natural to consider the normaliser for the Gauss decomposition on the level of the representations. For $n\geq 2$, consider the positive discrete series $\D_n^{+}$ of $\mathrm{SL}(2,\R)$. One of its most used realisations is given by \cite{K1986}
\begin{equation}
    \label{eq:positivediscreteseries}
    \D_n^{+}\begin{pmatrix}
        a & b\\ c& d
    \end{pmatrix}f(z)  = (cz+d)^{-n}f\left(\frac{az+b}{cz+d}\right) \quad \text{for } \begin{pmatrix}
        a & b\\ c& d
    \end{pmatrix} \in \mathrm{SL}(2,\R)
\end{equation}
on the space of the square-integrable holomorphic functions on the upper half-plane $\mathbb{H}^+ = \{z\in \C : \text{Im}\ z > 0\}$ with $\|f\|^2 = \int_{\mathbb{H}^+}|f(z)|^2y^{n-2}dxdy < \infty$ for $z = x+iy$. Note here that $\D_n^+(\exp(t\tilde E))f(z) = f(z+t)$, such that $\tilde E$ is diagonalised, in the sense of generalised eigenvectors, by the functions $h_p(z) = \exp(ipz)$ with eigenvalue $ip$. Importantly, these functions are only square-integrable (in the delta-function sense) on the upper half-plane for $p\geq0$. Therefore, the spectrum of $\tilde E$ in the positive discrete series only contains positive momenta, as was e.g. also calculated for the infinitesimal representations of the universal enveloping algebra in \cite{G2003}. Similarly, the spectrum of $\tilde F$ only contains negative momenta. The action for the negative discrete series $\D_n^-$ is the same as \eqref{eq:positivediscreteseries}, but now defined on the space of holomorphic functions on the lower half-plane. In particular, the spectrum of $\tilde E$ for the negative discrete series contains only negative momenta, while the spectrum of $\tilde F$ contains positive momenta. Physically, the positive momenta of $\tilde E$ can be interpreted to correspond to forward-directed trajectories, while the negative momenta correspond to backwards-directed trajectories. In the context of the AdS$_2$ patches, these would therefore correspond to particles on the right and left Poincaré wedges, respectively.\\

Recall from Section~\ref{sec:classicalGaussdecomp} that these momenta correspond to the Fourier basis for the Gauss decomposition. In particular, one expects the full spectrum of the parabolic elements, i.e. both positive and negative momenta, to appear. To get this full spectrum, one therefore always has to consider a direct sum of both the positive and negative discrete series $\D_n^+\oplus \D_n^-$. However, depending on the physical situation, one might expect or require the full set of eigenvalues to appear within a single irreducible representation. Again, this is where the normaliser comes into play. Namely, the action of $\tilde v$ on the space of functions is given by $\tilde v.f(z) \sim f(-z)$, i.e. it maps a function defined on the upper half-plane to a function on the lower half-plane. In particular, $\tilde v$ maps between the positive and negative discrete series, such that the direct sum $\D_n^+\oplus \D_n^-$ for $\mathrm{SL}(2,\R)$ corresponds to an irreducible representation for the normaliser.\footnote{Considering this specific combination of representations is also natural from the purely representation-theoretical point of view on our original group: it forms an {\it L-packet} for SL$(2, \mathbb{R}) \simeq \text{SU}(1,1)$ corresponding to discrete series \cite{AdamsBarbaschVogan2021}.} Therefore, the irreducible discrete series representations for the normaliser do contain the full spectrum of eigenvalues for the parabolic elements.\\

Next, we will consider the eigenstates of the Casimir operator in the infinitesimal action of the universal enveloping algebra. The action of the Casimir element \eqref{eq:classicalCasimir} in the Gauss decomposition can be calculated to be \cite{BMV2018}
\begin{equation}
    \label{eq:classicalCasimirAction}
    (-\Omega\triangleright f_{p_L, p_R})(\varphi) = \left(-\frac{1}{4}\partial_\varphi^2 - \frac{1}{2}\partial_{\varphi} + p_Lp_R e^{-2\varphi}\right)f_{p_L, p_R}(\varphi).
\end{equation}
Under the redefinition $f \to e^{\varphi}f$, this action reduces to the Liouville Hamiltonian\footnote{In our convention for the Casimir element, the Hamiltonian $\hat H$ is equal to minus the Casimir, i.e. $\hat H = -\Omega$.} with potential $V = p_Lp_R e^{-2\varphi}$. In the case where $p_Lp_R < 0$, this potential is unbounded from below, but the Casimir operator can be shown to have a self-adjoint extension where the eigenfunctions are given by the matrix elements of the discrete and principal unitary series, and the inner products between those eigenstates are exactly given by the Plancherel measure of the regular representations of $\mathrm{SU}(1,1)$ \cite{PR2016, MS2025, ALNW2018}. In the case $p_Lp_R>0$, it follows from our discussion on the discrete series that the Casimir operator cannot be diagonalised by matrix elements of the discrete series, as the positive (or negative) momentum states for $\tilde E$ and $\tilde F$ reside on the different half-planes. Therefore, only the principal unitary series (both spherical and non-spherical) should show up here. Indeed, for $p_Lp_R > 0$, the Casimir operator can be diagonalised via the Liouville eigenfunctions\footnote{Here, $\P^s$ refers to the principal unitary series of $\mathrm{SL}(2,\R)$ (see e.g. \cite{K1986}). In the context of the previous section, the notation $\P^s_{p_L, p_R}$ therefore corresponds to a matrix element for this irreducible representation.}
\begin{equation}
    \label{eq:principalunitaryeigenfunctions}
    \P^s_{p_L, p_R}(\varphi) = e^{-\varphi} K_{2is}(2\sqrt{p_L p_R} e^{-\varphi}),
\end{equation}
where $K_\alpha$ is the modified Bessel function of the second kind and $s\geq 0$ \cite{DVV1992, BMV2018, MS2025}. It can be easily verified that the corresponding eigenvalues are given by $\frac{1}{4}+s^2$, which indeed correspond to the principal unitary series. Moreover, the inner product between functions $\P^s_{p_L, p_R}$ is given by \cite{BMV2018, MS2025}
\begin{equation}
    \begin{split}
        \langle \P^s , \P^{s'}\rangle_{p_L, p_R} &= \int_{-\infty}^{\infty} d\varphi K_{2is}(2\sqrt{p_L p_R}e^{-\varphi})K_{2is'}(2\sqrt{p_L p_R}e^{-\varphi}) = \frac{\pi^2}{8s\sinh(2\pi s)}\delta(s - s').
    \end{split}
\end{equation}
Although on the right side one can read off (via \eqref{eq:Plancherelmeasureformula}) the correct $s\sinh 2\pi s$ measure that appears as the density of states in Schwarzian theories \cite{M2018, BMV2018, BMV2019, SW2017}, this is not the Plancherel measure, given by $s\tanh \pi s$ \cite{K2018}, that appears in the regular representation of $\mathrm{SL}(2,\R)$. On the other hand, the rescaled wavefunction $\tilde \P^s_{p_L, p_R} = \cosh(\pi s)\P^s_{p_L, p_R}$ would have the correct $s\tanh \pi s$ Plancherel measure, but not the one that corresponds to the Schwarzian theory. In fact, this rescaled wavefunction can be obtained from a mixed matrix elements of the (spherical) principal unitary series \cite{BMV2019}.  From a purely computational perspective, it does not matter whether one uses $\P^s_{p_L, p_R}$ or $\tilde \P^s_{p_L, p_R}$ in the calculation of the partition function and $n$-point functions, as observables ought to be independent of the specific normalisation of the wavefunction.\footnote{Indeed, one obtains the same results if one were to use the rescaled wavefunction $\tilde \P^s_{p_L, p_R}$ in the calculation of the partition function \eqref{eq:Liouvillepartitionfunction} and the $n$-point functions \eqref{eq:Liouville2pointfunction}, \eqref{eq:Liouville4pointfunction}, since $\langle \tilde 0 | \P^s\rangle$, the 3j-symbols \eqref{eq:3jsymbols} and the 6j-symbols \eqref{eq:6jsymbols} will be rescaled by the same $\cosh(\pi s)$ factor. Therefore, all the dynamics of Liouville QM can be derived from the representation theory of pure $\mathrm{SU}(1,1)$, and the normaliser is not strictly necessary in this specific case.} However, in the physical context of the $\mathrm{SL}(2,\R)$ gauge group formulation of JT-gravity, one does expect the density of states and the Plancherel measure to coincide \cite{Lin2018EntanglementJT,BMV2018, BMV2019}.\\

To resolve this issue, we claim that the functions $\P^s_{p_L, p_R}$ with the correct Plancherel measure instead correspond to matrix elements that appear in the regular representation of the normaliser $\mathrm{SL}(2,\R)\rtimes \Z_2$.\footnote{It was advocated in \cite{BMV2018, BMV2019} that Liouville QM is instead described by the semigroup $\mathrm{SL}^+(2,\R)$. However, we argue against this idea. Namely, one of their main arguments for this claim is that the correct Plancherel measure for Liouville QM can be obtained from a $q\to 1$ limit of a result by Ponsot and Teschner \cite{PT1999}. However, as explicitly stated in that same paper (Remark~5 and 6), their result does not have a classical analogue, making this $q\to 1$ limit (with $k=1$) ill-defined. Of course, it is unclear whether a notion of a Plancherel measure for the semigroup even exists; for instance, we are not aware of any sort of developed harmonic analysis theory which would allow one to make sensible statements about spectral decompositions of such (locally compact) semigroups. As is well-known, the usual derivation of the Plancherel theorem goes via passing to the $C^*$ group convolution algebra of a locally compact group, whereby $*$-representations of the algebra are put in bijective correspondence with unitary group representations. Sufficient control over the structural properties of this $C^*$ algebra is then what actually allows one to prove the Plancherel theorem in the locally compact group case. In the case of a semigroup, however, there is no obvious notion of unitarity even to start with, as semigroups do not have inverses, so that the current mathematical status of such conjectural Plancherel theory is not clear to us and likely requires additional assumptions or structure.} Namely, consider the space of square-integrable functions $L^2(\mathrm{SL}(2,\R)\rtimes \Z_2)$, on which the left and right actions of $\mathrm{SL}(2,\R)\rtimes \Z_2$ are given by $g\triangleright f(x) \coloneq f(xg)$ and $f(x)\triangleleft g \coloneq f(gx)$. Note that if we restrict this action to $\mathrm{SL}(2,\R)$, the space of square-integrable functions splits into a direct sum $L^{2}(\mathrm{SL}(2,\R))\oplus L^{2}(\mathrm{SL}(2,\R)\tilde v)$. Each of these two spaces can then be decomposed into its irreducible components, and both contain a direct integral of principal unitary series, which can, in turn, be realised on the spaces $L^2(\R)\oplus L^2(\R)$. This direct sum now has a natural interpretation of containing the Hilbert spaces for the left and right boundaries of the AdS$_2$ hyperboloid. Indeed, the action of $\tilde v$ maps between these two spaces, just as expected from the action of $\tilde v$ on AdS$_2$. We then argue\footnote{We remark here that we have not explicitly calculated the matrix elements for the normaliser of the classical group $\mathrm{SL}(2,\R)$, as, for the purposes of this paper, we are mostly interested in the quantum group case (from which the corresponding $q\to 1$ limit can then be taken). An explicit derivation based on the harmonic analysis of the classical group $\mathrm{SL}(2,\R)\rtimes \Z_2$, circumventing the reference to the quantum group case, therefore still has to be done.} that the Liouville wavefunctions $\P^s_{p_L, p_R}$ \eqref{eq:principalunitaryeigenfunctions} are given by the matrix elements corresponding to $\mathrm{SL}(2,\R)\tilde v$. Indeed, we will show in Section~\ref{sec:reductiondsSYK} that the quantum group equivalent of these matrix elements gives the eigenstates for the double-scaled SYK model. The $q\to 1$ limit then also gives the result for Liouville quantum mechanics. In fact, as we will soon see in Section~\ref{sec:reductiondsSYK}, in the quantum group case, the reduction to the double-scaled SYK model works {\it exclusively} on the level of the normaliser.

\subsection{The partition function and $n$-point functions}
\label{sec:LiouvillePartitionFunction}
We will now recap the calculation of the partition function and $n$-point functions for Liouville quantum mechanics. It was shown that these $n$-point functions are completely determined by the representation theory of $\mathrm{SU}(1,1)$ and given in terms of 3j- and 6j-symbols \cite{MTV2017}. For the derivation of the $n$-point functions, we follow the approach similar to \cite{BMV2018, BMV2019}. Importantly, however, \cite{BMV2018, BMV2019} used the method of `cutting' and `glueing' of the (Euclidean) spacetime to derive these $n$-point functions, which would require some extra care due to the non-compact nature\footnote{In particular, the group volume induced by its Haar measure is, of course, infinite.} of $\mathrm{SU}(1,1)$. Therefore, we phrase it here in a more representation-theoretic way.\\

Consider the functions $|\phi\rangle_{p_L, p_R}(\varphi') = e^{-2\varphi}\delta(\varphi - \varphi')$, which form a generalised basis for the subspace of functions with fixed eigenvalues of parabolic generators with $\langle \varphi|\varphi'\rangle_{p_L,p_R} = e^{-2\varphi}\delta(\varphi - \varphi')$. Under this basis, we have $\langle \varphi |f\rangle = f(\varphi)$ for any function $f_{p_L, p_R}\in L^2(\mathrm{SU}(1,1))$. As discussed in the previous section, the restriction to the functions with a fixed left and right action of the mixed parabolic elements corresponds to states on a two-sided Hilbert space, also known as \textit{wormhole states}, where the coordinate $\varphi$ is related to the geodesic length between two boundaries. Pictorially, a state $|\varphi\rangle_{p_L, p_R}$ is given by
\begin{equation}
    \label{eq:wormholestate}
    \begin{tikzpicture}[baseline=(current  bounding  box.center)]
        \begin{scope}
            \draw[thick, black] (1.2,-0.5) to (1.2,0.5);
            \draw[thick, black] (5.2,-0.5) to (5.2,0.5);
            \draw[thick, black, dotted] (1.2,0) to (5.2,0);
            
            \draw(0,0) node {$|\varphi\rangle_{p_L, p_R} = $};
            \draw(1.2,-0.8) node {\footnotesize $p_L$};
            \draw(5.2,-0.8) node {\footnotesize$p_R$};
            \draw(3.2, -0.3) node {\footnotesize $L = 2\varphi - 2\ln(\epsilon)$};
        \end{scope}
    \end{tikzpicture}
\end{equation}
where we have two boundaries with boundary conditions $p_L$ and $p_R$, separated by the bare geodesic length $L = 2\varphi - \ln(\epsilon)$, and $\epsilon$ is some renormalization parameter such that $L \to 0$ corresponds to $\varphi \to -\infty$. The variable $2\varphi$ is referred to as the renormalised geodesic length  \cite{HJ2020, M2018}. For the classical Lie group discussion in this section, we will assume that the geodesic length is indeed non-negative in that limit. As we will see in Section~\ref{sec:lengthpositivity}, in the von Neumann algebraic description of the quantum group, the length positivity will follow and does not need to be imposed by hand. One can think of the limiting states $|\varphi = -\infty\rangle_{p_L, p_R}$ and $\langle\varphi = -\infty|_{p_L, p_R}$ as the configurations where the boundaries touch, i.e. $L=0$. We will write $|\tilde 0\rangle_{p_L, p_R} = |-\infty\rangle_{p_L,p_R}$, where the tilde refers to the bare geodesic length. Pictorially, these states correspond to
\begin{equation}
    \begin{tikzpicture}[baseline=(current  bounding  box.center)]
        \begin{scope}
            \draw[thick, black] (1.5,0.1) to [bend right=20] (2,-0.1);
            \draw[thick, black] (2,-0.1) to [bend right=20] (2.5,0.1);
            \draw[draw=black, fill=black] (2,-0.1) circle (.05);
            
            \draw(0.2,0) node {$|\tilde 0 \rangle_{p_L, p_R} = $};
            \draw(2,-0.3) node {\footnotesize $L = 0$};
            \draw(1.5,0.3) node {\footnotesize $p_L$};
            \draw(2.5,0.3) node {\footnotesize $p_R$};
        \end{scope}
    \end{tikzpicture} \quad \text{and}\quad \begin{tikzpicture}[baseline=(current  bounding  box.center)]
        \begin{scope}
            \draw[thick, black] (1.5,-0.1) to [bend right=-20] (2,0.1);
            \draw[thick, black] (2,0.1) to [bend right=-20] (2.5,-0.1);
            \draw[draw=black, fill=black] (2,0.1) circle (.05);
            
            \draw(0.2,0) node {$\langle\tilde 0 |_{p_L, p_R} = $};
            \draw(2,0.3) node {\footnotesize $L = 0$};
            \draw(1.5,-0.3) node {\footnotesize $p_L$};
            \draw(2.5,-0.3) node {\footnotesize $p_R$};
        \end{scope}
    \end{tikzpicture}.
\end{equation}
As was noted in \cite{LMRS2023} and \cite{BMV2018, BMV2019}, the partition function can now be calculated by starting with a state of geodesic length $L = 0$, evolving this state over Euclidean time $\beta$, and closing up the circle by taking the inner product with the state of geodesic length $L = 0$ again. The Hamiltonian $\hat H$ that determines this time evolution is related to the Casimir element \eqref{eq:classicalCasimir}, up to a possible additional normalisation. Moreover, to reduce to Liouville QM, we will take the boundary conditions along the whole circle to be the same, and therefore consider $p_L = p = p_R$ \cite{DVV1992}.\footnote{Note that in this case we have $p_Lp_R > 0$. Therefore,  as discussed in the previous section, only the matrix elements for the principal unitary series show up as states in the Hilbert space for Liouville QM. Additionally, the operator algebra gets restricted to the operators that preserve the boundary conditions. Therefore, this `coset construction' with chosen boundary conditions explicitly breaks the `state-operator correspondence' (induced by the GNS map) that is present for the representation theory on the whole group.} From now on, we will not explicitly denote these boundary conditions. Therefore, the partition function is given by
\begin{equation}
    \mathcal{Z}(\beta) \coloneq \langle\tilde 0 | e^{-\beta \hat H} |\tilde 0\rangle=
    \begin{tikzpicture}[baseline={([yshift=-.5ex]current bounding box.center)}]
        \draw[black, thick] (0:0.75) arc (0:360:0.75);
        \draw[draw=black, fill=black] (-90:0.75) circle (0.07);
        \draw[draw=black, fill=black] (90:0.75) circle (0.07);

        \draw[black, ->] (1.1,-0.5) -- (1.1,0.5);

        \draw (-90:1) node {\footnotesize $L = 0$};
        \draw (90:1) node {\footnotesize $L = 0$};
        \draw (1.7,0) node {$e^{-\beta \hat H}$};
    \end{tikzpicture}.
\end{equation}
From the asymptotics of the modified Bessel functions, the Liouville wavefunctions \eqref{eq:principalunitaryeigenfunctions} approach a constant value\footnote{We remark here that $\lim_{\varphi \to -\infty} \P^s_{p_L, p_R}(\varphi) = 0$, such that $\langle \tilde 0 | \P^s\rangle = 0$. However, for the ratio of two functions, we have $\lim_{\varphi \to -\infty} \P^{s_1}_{p_L, p_R}(\varphi)/\P^{s_2}_{p_L, p_R}(\varphi) = 1$, such that the first-order term in the expansion of $\P^s_{p_L, p_R}$ around $\varphi = -\infty$ is independent of $s$. In the calculation of the partition and $n$-point functions, we appropriately normalise the results to obtain a finite answer. Note that this just corresponds to an overall rescaling of the partition function, which does not affect the physical observables.} (independent of $s$) in the limit $\varphi \to -\infty$. The partition function is therefore given by
\begin{equation}
    \label{eq:Liouvillepartitionfunction}
    \mathcal{Z}(\beta) = \int d\mu(\P^s)d\mu(\P^{s'})\langle \tilde 0 | \P^s\rangle \langle \P^s | e^{-\beta\hat  H} |\P^{s'}\rangle \langle \P^{s'} | \tilde 0\rangle \sim \int d\mu(\P^s)e^{-\beta \mathcal{C}(s)},
\end{equation}
where $\P^s$ is the Liouville wavefunction \eqref{eq:principalunitaryeigenfunctions} and $\mathcal{C}(s)$ is the corresponding Casimir eigenvalue. We have also left out the overall constant factor $|\langle \tilde 0|\P^s\rangle|^2$. This, of course, matches the known Schwarzian partition function \cite{DVV1992, M2018, SW2017}.\\

Let us now calculate the 2-point function corresponding to the discrete series, which is pictorially given by
\begin{equation}
    \langle \tilde 0| e^{-\beta \hat H}\O^\ell(\tau_1, \tau_2)| \tilde 0\rangle = \begin{tikzpicture}[baseline={([yshift=-.5ex]current bounding box.center)}]
        \draw[black, thick] (0:0.75) arc (0:360:0.75);
        \draw[draw=black, fill=black] (0:0.75) circle (0.07);
        \draw[draw=black, fill=black] (180:0.75) circle (0.07);
        \draw[black, dashed] (180:0.75) -- (0:0.75);
        
        \draw (0, 0.2) node {\footnotesize $\ell$};
    \end{tikzpicture}
\end{equation}
As we explained in the previous section, the GNS-map $\Lambda$ precisely gives the state-operator correspondence. Therefore, on an appropriate dense subset of $L^2(\mathrm{SU}(1,1)$, its inverse turns a square-integrable function into an operator acting on the square-integrable functions. The corresponding operator action is then just given by `pointwise multiplication' \eqref{eq:classicalLinftyaction}. For matrix elements, pointwise multiplication corresponds to taking a tensor product of representations, that is
\begin{equation}
    \label{eq:pointwisemultiplicationreps}
    \pi^{1}_{a,b}\cdot \pi^{2}_{c,d} = \left(\langle a|_{\pi^1}\otimes \langle c|_{\pi^2}\right)(\pi^1\otimes \pi^2)\left(|b\rangle_{\pi^1} \otimes |d\rangle_{\pi^2}\right),
\end{equation}
for $\pi^1$ and $\pi^2$ irreducible representations. The tensor product can, in turn, be decomposed into its irreducible components, i.e. $\pi^1\otimes \pi^2 \cong \int_{[\pi^1\otimes \pi^2]}^{\oplus} d\nu(\pi)\pi$, where $[\pi^1\otimes \pi^2]$ denotes the support (on the unitary dual) of the spectral measure $\nu$ for this tensor product decomposition. Note here that this measure $\nu$ is not assumed to be equal to the Plancherel measure. If an irreducible representation is (weakly) contained in the tensor product decomposition, we will denote this by $\pi \subset \pi^1\otimes \pi^2$. Moreover, the basis of such a representation can be written in terms of tensor products of the bases of $\pi^1$ and $\pi^2$, with coefficients given by \textit{Wigner 3j-symbols} or \textit{Clebsch-Gordan coefficients}. Precisely, for $\pi\subset \pi^1\otimes \pi^2$, the overlap between basis elements is given by
\begin{equation}
    \label{eq:3jsymbols}
    \begin{tikzpicture}[baseline={([yshift=-.5ex]current bounding box.center)}]
        \begin{scope}
            \draw[black, thick] (0,-0.5) to (0,0.5);
            \draw[black, thick] (-0.75,0) to (0,0);
            \draw[draw=black, fill=black] (0,0) circle (.05);

            \draw (0,0.65) node {\footnotesize $\pi$};
            \draw (0,-0.65) node {\footnotesize $\pi^2$};
            \draw (-1,0) node {\footnotesize $\pi^1$};
            \node[align=left, anchor=west] at (-0.05, -0.25) {\footnotesize $b$};
            \node[align=left, anchor=west] at (-0.05, 0.25) {\footnotesize $c$};
            \node at (-0.375, 0.15) {\footnotesize $a$};
        \end{scope}
    \end{tikzpicture} = \langle c|_\pi(|a\rangle_{\pi^1}\otimes |b\rangle_{\pi^2}) \eqcolon \left(\frac{d\mu(\pi)}{d\nu(\pi)}\right)^{\frac{1}{2}}\begin{bmatrix}
        \pi & \pi^1 & \pi^2\\
        c & a & b
    \end{bmatrix}\delta(a+b-c),
\end{equation}
where the delta function appears due to the conservation of momentum.\footnote{More precisely, suppose we have chosen a (generalised) basis $\{|a\rangle\}$ (in the dense subspace of smooth vectors of our irreducible representation) which diagonalises some $X\in \U$, i.e. $X|a\rangle = a|a\rangle$. On the tensor product, this action will be given by $\Delta(X)(|a\rangle_{\pi^1}\otimes |b\rangle_{\pi^2}) = (a+b)(|a\rangle_{\pi^1}\otimes |b\rangle_{\pi^2})$. Therefore, any overlap $\langle c|_\pi(|a\rangle_{\pi^1}\otimes |b\rangle_{\pi^2})$ can only be non-zero if the action of $X$ on $|c\rangle_\pi$ also has eigenvalue $a+b$.} In our definition of the 3j-symbol, we will always assume the (generalised) basis elements $|a\rangle_{\pi^1}$, $|b\rangle_{\pi^2}$ and $|c\rangle_\pi$ to be orthonormal. Moreover, we normalise the 3j-symbols with respect to the Plancherel measure $\mu$. We have also given a pictorial representation of this overlap, which can be thought of as a Feynman diagram vertex for the model's correlation functions. The multiplication of matrix elements \eqref{eq:pointwisemultiplicationreps} can now be written in terms of irreducible constituents and 3j-symbols as
\begin{equation}
    \label{eq:operatorproductexpansion}
     \pi^{1}_{a,b}\cdot \pi^{2}_{c,d} = \int_{[\pi^1\otimes \pi^2]}d\mu(\pi)\overline{\begin{bmatrix}
        \pi & \pi^1 & \pi^2\\
        a+c & a & c
    \end{bmatrix}}\begin{bmatrix}
        \pi & \pi^1 & \pi^2\\
        b+d & b & d
    \end{bmatrix} \pi_{a+c, b+d},
\end{equation}
which can be thought of as the operator product expansion for matrix elements. Here, one can note the appearance of the Plancherel measure in the decomposition due to our normalisation of the 3j-symbols.\\

It can be seen that in general, when we multiply two matrix elements by each other, the boundary conditions will change. For the two-point functions, however, we will require that the boundary conditions remain unchanged. Therefore, we will consider spinless matrix elements corresponding to the discrete series, i.e. $\D^\ell_{00}$ with $\ell \in \frac{1}{2}\Z_{\geq 0}$.\footnote{Since we are not considering the universal cover of AdS$_2$, the `conformal weights' can only take on integer/half-integer values. For a full continuous spectrum of conformal weights, one would instead have to consider the von Neumann algebraic group for the universal cover of the normaliser, i.e. $\widetilde{\mathrm{SU}(1,1)\rtimes \Z_2}$, for which the results in this section are expected to follow analogously.} Recall from the previous section that the discrete series correspond to particles on the Poincaré patch. Therefore, an insertion of the discrete series corresponds to an insertion of matter into the bulk of AdS$_2$. The bilocal operator corresponding to such an insertion at Euclidean time $\tau_{12} = \tau_1 - \tau_2$ is given by
\begin{equation}
    \O^\ell_{00}(\tau_{12}) \coloneq e^{\tau_{12}\hat H}\Lambda^{-1}(\D^\ell_{00})e^{-\tau_{12}\hat H}.
\end{equation}
Using the fact that $\Lambda^{-1}(\D^\ell_{00})|\P^{s}\rangle = |\D^\ell\otimes \P^s\rangle$ and the `operator product expansion' \eqref{eq:operatorproductexpansion}, it readily follows that the two-point function is given by
\begin{equation}
    \label{eq:Liouville2pointfunction}
    \begin{split}
        \langle \tilde 0| e^{-\beta \hat H}\O^\ell_{00}(\tau_{12})| \tilde 0\rangle&\sim \int d\mu(\P^{s_1})d\mu(\P^{s_2})e^{-(\beta - \tau_{12})\mathcal{C}(s_1)}e^{-\tau_{12} \mathcal{C}(s_2)}\langle \P^{s_1} | \Lambda^{-1}(\D^\ell_{00})|\P^{s_2}\rangle\\
        &\sim \int \prod_{n=1}^{2}\left\{d\mu(\P^{s_n})e^{-\beta_n \mathcal{C}(s_n)}\right\}\left|\begin{bmatrix}
            \P^{s_1} &  \D^\ell & \P^{s_2}\\
            p & 0 & p
        \end{bmatrix}\right|^2,
    \end{split}
\end{equation}
where $\beta_1 = \tau_{12}$ and $\beta_2 = \beta - \tau_{12}$. This is, of course, the same two-point function as found in \cite{BMV2018, BMV2019, MTV2017}.\\

Higher Euclidean time-ordered (i.e. non-crossing) correlators can simply be calculated by inserting extra operators $\O^\ell_{00}$ into the inner product, giving additional principal unitary series and 3j-symbols to integrate over. One could, in principle, now also allow operators with non-trivial spin, as long as the final state has a non-zero overlap with the vacuum state, i.e. contains a zero-spin matrix element in its expansion. Note that here we always take tensor products with additional discrete series, and thus a tensor product of two principal unitary series never appears. Therefore, we do not have to worry about possible multiplicities of tensor product decompositions in this context (see e.g. \cite{GroeneveltKoelink2001, VK1991} for the decomposition of tensor products). Physically, this means we only add matter content into the bulk, and never add any boundary particles.\\

Next, we will revisit the calculation of the out-of-time ordered correlator for the discrete series $\D^\ell$ and $\D^{\ell'}$, which is pictorially given by
\begin{equation}
    \label{eq:crossed4point}
    \langle \tilde 0|e^{-\beta \hat H}\O^{\ell_1}(\tau_1,\tau_3) \O^{\ell_2}(\tau_2, \tau_4)|\tilde 0\rangle = \begin{tikzpicture}[baseline={([yshift=-.5ex]current bounding box.center)}]
        \draw[black, thick] (0:0.75) arc (0:360:0.75);
        \draw[draw=black, fill=black] (30:0.75) circle (0.07);
        \draw[draw=black, fill=black] (150:0.75) circle (0.07);
        \draw[draw=black, fill=black] (-30:0.75) circle (0.07);
        \draw[draw=black, fill=black] (210:0.75) circle (0.07);
        \draw[black, dashed] (150:0.75) -- (-30:0.75);
        \draw[black, dashed] (210:0.75) -- (30:0.75);
        
        \draw (-0.3, 0.4) node {\footnotesize $\ell_1$};
        \draw (0.3, 0.4) node {\footnotesize $\ell_2$};
    \end{tikzpicture}
\end{equation}
Here, it can be seen that the ordering of the tensor product is different between the left and right boundaries. For a triple product of representations, there is an (associator) isomorphism $\pi^1\otimes (\pi^2\otimes \pi^3) \cong (\pi^1\otimes \pi^2)\otimes \pi^3$, which gives two distinct ways of writing out the basis elements of its irreducible components. In particular, let $\pi\subset \pi^1\otimes\pi^2\otimes \pi^3$, then there are $\pi^{23} \subset \pi^{2}\otimes \pi^3$ and $\pi^{12}\subset \pi^1\otimes \pi^2$ such that $\pi \subset \pi^1\otimes \pi^{23}$ and $\pi\subset \pi^{12} \otimes \pi^{3}$. We call $\pi^{23}$ and $\pi^{12}$ the intermediate representations. A basis for $\pi$ can now be constructed either through the intermediate representation $\pi^{23}$, or through the intermediate representation $\pi^{12}$. Namely, let $\{|a\rangle_{\pi^1}\}, \{|b\rangle_{\pi^2}\}$ and $\{|c\rangle_{\pi^3}\}$ be orthonormal generalised bases for $\pi^1, \pi^2$ and $\pi^3$ respectively. Then, using the 3j-symbols, we can write
\begin{equation}
    \label{eq:firsttripledecomposition}
    \begin{split}
        &|a\rangle_{\pi^1}\otimes |b\rangle_{\pi^2}\otimes |c\rangle_{\pi^3}\\
        &= \int\limits_{\mathclap{[\pi^2 \otimes \pi^3]}}d\nu(\pi^{23})\int\limits_{\mathclap{[\pi^1\otimes \pi^{23}]}} d\nu(\pi)\left(\frac{d\mu(\pi^{23})}{d\nu(\pi^{23})}\right)^{\frac{1}{2}}\left(\frac{d\mu(\pi)}{d\nu(\pi)}\right)^{\frac{1}{2}}  \begin{bmatrix}
            \pi^{23} & \pi^2 & \pi^3\\
            b+c & b & c
        \end{bmatrix} \begin{bmatrix}
            \pi & \pi^1 & \pi^{23}\\
            d & a & b+c
        \end{bmatrix} |d\rangle_{\pi}^{\pi^{23}}
    \end{split}
\end{equation}
and
\begin{equation}
    \label{eq:secondtripledecomposition}
    \begin{split}
        &|a\rangle_{\pi^1}\otimes |b\rangle_{\pi^2}\otimes |c\rangle_{\pi^3}\\
        &= \int\limits_{\mathclap{[\pi^1 \otimes \pi^2]}}d\nu(\pi^{12})\int\limits_{\mathclap{[\pi^{12}\otimes \pi^{3}]}} d\nu(\pi)  \left(\frac{d\mu(\pi^{12})}{d\nu(\pi^{12})}\right)^{\frac{1}{2}}\left(\frac{d\mu(\pi)}{d\nu(\pi)}\right)^{\frac{1}{2}}  \begin{bmatrix}
        \pi^{12} & \pi^1 & \pi^2\\
        a+b & a & b
    \end{bmatrix} \begin{bmatrix}
        \pi & \pi^{12} & \pi^{3}\\
        d & a+b & c
    \end{bmatrix} |d\rangle_{\pi}^{\pi^{12}},
    \end{split}
\end{equation}
where $d = a+b+c$. In other words, we just found two orthonormal bases for $\pi$ in terms of tensor products of basis elements of $\pi^1\otimes \pi^2\otimes \pi^3$, given by $|d\rangle_{\pi}^{\pi^{23}}$ and $|d\rangle_\pi^{\pi^{13}}$. These two bases then have to be related by a unitary transformation, given by the \textit{6j-symbol}. Importantly, this quantity only depends on the irreducible representations in question, and not on the particular basis elements. This unitary transformation between the two bases is given by
\begin{equation}
    \label{eq:6jsymbols}
    |d\rangle_\pi^{\pi^{12}} = \int_{[\pi^2\otimes \pi^3]}d\nu(\pi^{23})\left(\frac{d\mu(\pi^{12})}{d\nu(\pi^{12})}\right)^{\frac{1}{2}}\left(\frac{d\mu(\pi^{23})}{d\nu(\pi^{23})}\right)^{\frac{1}{2}}\left\{\begin{array}{ccc}
         \pi^1 & \pi^2 & \pi^{12} \\
         \pi^3 & \pi & \pi^{23}
    \end{array}\right\}|d\rangle^{\pi^{23}}_\pi,
\end{equation}
where we have also normalised the 6j-symbols with respect to the Plancherel measure. For a triple product of matrix elements, there are now two ways to decompose the two states in the matrix element. The out-of-time ordered correlators correspond to the case where we consider a different decomposition for the left and the right side. For an irreducible representation $\pi \subset \pi^{1}\otimes \pi^2\otimes \pi^3$, it follows that
\begin{equation}
    \label{eq:triplematrixproduct}
    \begin{split}
        \langle \pi_{xy}, \pi^{1}_{ab}\cdot \pi^{2}_{cd}\cdot \pi^{3}_{ef}\rangle = &\int d\mu({\pi^{12}})d\mu({\pi^{23}})\left\{\begin{array}{ccc}
         \pi^1 & \pi^2 & \pi^{12} \\
         \pi^3 & \pi & \pi^{23}
    \end{array}\right\}\\
    &\times \overline{\begin{bmatrix}
        \pi^{23} & \pi^2 & \pi^3\\
        c+e & c & e
    \end{bmatrix}}\overline{\begin{bmatrix}
        \pi & \pi^1 & \pi^{23}\\
        x & a & c+e
    \end{bmatrix}}  \begin{bmatrix}
        \pi^{12} & \pi^1 & \pi^2\\
        b+d & b & d
    \end{bmatrix} \begin{bmatrix}
        \pi & \pi^{12} & \pi^{3}\\
        y & b+d & f
    \end{bmatrix},
    \end{split}
\end{equation}
where $x = a+c+e$ and $y = b+d+f$, and where also the Plancherel measure appears.  On the level of the matrix elements, it does not matter how one chooses to decompose the different states in the triple tensor product. However, a fact to consider for the $n$-point functions is that the insertions of the discrete series occur at different Euclidean times, resulting in additional exponents of Casimir eigenvalues that correspond to the intermediate representations. It is exactly the insertion of operators at different times that distinguishes the out-of-time ordered correlators from the time-ordered ones. To calculate the crossed $4$-point function, we set $\pi^1 = \D^{\ell_1}$, $\pi^2 = \P^{s_1}$, $\pi^{3} = \D^{\ell_2}$, $\pi = \P^{s_3}$, $\pi^{12} = \P^{s_2}$ and $\pi^{23} = \P^{s_4}$.\footnote{This ordering of representations is the same as in the calculation of the the 6j-symbols by Groenevelt in Section~5 of \cite{G2006}, which were computed in the bases adapted to the one-parametric subgroup in an elliptic conjugacy class. As the 6j-symbols are basis independent, it gives the same result. Also, thus far, we have not made any explicit distinction between the positive and negative discrete series. The 6j-symbols in that paper that are relevant for us are computed for the tensor product $\D^+\otimes \P\otimes \D^-$ (or, equivalently, $\D^-\otimes \P\otimes \D^+$, related to it by an involutive automorphism of the algebra). We implicitly make the corresponding choice(s) in the calculation of the 4-point function.} Moreover, we assume that the matrix elements for the discrete representations are spinless, i.e. $a=b=e=f=0$. In addition to integrating over the intermediate representations, we also integrate over the initial and final representations, which appear due to the insertion of the resolution of identity before $| \tilde 0\rangle$ and after $\langle \tilde 0|$. From \eqref{eq:triplematrixproduct}, it then readily follows that the crossed 4-point function is given by
\begin{equation}
    \label{eq:Liouville4pointfunction}
    \begin{alignedat}{1}
    \begin{tikzpicture}[baseline={([yshift=-.5ex]current bounding box.center)}]
        \draw[black, thick] (0:0.75) arc (0:360:0.75);
        \draw[draw=black, fill=black] (30:0.75) circle (0.07);
        \draw[draw=black, fill=black] (150:0.75) circle (0.07);
        \draw[draw=black, fill=black] (-30:0.75) circle (0.07);
        \draw[draw=black, fill=black] (210:0.75) circle (0.07);
        \draw[black, dashed] (150:0.75) -- (-30:0.75);
        \draw[black, dashed] (210:0.75) -- (30:0.75);
        
        \draw (-0.3, 0.4) node {\footnotesize $\ell_1$};
        \draw (0.3, 0.4) node {\footnotesize $\ell_2$};
        \draw (-90:1) node {\footnotesize  $\beta_1$};
        \draw (0:1) node {\footnotesize  $\beta_2$};
        \draw (90:1) node {\footnotesize  $\beta_3$};
        \draw (180:1) node {\footnotesize  $\beta_4$};
    \end{tikzpicture} &\sim \int \prod_{n=1}^{4}\left\{d\mu(\P^{s_n})e^{-\beta_n \mathcal{C}(s_n)}\right\}\left\{\begin{array}{ccc}
         \D^{\ell_1} & \P^{s_1} & \P^{s_2} \\
         \D^{\ell_2} & \P^{s_3} & \P^{s_4}
    \end{array}\right\}\\
    &\times \overline{\begin{bmatrix}
        \P^{s_4} & \P^{s_1} & \D^{\ell_2}\\
        p & p & 0
    \end{bmatrix}}\overline{\begin{bmatrix}
        \P^{s_3} & \D^{\ell_1} & \P^{s_{4}}\\
        p & 0 & p
    \end{bmatrix}}\begin{bmatrix}
        \P^{s_2} & \D^{\ell_1} & \P^{s_{1}}\\
        p & 0 & p
    \end{bmatrix}\begin{bmatrix}
        \P^{s_3} & \P^{s_2} & \D^{\ell_{2}}\\
        p & p & 0
    \end{bmatrix}
    \end{alignedat}
\end{equation}
where $\sum_{n=1}^{4}\beta_n= \beta$. This is the same 4-point function as found in \cite{BMV2018, BMV2019, MTV2017}. It can be noted from the above computation that each Euclidean length $\beta_n$ between each vertex contributes a factor $e^{-\beta_n \mathcal{C}(s_n)}$, that each vertex with the boundary contributes a 3j-symbol, and that each crossing contributes a 6j-symbol. This is, in fact, more generally true, providing one with Feynman rules to compute general $n$-point functions in terms of the Casimir eigenvalues, and 3j- and 6j-symbols \cite{MTV2017}.\footnote{See e.g. a recent preprint \cite{CuiRozali2025splittinggluingsinedilatongravity} where interesting relations for 6j-symbols were obtained from analysing such glueing rules for higher-point functions, also in the $q$-deformed context.} Analogous Feynman rules were derived for the DSSYK model in \cite{BINT2019}. They therefore have a very similar representation-theoretical interpretation, but now adapted for the quantum group. An important change one should be aware of here is that the tensor product becomes braided (see e.g. \cite{QGAR, KasselQuantumGroups, CPAGtQG}), and that the product of matrix elements is no longer commutative. In Section~\ref{sec:sykpartitionfunction}, we will follow similar steps as in this section to revisit the calculation of the partition and $2$-point function for the $ q$-analogue of $\mathrm{SU}(1,1)$.

\section{The locally compact quantum group
\label{sec:quantumsu11}
$\mathrm{SU}_q(1,1)\rtimes \Z_2$}
It was directly calculated in \cite{BINT2019} that the $n$-point functions for the double-scaled SYK model \cite{K2015} are given in terms of 3j- and 6j-symbols corresponding to the quantum group $\mathrm{SU}_q(1,1)$. This suggests that, in line with the previous section, one can describe the double-scaled SYK model through the representation theory of the quantum group. Several attempts for such a description have been made \cite{BINN2023, BM2024, BMT2025, HVX2025QuantumSymmetry}, but none take advantage of the von Neumann algebraic description and the corresponding regular representations of the quantum group. Therefore, we will improve on this by considering the von Neumann algebraic description, as introduced by Kustermans and Vaes \cite{KV2000, KV2000_1, KV2003} (see also \cite{CK2010, TItQGaD, D2014} for overviews), of the quantum group $\mathrm{SU}_q(1,1)$, which was already conjectured in \cite{BINT2019} to provide the appropriate representation-theoretic framework for describing the dynamics of the model. Importantly, as will be discussed in Section~\ref{sec:quantumcoordinatealgebra}, this quantum group does not allow for a locally compact description, and one instead has to extend to its normaliser in $\mathrm{SL}(2,\C)$. The von Neumann algebraic quantum group for this normaliser has been constructed in \cite{KK2003}, and its dual in \cite{GKK2010}. We will discuss this construction in Section~\ref{sec:locallycompactquantumgroup} and discuss its regular representations in Section~\ref{sec:quantumregularrepresentations}. Then, in Section~\ref{sec:quantumGaussDecomp}, we construct its quantum Gauss decomposition and we derive the action of the Casimir operator in this decomposition in Section~\ref{sec:actionCasimiroperator}.

\subsection{The quantum coordinate algebra}
\label{sec:quantumcoordinatealgebra}
As is usual in non-commutative geometry, a deformation of a locally compact space is described through a deformation of the function algebra over that space. In this sense, a quantum group is defined as a deformation of the Hopf algebra of functions on a Lie group. In particular, consider the deformation parameter $0<q<1$, and define the \textit{quantum coordinate algebra} $\A_q(\mathrm{SU}(1,1))$ to be the Hopf $*$-algebra generated by $\alpha$ and $\gamma$ with relations \cite{QGAR}
\begin{equation}
    \label{eq:algebraicrelations}
    \alpha\gamma = q\gamma \alpha, \quad \alpha \gamma^* = q\gamma^*\alpha, \quad \gamma \gamma^* = \gamma^* \gamma, \quad \alpha \alpha^* - q^2 \gamma^* \gamma = 1 = \alpha^* \alpha - \gamma^* \gamma,
\end{equation}
with a coproduct given by
\begin{equation}
    \Delta(\alpha) = \alpha \otimes \alpha + q\gamma^* \otimes \gamma, \quad \Delta(\gamma) = \gamma\otimes \alpha + \alpha^*\otimes \gamma,
\end{equation}
and a counit $\epsilon(\alpha) = 1$, $\epsilon(\gamma) = 0$ and antipode $S(\alpha) = \alpha^*$, $S(\gamma) = -q\gamma$. It can be seen that this algebra is now non-commutative, and reduces to the classical coordinate algebra $\A(\mathrm{SU}(1,1))$ in the limit $q\to 1$. In the classical case, the generators $\alpha$ and $\gamma$ are functions $\mathrm{SU}(1,1)\to \C$ as defined in \eqref{eq:classicalAgenerators}. Therefore, for the quantum algebra, one could interpret $\alpha$ and $\gamma$ similarly as non-commuting functions over the quantum group $\mathrm{SU}_q(1,1)$. Similar to the classical case, the coproduct, counit, and antipode in the coordinate algebra can be thought of as describing the multiplication, unit, and inverse on $\mathrm{SU}_q(1,1)$.\\

Similarly, there is a deformation of the universal enveloping algebra. In particular, the \textit{quantum universal enveloping algebra} $\U_q(\mathfrak{su}(1,1))$ is the (associative, unital) algebra generated by $E,F,K$ and $K^{-1}$ subject to relations \cite{QGAR}
\begin{equation}
    \label{eq:quantums2lrelations}
    KK^{-1} = K^{-1}K = 1, \quad KE = qEK, \quad KF = q^{-1}FK, \quad EF - FE = \frac{K^{2} - K^{-2}}{q - q^{-1}}
\end{equation}
and with the $*$-structure $K^* = K$ and $E^* = -F$. Moreover, there is a natural Hopf algebra structure with coproduct $\Delta$, counit $\epsilon$ and antipode $S$ given by
\begin{equation}
    \begin{gathered}
        \Delta(K) = K\otimes K, \quad \Delta(E) = K\otimes E + E\otimes K^{-1}, \quad \Delta(F) = K\otimes F + F\otimes K^{-1}\\
        S(K) = K^{-1}, \quad S(E) = -q^{-1}E, \quad S(F) = -qF,\quad  \epsilon(K) = 1, \quad \epsilon(E) = \epsilon(F) = 0.
    \end{gathered}
\end{equation}
Here, it can be noted that the quantum universal enveloping algebra reduces to the classical one by defining $K = q^{H/2}$ and taking the limit $q\to 1$. Therefore, the Cartan element $K$ corresponds to an exponent of $H$, and is group-like\footnote{A group-like element of a Hopf algebra is an element $x$ such that the coproduct is given by $\Delta(x) = x\otimes x$.}. The Casimir element, i.e. generator of the centre, for the quantum universal enveloping algebra $\U_q(\mathfrak{su}(1,1))$ is given by
\begin{equation}
    \label{eq:CasimirElement}
    \Omega \coloneq \frac{q^{-1}K^{-2} + qK^2 - 2}{(q-q^{-1})^2} + FE.
\end{equation}
Similar to the classical case, there exists a non-degenerate dual pairing $\langle \cdot, \cdot\rangle \colon \U_q \times \A_q\to \C$ which satisfies the equalities \eqref{eq:dualpairing}, and is given by \cite{K1993}
\begin{equation}
    \langle K,\alpha\rangle = \sqrt{q}, \quad \langle K, \gamma\rangle = 0, \quad \langle E, \alpha\rangle = 0, \quad \langle E, \gamma\rangle = 0, \quad \langle F,\alpha\rangle = 0, \quad \langle F,\gamma\rangle = 1.
\end{equation}
Moreover, this duality induces a left and right action of $\U_q$ on $\A_q$ through \eqref{eq:leftrightdualaction}. In the classical case, these actions corresponded to taking derivatives along the direction of the Lie algebra elements, and we therefore interpret the action of the deformed algebra in a similar way, as `taking derivatives' on the quantum group manifold $\mathrm{SU}_q(1,1)$.\\

The description is, of course, again purely algebraic, and we would like to describe its analytical properties as well. In particular, as in the classical case, we would like to upgrade this algebra to a von Neumann algebra, which can be thought of as describing non-commutative measure theory. In the classical case, the commutative von Neumann algebra was given by essentially bounded functions on the corresponding space. In the quantum case, a natural choice would be a (non-commutative) von Neumann algebra that is generated by $\alpha$ and $\gamma$. To obtain such a von Neumann algebra, the algebraic objects need to be upgraded to operators, which can be achieved by considering representations of the coordinate algebra. It turns out, however, that these representations are not actually compatible with the coproduct \cite{W1991}. To gain some intuition behind this problem and its solution, we will still construct and discuss these representations first.\\

We define the \textit{$q$-interval} $[0,\infty)_q \coloneq \{q^k : k\in \Z\}$ and consider the Hilbert space $L^2_q([0,\infty))$ of `square-integrable' functions $f \colon [0,\infty)_q \to \C$ with inner product given by the \textit{$q$-integral}
\begin{equation}
    \langle f , g\rangle = \int_{0}^{\infty} \overline{f(x)}g(x)d_qx \coloneq (1-q)\sum_{x\in [0,\infty)_{q}}x\overline{f(x)}g(x),
\end{equation}
which reduces to the ordinary integral in the limit $q\to 1$. On this Hilbert space, there is an orthonormal basis $\{\delta_p\}_{p\in [0,\infty)_{q}}$, where $\delta_p$ is the function given by
\begin{equation}
    \delta_p (x) = \frac{1}{\sqrt{(1-q)p}}\delta_{p,x} \quad \text{for } x\in [0,\infty)_q,
\end{equation}
and can be interpreted as a $q$-deformation of the Dirac-delta function. Let us now construct the irreducible representations of $\A_q(\mathrm{SU}(1,1))$ on this Hilbert space. Since the group is non-compact, the actions of the algebra generators will generally be given by unbounded operators. Therefore, the representation is only well-defined on some dense subset of  $L^2_q([0,\infty))$, which for our purposes will be given by all functions $[0,\infty)_q\to \C$ which are non-zero on only finitely many basis elements, i.e. are compactly supported. For $\theta \in [0,2\pi)$, there are (mutually inequivalent) irreducible $*$-representations $\pi_\theta$ of $\A_q(\mathrm{SU}(1,1))$ on the compactly supported functions in $\H_\theta \coloneq L^2_{q^2}([0,\infty))$ given by \cite{W1991}
\begin{equation}
    \label{eq:su11representation}
    \pi_\theta(\alpha)\delta_x = e^{i\theta}\sqrt{1 + x}~\delta_{q^{-2}x}, \quad \pi_\theta(\gamma)\delta_x = e^{i\theta}\sqrt{x}~\delta_x
\end{equation}
for $x\in [0,\infty)_{q^2}$.\footnote{Note here that a representation of $\A_q(\mathrm{SU}(1,1))$ is defined on the space of functions over the $q^2$-interval. This inconsistency in the power of $q$ is due to a historical difference in convention between special functions and quantum groups. Moreover, we only consider representations such that the spectrum of the spherical element is given by $\sigma(\gamma^*\gamma) = q^{2\Z}$. In principle, there is a bigger family of representations with $\sigma(\gamma^*\gamma) = q^{2\beta}q^{2\Z}$ for any $\beta \in (-\frac{1}{2},\frac{1}{2})$, where $\beta$ corresponds to a character of the fundamental group of the corresponding symplectic leaf $\Sigma_\theta$ \eqref{eq:symplecticleafptheta} \cite{K1993}.} Here, we choose a particular convention (which one might call the `polarisation') for the representations, requiring that $(\pi_\theta(\gamma^*\gamma)f)(x) = xf(x)$ for a function $f\in L^2_{q^2}([0,\infty))$. Therefore, one may interpret $\gamma^*\gamma$ as a position operator. Moreover, it can be verified that this operator is invariant under the left and right action of the Cartan element $K$, i.e. $K\triangleright \gamma^*\gamma = \gamma^*\gamma = \gamma^*\gamma \triangleleft K$. In particular, this element generates the $*$-subalgebra of left and right $K$-invariant elements, i.e.\footnote{Here, $\langle \gamma^*\gamma \rangle$ denotes the algebra generated by $\gamma^*\gamma$, i.e. all polynomials in $\gamma^*\gamma$ with complex coefficients.}
\begin{equation}
    \A_q(\mathrm{U}(1)\backslash\mathrm{SU}(1,1) /\mathrm{U}(1)) = \langle \gamma^*\gamma\rangle.
\end{equation}
Therefore, the position operator $\gamma^*\gamma$ is a \textit{spherical element} for the Cartan decomposition of the quantum group $\mathrm{SU}_q(1,1)$. \\

These representations can be interpreted to be obtained from geometric quantisation of a symplectic leaf of $\mathrm{SU}(1,1)$. Indeed, the deformation of the coordinate algebra \eqref{eq:algebraicrelations} corresponds to a (formal) deformation of the Poisson structure on $\mathrm{SU}(1,1)$, i.e. a commutator bracket $[\cdot , \cdot]$ such that $h^{-1}[a,b]\text{ mod } h = \{a\text{ mod } h, b\text{ mod }h\}$ for $a,b \in \A_q\otimes \C[[h]]$ and $q = \exp(-h)$, where $\{\cdot, \cdot\}$ denotes the `standard' Poisson bracket on $\mathrm{SU}(1,1)$ \cite{K1994}. The $h\to 0$ limit of the kernel of an irreducible representation then corresponds to a maximal Poisson ideal, and therefore a symplectic leaf, of $\mathrm{SU}(1,1)$ \cite{LevendorskiiSoibelman1991, VaksmanSoibelman1988}, giving a one-to-one correspondence between the symplectic leaves and the irreducible representations of the quantum coordinate algebra. For $\pi_\theta$, this kernel is given by $\ker \pi_{\theta} = \langle \gamma - e^{2i\theta}\gamma^*\rangle$, with the corresponding symplectic leaf consisting of the elements in $\mathrm{SU}(1,1)$ with $\text{arg } \gamma = \theta$, i.e.
\begin{equation}
    \label{eq:symplecticleafptheta}
    \Sigma_\theta \coloneq \left\{\begin{bmatrix}
        e^{i\phi}\cosh \frac{\rho}{2} & e^{-i\theta}\sinh\frac{\rho}{2}\\
        e^{i\theta}\sinh \frac{\rho}{2} & e^{-i\phi}\cosh \frac{\rho}{2}
    \end{bmatrix} : \phi \in [0,2\pi) \text{ and } \rho > 0\right\}\subset \mathrm{SU}(1,1).
\end{equation}
The geometric interpretation of the representation $\pi_\theta$ is now as follows. Firstly, it can be interpreted as a geometric quantisation (see e.g. \cite{WoodhouseGeometricQuantisation}) of the symplectic leaf $\Sigma_\theta$, with the functions $\alpha,\gamma$ acting as operators on the prequantum Hilbert space, and a chosen polarisation such that $\gamma^*\gamma$ acts as the position operator. In addition, the symplectic leaf itself, and therefore the coordinates, are quantised, i.e. $\sinh\rho/2 \in q^{\Z}$. Thus, the representations $\pi_\theta$ of $\A_q(\mathrm{SU}(1,1))$ correspond to a `double quantisation' (that is, quantisation along with a $q$-deformation) of the two-dimensional symplectic leaves of $\mathrm{SU}(1,1)$. Note here an important difference with the classical case: the irreducible $*$-representations of the classical coordinate algebra $\mathcal{A}(\textrm{SU}(1,1))$ are all one-dimensional, corresponding to evaluations at a given point, whereas the irreducible $*$-representations of the quantised coordinate algebra are infinite-dimensional. We see that, in this optics, $q$-deformation tends to glue ordinary points into non-local `fat points', corresponding to infinite-dimensional irreducible representations of the coordinate algebra.\\

Let us now comment on the incompatibility of the representations on the level of the Hopf algebra. In particular, one could ask whether $(\pi_\theta\otimes \pi_\phi)\circ \Delta$ is again a well-defined representation of $\A_q(\mathrm{SU}(1,1))$, that is, whether there exists a dense subset of the Hilbert space that is closed under the action of the coordinate algebra. Such a dense subset usually corresponds to certain boundary conditions on the functions. Therefore, we could equivalently say that the action of the coordinate algebra should not deform the boundary conditions. Since, according to the choice we made at the start, we consider functions over the spectrum of the position operator $\gamma^*\gamma$, these boundary conditions correspond to self-adjoint extensions of $((\pi_\theta\otimes \pi_\phi)\circ \Delta)(\gamma^*\gamma)$. This operator is given by a doubly-infinite Jacobi operator and its self-adjoint extensions are parametrised by an angle $\psi \in [0,2\pi)$ \cite{K2001}, with corresponding boundary conditions given by \cite{K1994}
\begin{equation}
    \label{eq:poincareboundaryconditions}
    \D_\psi = \{f\in L^2_{q^{2}}([0,\infty)) : \cos{\psi}\cdot f(\infty) +\sin{\psi}\cdot (D_{q^{2}}f)(\infty) = 0\},
\end{equation}
where $f(\infty) \coloneq \lim_{n\to\infty}f(q^{-2n})$ and $(D_{q^2}f)(x) \coloneq \frac{f(x) - f(qx)}{x - qx}$ is the $q$-derivative. Unfortunately, these boundary conditions are deformed by the action of other elements in the coordinate algebra, for example, $(\pi_\theta\otimes \pi_\varphi)(\Delta(\alpha))(\D_{\psi}) \not\subset \D_\psi$. Therefore, the symmetric operator $\gamma^*\gamma$ cannot be extended to a self-adjoint operator in a way that is compatible with the action of $\alpha$, making the tensor product of representations ill-defined.\footnote{It would be interesting to understand a stringy interpretation of this argument better, e.g. in terms of brane boundary conditions becoming `fuzzy' upon $q$-deformation.} This is the essence of the no-go theorem by Woronowicz \cite{W1991}, which has the consequence that the quantum group $\mathrm{SU}_q(1,1)$ does not exist as a locally compact quantum group. This inconsistency can be regarded as a problem with defining the group multiplication on the naive quantum group manifold $\mathrm{SU}_q(1,1)$. \\

As pointed out in \cite{K1994}, the symplectic geometric interpretation of this negative result is the fact that the local \textit{dressing action}\footnote{Recall that the Poisson structure on $\mathrm{SU}(1,1)$ induces a Lie bracket on $\mathfrak{g}^*$. Moreover, $\mathfrak{g}^*$ can be realised as the Lie algebra of left-invariant differential 1-forms on $\mathrm{SU}(1,1)$. The bundle map $B \colon T^*\mathrm{SU}(1,1)\to T\mathrm{SU}(1,1)$ then induces the linear map $\lambda \colon \mathfrak{g}^* \to \mathcal{X}(\mathrm{SU}(1,1))$, where $\lambda(\xi)$ is referred to as the left dressing vector field corresponding to $\xi\in \mathfrak{g}^*$. Integrating $\lambda$ gives a local action $\mathrm{SU}(1,1)^*$ of $\mathrm{SU}(1,1)$, which is called the left {\it local dressing action} \cite{CPAGtQG, K1994}. The right (local) dressing action is defined analogously.} of $\mathrm{SU}(1,1)^*$ on $\mathrm{SU}(1,1)$ cannot be extended to a global one. Explicitly, let $\H^*_\theta$ with basis $\{\delta^*_x\}$ be the dual of $\H_\theta$ such that $\langle \delta^*_x, \delta_y\rangle = \delta_{x,y}$. We can define the dual representation of $\A_q$ on $\H_\theta^*$ via $\langle \pi^*(a)\delta^*_x, \delta_y\rangle \coloneq \langle \delta^*_x, \pi(S(a))\delta_y\rangle$ for $a\in \A_q$. It can then be easily verified that $\pi^*_\theta \cong \pi_{\theta + \pi}$. In particular, for the tensor product of representations, we have $\pi_\theta\otimes \pi_{\theta+\pi} \cong \pi_\theta\otimes \pi^*_{\theta}$, which defines an action on $\H_\theta\otimes \H_\theta^*$ corresponding to the quantum adjoint action and reduces to the local dressing action in the classical $q\to 1$ limit \cite{K1994}. The above negative result then corresponds to the fact that the local dressing action of $\mathrm{SU}(1,1)^*$ on $\mathrm{SU}(1,1)$ does not extend to a global one -- $\mathrm{SU}(1,1)\cdot\mathrm{SU}(1,1)^*$ is not dense in $\mathrm{SL}(2,\C)$ -- thus making the tensor product of representations ill-defined. The same argument can be extended to any tensor product of representations $\pi_\theta\otimes \pi_\phi$.\\

The solution by Korogodsky \cite{K1994} was to extend $\mathrm{SU}(1,1)$ to its normaliser in $\mathrm{SL}(2,\C)$. Namely, for this normaliser, the dressing action does exist globally, which therefore also suggests that the tensor product of irreducible representations of the coordinate algebra of this normaliser may be well-defined. Indeed, in such an extension, the boundary conditions \eqref{eq:poincareboundaryconditions} will instead be given by smoothness conditions, which may remain unaltered under the action of the coordinate algebra. Let us make this explicit. Recall from \eqref{eq:definitionnormalisersu11} that the normaliser of $\mathrm{SU}(1,1)$ in $\mathrm{SL}(2,\C)$ is given by
\begin{equation}
    \begin{split}
        \mathrm{SU}(1,1)\rtimes\Z_2 &=  \mathrm{SU}(1,1)\cup \mathrm{SU}(1,1)v\\
        &= \left\{\begin{bmatrix}
            \alpha & e \bar \gamma\\
            \gamma & e \bar \alpha
        \end{bmatrix} : \alpha, \gamma \in \C, e \in \{\pm 1\} \text{ and } |\alpha|^2 - |\gamma|^2 = e\right\},
    \end{split} 
\end{equation}
where we have now explicitly written out both parts in the normaliser in matrix form, by introducing an extra variable $e \in \{\pm 1\}$. In particular, the connected component of the identity corresponds to $e = +1$, while the connected component of $v$ corresponds to $e = -1$.\\

Thus, to quantise the coordinate algebra of the normaliser, we introduce an additional element $e$ satisfying 
\begin{equation}\label{eq:e-relations}
    e^2 = 1, \quad \Delta(e) = e\otimes e, \quad S(e) = e, \quad \varepsilon(e) = 1, \quad e^* = e
\end{equation}
and $ae = ea$ for all $a\in \A_q(\mathrm{\mathrm{SU}}(1,1))$. We also define its dual pairing with the universal enveloping algebra as $\langle X, e\rangle = \epsilon(X)$ for $X\in \U_q(\mathfrak{su}(1,1))$.
The coordinate algebra of the normaliser $\A_q(\mathrm{SU}_q(1,1)\rtimes \Z_2)$ is then generated by $\alpha,\gamma,e$, subject to \eqref{eq:e-relations}, and the following relations/comultiplication:\footnote{In particular, the coordinate algebra for the normaliser is obtained from the coordinate algebra of $\mathrm{SL}_q(2,\C)$ (together with the specified element $e$) by imposing the $*$-structure $\alpha^* = e\delta$, $\beta^* = qe\gamma$, $\gamma^* = q^{-1}e\beta$, $\delta^* = e\alpha$ and $e^* = e$ \cite{KK2003}.}
\begin{equation}
    \begin{gathered}
        \alpha \gamma = q\gamma \alpha, \quad \alpha \gamma^* = q\gamma^* \alpha, \quad \gamma \gamma^* = \gamma^* \gamma, \quad \alpha \alpha^* - q^2\gamma^* \gamma = e = \alpha^* \alpha - \gamma^* \gamma,\\
        \Delta(\alpha) = \alpha \otimes \alpha + qe\gamma^* \otimes \gamma, \quad \Delta(\gamma) = \gamma\otimes \alpha + e\alpha^* \otimes \gamma.
    \end{gathered}
\end{equation}
Moreover, the antipode becomes $S(\alpha) = e\alpha^*$ and $S(\gamma) = -q\gamma$, while the counit remains the same.\\

The representations of this extended algebra now act on a function space with an extended domain. Namely, let us consider the $q$-interval $(-\infty, -1)_q \coloneq \{-q^{-k}: k\in \Z_{>0}\}$ and define $I_q = (-\infty, -1)_q\cup [0,\infty)_q$. We consider the Hilbert space of square-integrable functions $f\colon I_q \to \C$ over this domain, i.e. $\H_q\coloneq L^2_q(I_q) = L^2_q((-\infty, -1))\oplus L^2_q([0,\infty))$, with inner product
\begin{equation}
    \langle f, g\rangle = \int_{I_q} \overline{f(x)}g(x) d_qx \coloneq (1-q)\sum_{x\in I_q}|x|\overline{f(x)}g(x).
\end{equation}
Similar to before, there is an orthonormal basis $\{\delta_p\}_{p\in I_q}$, where $\delta_p$ is given by the function
\begin{equation}
    \label{eq:qdeltafunction}
    \delta_p(x) = \frac{1}{\sqrt{(1-q)|p|}}\delta_{p,x} \quad \text{for } x\in I_q.
\end{equation}
The irreducible representations of the coordinate algebra $\A_q(\mathrm{SU}_q(1,1)\rtimes \Z_2)$ are now again defined on the compactly supported functions. In particular, for each $\theta \in [0,2\pi)$, there is a $*$-representation of $\A_q(\mathrm{SU}_q(1,1)\rtimes \Z_2)$ on the compactly supported functions in $\H_{q^2}$ given by \cite{KK2003}\footnote{We use a different convention than Koelink and Kustermans in \cite{KK2003}. We get the same representation under the redefinition $\delta_{\pm q^{2k}} \to e^{ik\theta}\delta_{\pm q^{2k}}$ and $x \to \sign{p}p^{-2}$.}
\begin{equation}
    \label{eq:repsNormalizer}
    \begin{split}
        \pi_\theta(\alpha)\delta_x &= e^{i\theta}\sqrt{\sign{x} + |x|}~\delta_{q^{-2}x},\\
        \pi_\theta(\gamma)\delta_x &= e^{i\theta}\sign{x}\sqrt{|x|}~\delta_x,\\
        \pi_\theta(e)\delta_x &= \sign{x}\delta_x,
    \end{split}
\end{equation}
for $x\in (-\infty, -1)_{q^2}\cup [0,\infty)_{q^2}$. If we restrict the action to just the functions in $L^2_{q^2}([0,\infty))$, we get back the same action as in \eqref{eq:su11representation}. It is important to note that this representation is not irreducible, as $L^2_{q^2}((-\infty, -1))$ and $L^2_{q^2}([0,\infty))$ do not talk to each other through the action of the coordinate algebra. Also, $(\pi_\theta(e\gamma^*\gamma)f)(x) = xf(x)$ for a function $f\in \H_{q^2}$, such that now $e\gamma^*\gamma$ has become the position operator. This is also expected, as $\gamma^*\gamma$ is a positive operator and thus can never have negative elements in its spectrum, which is resolved by the additional multiplication by $e$.\\

\begin{figure}
    \centering
    \includegraphics[width = 0.4\textwidth]{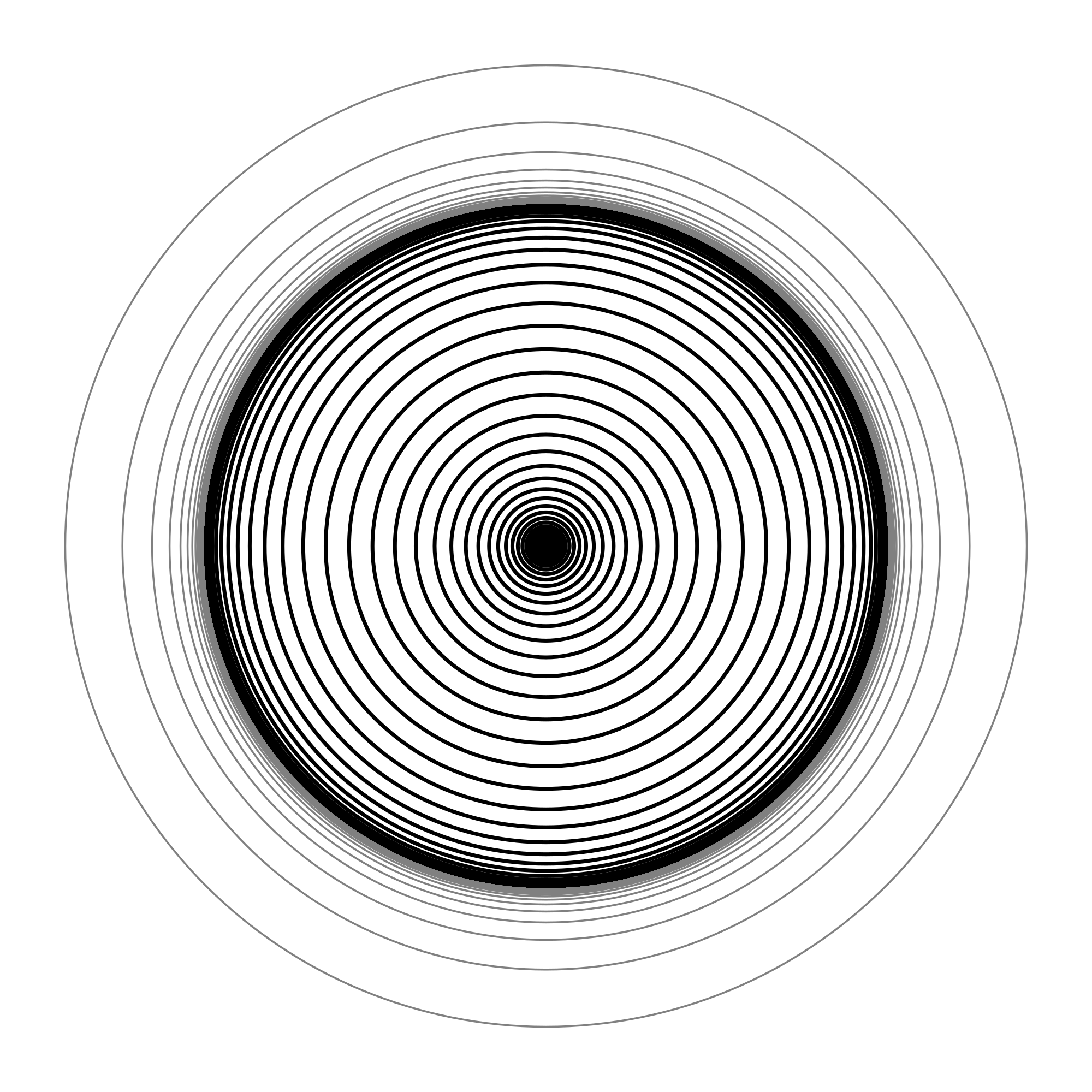}
    \caption{The extended quantum Poincaré disk, where the points $z(x,\theta) \coloneq e^{i\theta}(1+x^{-1})^{-1/2}$ for $x\in I_{q^2}$ and $\theta \in [0,2\pi)$ are plotted. The extension of the Poincaré disk is given in grey. This picture corresponds to a slicing through all the two-dimensional symplectic leaves of $\mathrm{SU}_q(1,1)\rtimes \Z_2$.}
    \label{fig:extendedquantumPoincaredisk}
\end{figure}

The coordinates $\theta$ and $x$ can be interpreted as coordinates on the \textit{extended quantum Poincaré disk}. That is, we consider the points $z(x,\theta) \coloneq e^{i\theta}(1+x^{-1})^{-1/2} \in \C$ for $x\in I_{q^2}$ and $\theta \in [0,2\pi)$, which are plotted in Figure~\ref{fig:extendedquantumPoincaredisk}.
Note that in the quantum case, the extended disk has a `maximal ring'. That is, the radial coordinate has a maximal value which corresponds to the end-point $-q^{-1}\in (-\infty, -1)_q$. The limits $x\to \infty$ and $x\to -\infty$ for $x\in I_{q^2}$ correspond to the limit towards the unit circle $|z| = 1$ from the inside and outside of the disk, respectively. Also, in the $q\to 1$ limit, the extended quantum Poincaré disk becomes dense in $\C$. We emphasise here that Figure~\ref{fig:extendedquantumPoincaredisk} does not correspond to just one symplectic leaf of $\mathrm{SU}(1,1)\rtimes \Z_2$. On the contrary, for every $\theta\in [0,2\pi)$, the radial slice at that angle corresponds to one symplectic leaf of $\mathrm{SU}(1,1)$ for $|z| <1$ and one symplectic leaf of $\mathrm{SU}(1,1)v$ for $|z| > 1$. Therefore, Figure~\ref{fig:extendedquantumPoincaredisk} should be interpreted as a slice through all the 2-dimensional symplectic leaves of $\mathrm{SU}_q(1,1)\rtimes \Z_2$, where the slicing is orthogonal to the coordinate $\phi$ as given in \eqref{eq:symplecticleafptheta}.\\

We would now like to verify that the above representations are indeed compatible with the coproduct. Since the representation is now defined on the space of functions over the spectrum of $e\gamma^*\gamma$, we should correspondingly consider the self-adjoint extensions of $(\pi_\theta \otimes \pi_\phi)(\Delta(e\gamma^*\gamma))$. This operator is a direct sum of an unbounded Jacobi operator and a doubly infinite Jacobi operator, which both have deficiency indices $(1,1)$. The deficiency indices of $(\pi_\theta \otimes \pi_\phi)(\Delta(e\gamma^*\gamma))$ are therefore $(2,2)$, which allows for more freedom in the choice of the self-adjoint extension, with the boundary conditions becoming `fuzzier'. In particular, there are now self-adjoint extensions parametrised by $\lambda \in \T$ with smoothness conditions \cite{KK2003} 
\begin{equation}
    \label{eq:smoothnessConditionsnormalizer}
    \D_\lambda = \{f \in \H_{q^2} : f(\infty) = \lambda f(-\infty) \text{ and } (\D_{q^2} f)(\infty)) = \lambda (D_{q^{2}}f)(-\infty)\},
\end{equation}
where $f(-\infty) \coloneq \lim_{n\to\infty}f(-q^{-2n})$. The important insight is now that the action of the quantum group $\A_q(\mathrm{SU}(1,1)\rtimes\Z_2)$ deforms the functions at $\infty$ and $-\infty$ in a similar way, such that the smoothness conditions \eqref{eq:smoothnessConditionsnormalizer} remain satisfied, rendering the tensor product of representations of the normaliser to be well-defined \cite{K1994}. From Figure~\ref{fig:extendedquantumPoincaredisk}, it can be noted that \eqref{eq:smoothnessConditionsnormalizer} corresponds to a (discretised version of a) smoothness condition along the unit circle, i.e. the boundary of the Poincaré disk, thus connecting the inner and outer disks along the Poincaré horizon.

\subsection{The von Neumann algebraic quantum group}
\label{sec:locallycompactquantumgroup}
Now that we have well-defined representations of $\A_q(\mathrm{SU}(1,1)\rtimes\Z_2)$, we are ready to introduce the corresponding locally compact quantum group, generalising (and $q$-deforming) the example of Section~\ref{sec:classicalLocallyCompactGroup}. Different theories for locally compact quantum groups have been developed by Masuda, Nakagami and Woronowicz \cite{MNW2003}, and by Kustermans and Vaes \cite{KV2000, KV2000_1}. We will follow the approach of the latter, in which a locally compact quantum group is defined as follows: Consider a von Neumann algebra $M$ together with a unital normal $*$-homomorphism $\Delta \colon M \to M\otimes M$ such that $(\Delta \otimes \id)\circ \Delta = (\id \otimes \Delta)\circ \Delta$. If there, moreover, exist two normal, semi-finite, faithful (n.s.f.)\footnote{A n.s.f. weight is a map $\phi \colon M^+\to [0,\infty]$ that is $\sigma$-weakly lower semi-continuous (normal), such that $\M_\phi^+ = \{x \in M^+ : \phi(x) < \infty\}$ generates $M$, i.e. $(\M_\phi^+)' = M$ (semi-finite), and $\phi(x) = 0$ implies $x = 0$ (faithful) \cite{CK2010}. One can think of it as a sufficiently well-behaved (e.g. with respect to taking limits) measure on our `non-commutative measure space'.} weights $\phi,\psi$ on $M$ such that\footnote{Here, $M_*$ denotes the predual of $M$ (see footnote~\ref{footnote:predualdefinition}).}
\begin{equation}
    \begin{alignedat}{3}
        \phi((\omega \otimes \id) \Delta(x)) &= \phi(x)\omega(1), \quad &&\text{for all } \omega \in M_*^+, x\in \M_\phi^+, \quad && \text{(left invariance)}, \\
        \psi((\id \otimes \omega) \Delta(x)) &= \psi(x)\omega(1), \quad &&\text{for all } \omega \in M_*^+, x\in \M_\psi^+, \quad && \text{(right invariance)},
    \end{alignedat}
\end{equation}
then $(M,\Delta)$ is called a\textit{ von Neumann algebraic quantum group}.\footnote{One might be surprised that there is no mention of the antipode and the counit in the definition of the von Neumann algebraic quantum group. However, it turns out that the antipode and counit can be obtained from the invariant Haar weights. Moreover, although the antipode $(Sx)(g) = x(g^{-1})$ for the classical group is a bounded operator, this is not true in general. Namely, for general locally compact quantum groups, the antipode can only be defined on a $\sigma$-strong-$*$ dense set, which can be considered as a slight weakening of the invertibility property for elements of a usual group \cite{CK2010}.} Although this definition is rather technical, one can recognise the same ingredients as in Section~\ref{sec:classicalLocallyCompactGroup}, where the von Neumann algebra was given by the essentially bounded functions and $\phi$ and $\psi$ by the Haar weight. Since $\mathrm{SU}(1,1)$ is unimodular, one has $\phi=\psi$, which, as we will see, remains true for the quantum deformation. For the rest of this paper, it is not necessary to understand all the technical details of this definition, so we will not discuss them. We refer to \cite{KV2000, KV2000_1, KV2003, D2014, CK2010, TItQGaD} for full overviews.\\

For constructing the von Neumann algebraic quantum group of $\mathrm{SU}_q(1,1)\rtimes \Z_2$, we follow the work of Koelink and Kustermans \cite{KK2003}. Firstly, it should be noted that the representations $\pi_\theta$ given in \eqref{eq:repsNormalizer} are too small to be faithful. Instead, we get a faithful representation by taking the direct integral $\pi = (2\pi)^{-1}\int_0^{2\pi}\pi_\theta d\theta$ \cite{KS2001}. Geometrically, this gives a representation of the coordinate algebra over all the quantised 2-dimensional symplectic leaves of $\mathrm{SU}_q(1,1)\rtimes \Z_2$. The corresponding Hilbert space for this representation is then isomorphic to
\begin{equation}
    \label{eq:directSumHilbertspace}
    \H \coloneq \int_0^{2\pi}\frac{d\theta}{2\pi}\H_{q^2} \cong L^2(\T)\otimes\H_{q^2}.
\end{equation}
Let $\{\zeta^m\}_{m\in\Z}$ be the canonical orthonormal basis of $L^2(\T)$. The faithful $*$-representation $\pi$ of $\A_q(\mathrm{SU}(1,1)\rtimes\Z_2)$ on this Hilbert space can then be explicitly written as
\begin{equation}
    \label{eq:actioncoordinatealgebradirectsum}
    \begin{split}
        \pi(\alpha)(\zeta^m\otimes \delta_x) &= \sqrt{\sign{x} + |x|}~\zeta^{m+1}\otimes \delta_{q^{-2}x},\\
        \pi(\gamma)(\zeta^m \otimes \delta_x) &= \sign{x}\sqrt{|x|}~\zeta^{m+1}\otimes \delta_x,\\
        \pi(e)(\zeta^{m}\otimes \delta_x) &= \sign{x}\zeta^m\otimes \delta_x,
    \end{split}
\end{equation}
on basis elements $\zeta^m\otimes \delta_x \in L^{2}(\T)\otimes \H_{q^2}$. To lighten up the notation a bit, in what follows we will not explicitly write $\pi$, but simply denote the action of an element $a\in \A_q(\mathrm{SU}(1,1)\rtimes\Z_2)$ on $\H$ as $a.(\zeta^m\otimes \delta_x) \coloneq \pi(a)(\zeta^m\otimes \delta_x)$. One might now suggest that the appropriate von Neumann algebra for the quantum group is the one generated by $\alpha, \gamma$ and $e$. However, the subspaces $L^2_{q^2}((-\infty, -1))$ and $L^2_{q^2}([0,\infty))$ form irreducible representations of the coordinate algebra, and, therefore, `do not talk' to each other. This will have the consequence that the associated von Neumann algebra is not closed under comultiplication \cite{KK2003}. Thus, in order to resolve this, we introduce an additional generator $u\in \B(\H)$ given by
\begin{equation}
    u.(\zeta^m\otimes \delta_x) = \zeta^m\otimes \delta_{-x},
\end{equation}
where $\delta_x = 0$ if $x\not\in I_{q^2}$, which maps between $L^2_{q^2}((-\infty, -1))$ and $L^2_{q^2}([0,\infty))$. It can be seen here that $u$ plays a similar role for the quantum group as $v$ has for the classical normaliser, since the action of $u$ corresponds to a reflection with respect to $x = 0$.\\

The von Neumann algebra of `essentially bounded operators' on $\mathrm{SU}_q(1,1)\rtimes \Z_2$, denoted by $L^\infty_q(\mathrm{SU}(1,1)\rtimes\Z_2)$, is now defined to be the von Neumann algebra on $\H$ generated\footnote{Via the so called {\it affiliation construction} of Woronowicz \cite{W1991}, which allows one to describe a von Neumann algebra that is `generated' by unbounded operators.} by $\alpha, \gamma, e$ and $u$. Explicitly, it is given by the commutant\footnote{The commutant of a subset $\A\subseteq \B(\H)$ is the von Neumann algebra $\A' \coloneq \{x\in \B(\H) : xa = ax \text{ for all } a\in \A\}$. Moreover, let $A \colon \D(A)\to \H$ and $B\colon \D(B)\to \H$ be two (densely defined) unbounded operators, then $B$ is an extension of $A$, denoted by $A\subseteq B$, if $\D(A)\subseteq \D(\B)$ and $B|_{\D(A)} = A$.}
\begin{equation}
    L^\infty_q(\mathrm{SU}(1,1)\rtimes\Z_2) \coloneq \{A \in \B(\H) : Ax \subseteq xA \text{ and } Ax^* \subseteq x^*A \text{ for } x\in \{\alpha, \gamma, e, u\}\}',
\end{equation}
which can be shown to be isomorphic to $L^{\infty}(\T)\otimes\B(L^2(I_{q^2}))$  \cite{KK2003}. The important result by Koelink and Kustermans in that same paper is that there exists a coproduct
\begin{equation}
    \Delta \colon L^\infty_q(\mathrm{SU}(1,1)\rtimes\Z_2) \to L^\infty_q(\mathrm{SU}(1,1)\rtimes\Z_2)\otimes L^\infty_q(\mathrm{SU}(1,1)\rtimes\Z_2)
\end{equation}
which is well-defined and coassociative. For the purposes of this paper, we will not need the explicit (and quite bulky!) expression of this coproduct. It can be noted that this coproduct,  when extended to the unbounded operators $\alpha, \gamma$ and $e$, coincides with the coproduct on $\A_q(\mathrm{SU}(1,1)\rtimes\Z_2)$.\\

For $n,m\in \Z$ with $n\equiv m\text{ mod } 2$ and $x\in I_{q^2}$ such that $\pm q^{n+m}x \in I_{q^2}$, let us now define the map
\begin{equation}
    \Phi^{\pm}(n,x,m)\zeta^r\otimes \delta_y = \delta_{\pm q^{n+m}x}(y) \zeta^{r+m}\otimes \delta_x,
\end{equation}
where $\delta_p$ is the $q^2$-delta function as defined in \eqref{eq:qdeltafunction}. It turns out that these matrix elements form a ($\sigma$-weakly dense) topological basis for our von Neumann algebra $L^\infty_q(\mathrm{SU}(1,1)\rtimes\Z_2)$ \cite{KK2003}. The elements $\Phi^{-}(n,x,m)$ correspond to operators that map between $L^2_{q^2}([0,\infty))$ and $L^2_{q^2}((-\infty, -1))$. In the context of symplectic leaves as described around \eqref{eq:symplecticleafptheta}, these elements map between the quantisations of symplectic leaves of $\mathrm{SU}(1,1)$ and $\mathrm{SU}(1,1)v$. As we will see in Section~\ref{sec:reductiondsSYK}, it's exactly these elements that appear in the reduction to the double-scaled SYK model, and therefore also to Liouville QM in the $q\to 1$ limit, which shows the point of considering the normaliser for treating these theories.\\

Following \cite{KK2003}, we will now put the von Neumann algebra $L^\infty_q(\mathrm{SU}(1,1)\rtimes\Z_2)$ (with the coproduct $\Delta$) into the framework of von Neumann algebraic quantum groups, by presenting the construction of an appropriate (left- and right-invariant) Haar weight, whose GNS space will reproduce the space of square-integrable functions with the basis introduced above (and, of course, with an action of $L^\infty_q(\mathrm{SU}(1,1)\rtimes\Z_2)$). First, we define the `square-integrable functions' on $\mathrm{SU}_q(1,1)\rtimes \Z_2$ as the Hilbert space\footnote{Note that the operators $\Phi^{\pm}(n,x,m)$ correspond to maps $\H_{q^2}\to \H_{q^2}$ with an additional phase shift. The corresponding Hilbert space (after GNS construction) should reflect this same structure, and is therefore given by $ L^2(\T)\otimes \H_{q^2}\otimes \H_{q^2}$. Additional intuition for why tensoring with another Hilbert space $\H_{q^2}$ is needed here comes from comparing to how the classical $\mathrm{SU}(1,1)$ Lie group can be thought of as a circle fibration over the hyperbolic disk.}
\begin{equation}
    L^2_{q}(\mathrm{SU}(1,1)\rtimes\Z_2) \coloneqq L^2(\T)\otimes \H_{q^2}\otimes \H_{q^2}.
\end{equation}
For $n\equiv m \text{ mod }2$ and $x\in I_{q^2}$ such that $\pm q^{n+m}x \in I_{q^2}$, the elements
\begin{equation}
    \label{eq:Lq2Cartanbasis}
    \delta^\pm_{n,x,m} \coloneq \zeta^m\otimes \delta_x \otimes \delta_{\pm q^{n+m}x},
\end{equation}
form an orthonormal basis for $L^2_{q}(\mathrm{SU}(1,1)\rtimes\Z_2)$. That is, they satisfy 
\begin{equation}
    \langle \delta^{\epsilon}_{n,x,m},\delta^{\epsilon'}_{n',x',m'}\rangle = \delta_{\epsilon,\epsilon'}\delta_{n,n'}\delta_{x,x'}\delta_{m,m'}.
\end{equation}
Moving on to the constructing an appropriate Haar weight $\phi$, we note that the trace on $L^\infty_q(\mathrm{SU}(1,1)\rtimes\Z_2)$ is given by $\text{Tr}(a) = \sum_{x\in I_{q^2}}\langle a.\zeta^0\otimes \delta_x, \zeta^0\otimes \delta_x\rangle$.
The correct Haar weight that one needs for the construction is actually just a slight `twist' of this trace, given by\footnote{Although the trace is already an n.s.f. weight, it is not left- and right-invariant and therefore not compatible with the additional symmetry. Unlike in the classical case, the correct Haar weight for the quantum group is therefore not tracial \cite{KV2000_1}.}
\begin{equation}
    \phi(a) \coloneq (1-q^2)\text{Tr}(|\gamma| a|\gamma|),
\end{equation}
which indeed was shown to be left- and right-invariant in \cite{KK2003}. Let us now spell out the GNS construction for this Haar weight.\footnote{We remind that, according to the GNS construction for a n.s.f. weight $\phi$ on a von Neumann algebra $M$, the Hilbert space is constructed as the completion of $\{x\in M : \phi(x^*x) < \infty\} / \{x\in M : \phi(x^*x) = 0\}$, where additionally, for a n.s.f (and thus faithful) weight, the ideal $\{x\in M : \phi(x^*x) = 0\}$ is zero. In particular, this gives rise to a natural normal, non-degenerate and faithful $*$-representation of $M$, in addition to the natural GNS map \cite{CK2010}. Note that the approach in \cite{KK2003} is somewhat reversed: a Hilbert space is a priori given, and is then shown to correspond to a GNS construction for $\phi$. This is the route we follow as well in our discussion of the GNS construction here.} That is, define:
\begin{enumerate}[label = (\alph*)]
    \item The subspace $\mathcal{N}_{\phi} = \{x\in  L^\infty_q(\mathrm{SU}(1,1)\rtimes\Z_2) : \phi(x^*x) < \infty\}$;
    \item The linear map $\Lambda \colon \mathcal{N}_{\phi} \to L^2_{q}(\mathrm{SU}(1,1)\rtimes\Z_2)$, $
        a \mapsto \sqrt{1-q^2}\sum_{x \in I_{q^2}}(a|\gamma|\otimes \id_{\H_{q^2}}).\delta^{+}_{0, x, 0}$;
    \item The unital $*$-homomorphism $\pi \colon L^\infty_q(\mathrm{SU}(1,1)\rtimes\Z_2) \to \B(L^2_{q}(\mathrm{SU}(1,1)\rtimes\Z_2))$, $a\mapsto a\otimes \id_{\H_{q^2}}$.
\end{enumerate}
The triple $(L^2_{q}(\mathrm{SU}(1,1)\rtimes\Z_2), \pi, \Lambda)$ gives the GNS construction for the Haar weight $\phi$ \cite{KK2003}. That is, similar to Section~\ref{sec:classicalLocallyCompactGroup}, it satisfies the following properties:
\begin{enumerate}[label = (\alph*)]
    \item $\Lambda(\mathcal{N}_{\phi})$ is dense in $L^2_{q}(\mathrm{SU}(1,1)\rtimes\Z_2)$;
    \item $\langle \Lambda(a), \Lambda(b)\rangle = \phi(a^*b)$ for all $a,b\in \mathcal{N}_\phi$;
    \item $\pi(x)\Lambda(a) = \Lambda(xa)$ for all $x\in L^\infty_q(\mathrm{SU}(1,1)\rtimes\Z_2)$ and $a\in \mathcal{N}_{\phi}$.
\end{enumerate}
The upshot is that the pair $(L^{\infty}_q(\mathrm{SU}(1,1)\rtimes\Z_2), \Delta)$ with Haar weight $\phi$ is now a von Neumann algebraic quantum group \cite{KK2003}. It can be easily verified that $\Lambda(\Phi^{\pm}(n,x,m)) = \delta^{\pm}_{n,x,m}$, such that the GNS-map $\Lambda$ indeed maps the basis elements of the essentially bounded operators to the basis elements of the square integrable functions.  Moreover, it can be noted that $\Lambda$ is invertible on the compactly supported functions. Physically, this map therefore gives a state-operator correspondence. The $*$-homomorphism $\pi$ defines a left action of $L^\infty_q(\mathrm{SU}(1,1)\rtimes\Z_2)$ on $ L^2_q(\mathrm{SU}(1,1)\rtimes\Z_2)$, given simply by $a.v = \pi(a)v$. Similarly, we can define a right action by $v.a = \pi(a^*)v$ for $a\in L^{\infty}_q(\mathrm{SU}(1,1)\rtimes\Z_2)$ and $v\in L^{2}_q(\mathrm{SU}(1,1)\rtimes\Z_2)$ with corresponding GNS-map $\Lambda^*(a) \coloneq \Lambda(a^*)$. Note that the left and right actions are compatible with the inner product, i.e. $\langle v, a.w\rangle = \langle v.a, w\rangle$ for $v,w\in  L^2_q(\mathrm{SU}(1,1)\rtimes\Z_2)$.\\

Although the von Neumann algebraic setting gives a precise and formal framework to define the locally compact quantum group, some intuition is lost on the way: the coordinate functions we used to generate this von Neumann algebra were unbounded (as is common for many physically interesting operators), and thus are themselves not part of it. So, in order to get a bit more intuitive picture, let us extend the $*$-homomorphism and the map $\Lambda$ in a natural way to also include the unbounded operators in $\A_q(\mathrm{SU}(1,1)\rtimes\Z_2)$. In particular, it follows from \eqref{eq:actioncoordinatealgebradirectsum} that the (unbounded) action of $\A_q(\mathrm{SU}(1,1)\rtimes\Z_2)$ on the compactly supported functions in $L^2_{q}(\mathrm{SU}(1,1)\rtimes\Z_2)$ is given by
\begin{equation}
    \label{eq:actionsofSUonGNSspace}
    \begin{split}
        \alpha. \delta_{n,x,m}^{\pm} &= \sqrt{\sign{x}+|x|}~\delta_{n+1, q^{-2}x, m+1}^{\pm},\\
        \gamma. \delta_{n,x,m}^{\pm} &= \sign{x}\sqrt{|x|}~\delta_{n-1, x, m+1}^{\pm},\\
        e . \delta_{n,x,m}^{\pm} &= \sign{x} \delta_{n,x,m}^{\pm}.
    \end{split}
\end{equation}
Note here that $\left(e\gamma^*\gamma\right) . \delta^\pm_{n,x,m}  = x\delta^{\pm}_{n,x,m}$, such that its spectrum is given by $\sigma(e\gamma^*\gamma) = I_{q^2}\cup\{0\}$.\\

With the above action in mind, we define the set of formal infinite linear combinations of basis elements
\begin{equation}
    \F_q(\mathrm{SU}(1,1)\rtimes\Z_2) \coloneq \left\{\sum_{n,x,m, \epsilon} f_{n,x,m}^{\epsilon}\delta_{n,x,m}^{\epsilon} : f_{n,x,m}^{\epsilon} \in \C\right\}.
\end{equation}
which corresponds to functions over $\mathrm{SU}_q(1,1)\rtimes \Z_2$ which are not necessarily square-integrable.\footnote{One way to think about them is as distributions or generalised functions, dual to the compactly supported functions considered above. These can be given an appropriate topology (modelled on the usual case of distributions) to make the formally extended maps considered here continuous. We do not dwell on this issue further, since this is not the main focus here.} This space of functions splits into the direct sum $\F_q = \F^+_q\oplus \F^-_q$, where $\F_q^+$ and $\F_q^{-}$ are spanned by $\delta^+_{n,x,m}$ and $\delta^-_{n,x,m}$, respectively. The GNS-map $\Lambda$ can now be formally extended to the map
\begin{equation}
    \tilde \Lambda \colon \A_q(\mathrm{SU}(1,1)\rtimes\Z_2) \to \F_q^+(\mathrm{SU}(1,1)\rtimes\Z_2), \quad \tilde \Lambda(a) = \sqrt{1-q^2}\sum_{x\in I_{q^2}}(a|\gamma|\otimes \id).\delta^{+}_{0,x,0}
\end{equation}
and analogously we extend $\Lambda^*$ to $\tilde \Lambda^*$. This trick allows us to turn the elements in $\F_q^\pm$ into actual functions over the locally compact quantum group. Namely, the coordinates $n,m$ in a basis element $\delta^{\pm}_{n,x,m}$ correspond to Fourier components, while the coordinate $x$ corresponds simply to an evaluation at that point on the $q$-interval $I_{q^2}$. We therefore consider the basis elements as functions $\delta^\pm_{n,y,m}\colon [0,2\pi)\times I_{q^2}\times [-2\pi,2\pi) \to \C$ given by
\begin{equation}
    \delta^\pm_{n,y,m}(\theta, x, \psi) \coloneq e^{\frac{i}{2}(n\theta +m\psi)}\delta_{\pm y q^{n+m}}(x),
\end{equation}
where $\delta_y$ is the $q^2$-delta function as defined in $\eqref{eq:qdeltafunction}$. Note here that the domains of $\theta$ and $\psi$ are different because of the additional constraint $n\equiv m \text{ mod }2$. Moreover, the parametrisation exactly corresponds to the Cartan decomposition of $\mathrm{SU}(1,1)$ as given in \eqref{eq:classicalCartanDecomp}. In this way, one can interpret the elements in $\F^+_q$ as functions $\mathrm{SU}_q(1,1)\rtimes \Z_2\to \C$, and we see that the `quantum group manifold' is now parametrised by the (partially) discretised coordinates. That is, the operator-algebraic description allows us to describe the non-commutative group manifold as a discretised (but commutative) `manifold' through the spectra of the non-commutative operators.\\

Similarly, through the GNS-map $\tilde \Lambda$, the elements in the coordinate algebra can be mapped to functions on the discretised coordinates. For example, for the generators $\alpha$ and $\gamma$ in $\A_q(\mathrm{SU}_q(1,1)\rtimes \Z_2)$, this gives
\begin{equation}
    \label{eq:coordinateAlgebraFunctions}
    \begin{split}
        \tilde \Lambda(\alpha)(\theta, x, \psi) &= e^{i\frac{\theta + \psi}{2}}\sqrt{\sign{x} + |x|},\\
        \tilde \Lambda(\gamma)(\theta, x, \psi) &= e^{i\frac{\psi-\theta}{2}} \sign{x}\sqrt{|x|}.
    \end{split}
\end{equation}
In the $q\to 1$ limit and for $x = \sinh^2\rho/2$ with $x\geq 0$, we recognise the same expressions as in the classical Cartan decomposition \eqref{eq:classicalCartanDecomp}. For $x<-1$ and $x = -\cosh^2\rho/2$, these correspond to functions over $\mathrm{SU}(1,1)v$. For arbitrary $f,g\in \F_q^{+}$, the inner product is given by
\begin{equation}
    \langle f , g\rangle = \int_{0}^{2\pi}\frac{d\theta}{2\pi}\int_{-2\pi}^{2\pi}\frac{d\psi}{4\pi} \int_{I_{q^2}}d_{q^2}x~\overline{f(\theta, x, \psi)}g(\theta, x, \psi),
\end{equation}
which, in the limit $q\to 1$, reduces to the Haar measure of $\mathrm{SU}(1,1)\rtimes \Z_2$. Indeed, for $x\geq 0$ and $x = \sinh^2 \rho/2$, this is exactly the Haar measure as given in \eqref{eq:classicalCartanHaarmeasure}.\\

Let us also comment on the functions in $\F^-_q$. On the level of the algebra, these functions correspond to the elements in $L_q^{\infty}(\mathrm{SU}(1,1)\rtimes \Z_2)$ which contain a non-trivial factor of $u$. The functions in $\F_q^-$ corresponding to $u\alpha$ and $u\gamma$ are given by
\begin{equation}
    \begin{array}{ll}
        \tilde \Lambda(u\alpha)(\theta, x, \psi) &= e^{i\frac{\theta + \psi}{2}}\sqrt{\sign{x} + |x|}  \\
        \tilde \Lambda(u\gamma)(\theta, x, \psi) &= e^{i\frac{\psi-\theta}{2}} \sign{x}\sqrt{|x|} 
    \end{array} \qquad \text{for } |x| \geq 1
\end{equation}
and they are zero for $|x| < 1$. Note that these are the same functions as in \eqref{eq:coordinateAlgebraFunctions}, but with the additional constraint $|x|\geq 1$. Therefore, classically, one can think of $u\alpha$ and $u\gamma$ as functions over a restricted subspace of $\mathrm{SU}(1,1)$. If we write the coordinates in the Gauss decomposition \eqref{eq:classicalGaussDecomposition}, this subspace corresponds to a restriction on the coordinate $\varphi$, which, as we will see in Section~\ref{sec:lengthpositivity}, gives rise to length positivity.

\subsection{The regular representations}
\label{sec:quantumregularrepresentations}
We can now also define the action of the universal enveloping algebra $\U_q(\mathfrak{su}(1,1))$ on the square-integrable functions $L^{2}_q(\mathrm{SU}(1,1)\rtimes\Z_2)$. Similar to the coordinate algebra, these actions are given by unbounded operators and are therefore defined on the compactly supported functions. Inspired by \cite{MMNSU19902} and \cite{GKK2010}\footnote{In particular, it can be seen that our left action is very similar to the one in Definition~4.1 in \cite{GKK2010}, but adapted for our convention of basis elements. However, that paper didn't consider the right action, and also the compatibility with the action of the coordinate algebra $\A_q$ and the GNS map $\Lambda$ was not explicitly discussed}, we define the left action to be given by
\begin{equation}
    \label{eq:leftregularaction}
    \begin{split}
        K.\delta_{n,x,m}^{\pm} &= q^{\frac{m}{2}} \delta_{n,x,m}^{\pm},\\
        (q-q^{-1})E. \delta_{n,x,m}^{\pm} &= -q^{\frac{m+1}{2}}\sqrt{1 \pm q^{-n-m-2}x^{-1}}\delta_{n,x,m+2}^{\pm} + q^{-\frac{m+1}{2}}\sqrt{1 + x^{-1}}\delta_{n, q^{-2}x, m+2}^{\pm},\\
        (q-q^{-1})F. \delta_{n,x,m}^{\pm} &= q^{\frac{m-1}{2}}\sqrt{1 \pm q^{-n-m}x^{-1}}\delta_{n,x,m-2}^{\pm} - q^{-\frac{m-1}{2}}\sqrt{1 + q^{-2}x^{-1}}\delta_{n,q^2x, m-2}^{\pm},
    \end{split}
\end{equation}
and the right action to be given by
\begin{equation}
    \label{eq:rightregularaction}
    \begin{split}
        \delta_{n,x,m}^{\pm} . K &= q^{-\frac{n}{2}}\delta_{n,x,m}^{\pm},\\
        (q-q^{-1})\delta_{n,x,m}^{\pm}.E &= \mp q^{\frac{n+1}{2}}\sqrt{1 \pm q^{-n-m-2}x^{-1}}\ e.\delta_{n+2, x, m}^{\pm}+q^{-\frac{n+1}{2}}\sqrt{1+x^{-1}}\ e.\delta_{n+2, q^{-2}x, m}^{\pm},\\
        (q-q^{-1})\delta_{n,x,m}^{\pm}.F &= \pm q^{\frac{n-1}{2}}\sqrt{1 \pm q^{-n-m}x^{-1}}\ e.\delta_{n-2,x,m}^{\pm}-q^{-\frac{n-1}{2}}\sqrt{1 + q^{-2}x^{-1}}\ e.\delta_{n-2, q^2x, m}^{\pm}.
    \end{split}
\end{equation}
Note here that the Cartan element $K$ has diagonal left and right actions on this basis, corresponding to the Cartan decomposition. These actions reduce to \eqref{eq:classicalCartanactions} in the limit $q\to 1$. Moreover, these actions have been constructed in such a way that they satisfy the following properties:
\begin{enumerate}[label = (\alph*)]
    \item The left action and right action of $\U_q(\mathfrak{su}(1,1))$ commute with each other;
    \item The action of $\A_q(\mathrm{SU}(1,1)\rtimes\Z_2)$ on $L_q^2(\mathrm{SU}(1,1)\rtimes\Z_2)$ is an equivariant representation with respect to the left and right action of $\U_q(\mathfrak{su}(1,1))$, that is\footnote{Here, we use Sweedler notation for the coproduct, that is $\Delta(X) = \sum X_{(1)}\otimes X_{(2)}$. Moreover, in the classical limit where the coproduct is given by $\Delta(X) = X\otimes 1 + 1\otimes X$, the relations reduce to $X\triangleright a = [X,a]$ and $a\triangleleft X = [a,X]$, thus motivating these relations.}
    \begin{equation}
            X.(a.f) = \sum(X_{(1)}\triangleright a).(X_{(2)} . f) \quad \text{and}\quad  (f. a).X = \sum (f.X_{(1)}).(a\triangleleft X_{(2)});
    \end{equation}
    \item $X . \tilde \Lambda(a) = \tilde \Lambda(X \triangleright a)$ and $\tilde \Lambda^*(a).X = \tilde  \Lambda^*(a \triangleleft X)$;
\end{enumerate}
for all $X \in \U_q(\mathfrak{su}(1,1))$, $a\in \A_q(\mathrm{SU}(1,1)\rtimes\Z_2)$ and $f\in L_q^2(\mathrm{SU}(1,1)\rtimes\Z_2)$ compactly supported. These properties can be verified by explicit calculation. The compatibility relations now allow us to define the left and right action of $\U_q(\mathfrak{su}(1,1))$ on the compactly supported functions in $L^\infty_q(\mathrm{SU}(1,1)\rtimes \Z_2)$ given by
\begin{equation}
    X \triangleright a \coloneq (\Lambda)^{-1}(X.\Lambda(a)), \quad a\triangleleft X \coloneq (\Lambda^*)^{-1}(\Lambda^*(a).X),
\end{equation}
for $X\in \U_q(\mathfrak{su}(1,1))$ and $a\in L^\infty_q(\mathrm{SU}(1,1)\rtimes \Z_2)$ with $a$ compactly supported. In summary, all the left and right actions are compatible with each other and with the GNS-map $\Lambda$, and we get the following picture:
\begin{equation}
    \label{eq:compatibilitypicturequantumgroups}
    \begin{tikzpicture}[baseline={([yshift=-.5ex]current bounding box.center)}]
        \node  (a) {$L_q^\infty(\mathrm{SU}(1,1)\rtimes\Z_2)$\strut};
        \node[right=2mm of a] (b) {};
        \node[right=2mm of b] (c) {$\U_q(\mathfrak{su}(1,1))$\strut};
        \node[below=7mm of b] (d) {$L_q^2(\mathrm{SU}(1,1)\rtimes\Z_2)$.\strut};

        \draw[->, transform canvas = {yshift=-1mm, xshift = 1mm}] (a.0) arc (-150:150:2.5mm);
        \draw[->, transform canvas = {yshift=1mm, xshift = -4mm}] (d.120) arc (-30:270:2.5mm);
        \draw[->, transform canvas = {yshift=-0.5mm, xshift = 1mm}] (d.30) arc (-100:200:2.5mm);
        \draw[->] (a.south) to [out=-90, in = 180] node[left=0mm] {$\Lambda$} (d.west) ;
    \end{tikzpicture}
\end{equation}

In the previous section, we upgraded the coordinate algebra $\A_q$ from a purely algebraic object to its von Neumann algebraic incarnation, and one can now do the same for the universal enveloping algebra $\U_q$. In particular, the von Neumann algebraic counterpart of this universal enveloping algebra is given by the (quantum) \textit{Pontryagin dual}. The Pontryagin dual here denotes the full, non-abelian unitary dual of the operator algebraic quantum group $L^\infty_q(\mathrm{SU}(1,1)\rtimes \Z_2)$, and $\hat M_q$ should be thought of as the space of `essentially bounded functions' on that unitary dual. This Pontryagin dual von Neumann algebraic quantum group can be readily described by using a special unitary operator, called \textit{multiplicative unitary} $W\in \B(\K\otimes \K)$, where $\K \coloneq L^2_q(\mathrm{SU}(1,1)\rtimes \Z_2)$, 
which in fact encodes most of the structure of the original von Neumann algebraic quantum group $M$ \cite{CK2010}: for example, the coproduct on $L^{\infty}_q(\mathrm{SU}(1,1)\rtimes \Z_2)$ is given by $\Delta(x) = W^*(1\otimes x)W$. The Pontryagin dual\footnote{For a locally compact abelian group $G$, it can be shown from here that $\hat M \cong L^{\infty}(\hat G)$, with $\hat G$ the dual group of irreducible unitary representations \cite{CK2010}. This definition is therefore a far-reaching generalisation of the ordinary Pontryagin dual for locally compact abelian groups.} is then defined as the von Neumann algebra $\hat M_q \coloneq \overline{\{(\omega\otimes \iota)(W) : \omega \in \B(\K)_*\}}\subset \B(\K)$ with coproduct $\hat \Delta(\hat x) = \Sigma W(\hat x\otimes 1)W^*\Sigma$, where $\Sigma$ is the flip operator $\Sigma\colon x\otimes y\to y\otimes x$ and $\hat x\in \hat M_q$. A remarkable fact is that $\hat M_q$ is again a von Neumann algebraic quantum group \cite{KV2003}. The intuition here is as just in classical Lie group case: $\hat{M}_q$ should be thought of as an appropriate analogue of the group convolution algebra of (Hecke) integral operators acting on the functions on the group, whereas the subalgebra $\U_q(\mathfrak{su}(1,1))$ should be thought of as representing `convolutions with the distributions supported at the quantum group unit element'.\\

Of course, the generators $E,F,K$ of the universal enveloping algebra  cannot themselves belong to $\hat{M}_q$ (after all, as a von Neumann algebra, it can only contain bounded operators), but it was shown in \cite{GKK2010} that these generators are affiliated with $\hat M_q$, which is `as close as an unbounded operator can get' to belonging to / generating a von Neumann algebra. However, since the quantum group in consideration is the normaliser $\mathrm{SU}_q(1,1)\rtimes \Z_2$, these operators are not enough to fully generate the dual (by affiliation construction). In particular, it is expected for the normaliser that the (symmetrised) Casimir operator $\tilde \Omega \coloneq \frac{1}{2}((q-q^{-1})^2\Omega + 2)$ does not commute with all elements of $\hat M_q$. Indeed, it can be shown  \cite{GKK2010} that there is a $\Z_2$ grading given by $\hat M_q = \hat M^+_q \oplus \hat M^-_q$ such that $a\tilde \Omega \subseteq \tilde \Omega a$ for $a\in \hat M_q^{+}$ and $b\tilde \Omega \subseteq -\tilde \Omega b$ for $b\in \hat M_q^{-}$. Therefore, there are additional generators $U_0^{+-}, U_0^{-+} \in \hat M_q^{-}$ for the Pontryagin dual that map eigenvectors of $\tilde \Omega$ with eigenvalue $\lambda$ to eigenvectors with eigenvalue $-\lambda$ \cite{GKK2010}. By blending in these additional generators, one could replace $\U_q(\mathfrak{su}(1,1))$ with $\hat M_q$ in \eqref{eq:compatibilitypicturequantumgroups}, but we won't need it for the purposes of this paper.\\

Lastly, it was shown in \cite{GKK2010} that the left regular representation decomposes into a direct integral of the principal unitary series, the discrete series and the strange series.\footnote{We refer to \cite{BK1993} for the classification of irreducible representations of the quantum universal enveloping algebra $\U_q(\mathfrak{su}(1,1))$.} The appearance of the principal unitary and the discrete series is, of course, expected from the regular representations of classical $\mathrm{SU}(1,1)$ \cite{K2018, SL2R}. The appearance of the strange series can be thought of as an additional novel, inherently quantum, effect.

\subsection{The quantum Gauss decomposition}
\label{sec:quantumGaussDecomp}
In order to describe the reduction to the double-scaled SYK model, we need to derive the {\it quantum Gauss decomposition} of the (von Neumann algebraic) quantum group, in parallel to what was done in Section~\ref{sec:classicalGaussdecomp} for the classical case. We follow an approach similar to that of \cite{K1993, KS2001}. Most calculations for this section are rather technical and are therefore deferred to Appendix~\ref{sec:calcsqgactions} to maintain readability. Moreover, we will frequently use results for special functions as given in Appendix~\ref{sec:specialfunctions}.\\

To begin with, let us consider the quantum equivalent of the parabolic elements. We will look for them among the special class of elements given by
\begin{equation}
    \label{eq:deftwistedprimitiveelement}
    Y_s \coloneq q^{\frac{1}{2}}E - q^{-\frac{1}{2}}F - (s+s^{-1})\frac{K - K^{-1}}{q - q^{-1}}
\end{equation}
for $s\in \R \setminus \{0\}$. These are the so-called \textit{twisted primitive elements}, whose coproduct reads $\Delta(Y_s) = K\otimes Y_s + Y_s \otimes K^{-1}$ and thus closely resembles the coproduct of a classical Lie algebra element. It can be seen that, with appropriate scaling of parameters, these reduce to the parabolic elements \eqref{eq:classicalParabolicElements} in the classical limit, namely $\lim_{q\to 1}Y_{q^{n}} = 2i\tilde E$ and $\lim_{q\to 1} Y_{-q^{m}} = 2i \tilde F$. Moreover, in the limit $s\to \infty$, we define
\begin{equation}
    Y_{\infty} \coloneq \lim_{s\to\infty}s^{-1}(q-q^{-1})Y_s = K^{-1} - K,
\end{equation}
which, in the classical limit, corresponds to the classical Cartan element. Unlike the classical case, however, these twisted primitive elements are not self-adjoint. Therefore, we instead consider the slightly modified elements $Y_sK$ and $K^{-1}Y_s$, which are self-adjoint. For future convenience, we also introduce the symmetrised (in the sense of having a symmetric spectrum around $0$) twisted primitive elements, given by
\begin{equation}
    \label{eq:normalizedTwistedPrimitives}
    \tilde Y_t^{L} \coloneq \frac{1}{2}\left[(q-q^{-1})K^{-1}Y_t + (t+t^{-1})\right], \quad \tilde Y_s^{R} \coloneq \frac{1}{2}\left[(q-q^{-1})Y_sK - (s+s^{-1})\right].
\end{equation}
The left and right actions of these versions of twisted primitive elements will correspond to symmetric (i.e. formally Hermitian) operators in the left and right regular representations.\\

The quantum Gauss decomposition\footnote{Throughout the paper, we will occasionally refer to the decomposition with general values of $s,t$ as the `quantum Gauss decomposition'.} now corresponds to functions that have diagonal left action of $\tilde Y_t^L$ and diagonal right action of $\tilde Y_s^R$ for $t = q^{n}$ and $s = -q^{m}$. Note here that, unlike the corresponding parabolic elements in the classical case, the variables $s$ and $t$ can take on any power of $q$. In fact, as was shown in \cite{S2003}, diagonalisation with respect to the twisted primitive elements gives the quantum group a dynamical structure, and the variables $s$ and $t$ are the corresponding dynamical parameters. In $\A_q(\mathrm{SU}(1,1)\rtimes\Z_2)$, we define the elements
\begin{equation}
    \label{eq:coordinateAlgebraGauss}
    \alpha_{st} = \alpha + qt^{-1}e\gamma^* - s^{-1}\gamma - s^{-1}t^{-1}e\alpha^*, \quad \gamma_{st} = -s^{-1}\alpha - qs^{-1}t^{-1}e\gamma^* + \gamma + t^{-1}e\alpha^*,
\end{equation}
on which $\tilde Y_t^{L}$ acts diagonally from the left and $\tilde Y_s^{R}$ diagonally acts from the right \cite{K1993, KS2001}.\footnote{We use a different convention than \cite{KS2001}. In particular, we consider slightly different elements when looking at the left and right action of the twisted primitive elements, and use a different normalisation such that the limit $s,t \to \infty$ of these elements is well-defined.} Here, it can be seen that $\alpha_{\infty, \infty} = \alpha$ and $\gamma_{\infty,\infty} = \gamma$, which is also expected as the limit $s,t\to \infty$ corresponds to the Cartan decomposition. Moreover, these elements satisfy dynamical commutation relations \cite{KS2001}. For example, we have $\alpha_{q^{-1}s, qt}\gamma_{st} = q\gamma_{qs, qt}\alpha_{st}$. The dynamical structure reveals itself in the way the parameters change values upon multiplication of the corresponding algebra elements.\\

Next, we can consider the generalisation of the spherical element $e\gamma^*\gamma$, i.e. the position operator, for the Gauss decomposition. That is, consider an element on which $Y_t$ and $Y_s$ act trivially from the left and the right, respectively. We define \cite{KS2001}:\footnote{\label{footnote:sphericalelement}We remark here that the spherical element in \cite{KS2001} was defined as an element in $\mathrm{SL}_q(2,\C)$. Thus, the corresponding spherical element of the normaliser is obtained by imposing the $*$-structure of the latter, as opposed to the one of $\mathrm{SU}_q(1,1)$.}
\begin{equation}
    \label{eq:sphericalelement}
    \begin{split}
        \rho_{st} \coloneq e\gamma^*_{s,qt} \gamma_{s,qt} - s^{-2} - q^{-2}t^{-2}.
    \end{split}
\end{equation}
Here, it can be noted that $\rho_{\infty, \infty} = e\gamma^*\gamma$, i.e. it reduces to the spherical element for the Cartan decomposition in the limit $s,t\to \infty$. Moreover, it is an \textit{$(s,t)$-spherical element}, i.e. $Y_t\triangleright \rho_{st} = 0 = \rho_{st}\triangleleft Y_s$. In particular, by a straightforward adaptation of the argument from \cite{KS2001} (see footnote~\ref{footnote:sphericalelement}) one sees that $\rho_{st}$ (together with the trivial spherical element $e$) generates the subalgebra of $\A_q(\mathrm{SU}(1,1)\rtimes\Z_2)$ of $(s,t)$-spherical elements, i.e.
\begin{equation}
    \A_q((Y_s) \backslash \mathrm{SU}(1,1)\rtimes\Z_2 /(Y_t))  = \langle \rho_{st},e\rangle,
\end{equation}
where $(Y_s)$ denotes the Lie group generated by the classical limit of $Y_s$. The elements $\alpha_{st}$, $\gamma_{st}$ and $\rho_{st}$ satisfy the relations
\begin{equation}
    \label{eq:sphericalElementcommutation}
    \alpha_{s,qt}\rho_{st} = q^2\rho_{qs,qt}\alpha_{s,qt} \quad \text{and}\quad \gamma_{s,qt}\rho_{st} = \rho_{q^{-1}s, qt}\gamma_{s,qt}.
\end{equation}
Classically, i.e. in the $q\to 1$ limit, it can be calculated that $\rho_{-1,1}(g(\gamma_L, \varphi, \gamma_R)) = -2+4e^{2\varphi}$ in the Gauss decomposition \eqref{eq:classicalGaussDecomposition}. Therefore, the spherical element $\rho_{st}$ corresponds to the coordinate $\varphi$, and in the quantum case, a deformation of this coordinate. Moreover, the eigenvalues of the left and right action of $Y_tK$ and $K^{-1}Y_s$ will correspond to the quantised momenta $p_L$ and $p_R$.\\

Thus, we are interested in finding a basis of $L^2_q(\mathrm{SU}(1,1)\rtimes\Z_2)$ which diagonalises the left action of $\tilde Y_t^L$, the right action of $\tilde Y_s^{R}$ and the left/right action of $\rho_{st}$. This diagonalisation has to be done in several steps. As a start, we will derive the \textit{quantum Iwasawa decomposition}, in which we diagonalise the basis first with respect to $\rho_{\infty, t}$ and then with respect to the left action of $\tilde Y_t^{L}$. Afterwards, we will diagonalise with respect to $\rho_{st}$ and the right action of $\tilde Y_s^{R}$ to obtain the quantum Gauss decomposition. This derivation will therefore take four different diagonalisation steps.\\

Using the left and right action of the universal enveloping algebra in \eqref{eq:leftregularaction} and \eqref{eq:rightregularaction}, the actions of $\tilde Y_t^L$, $\tilde Y_s^R$ On the basis $\{\delta^{\pm}_{n,x,m}\}$, as given in \eqref{eq:Lq2Cartanbasis}, can be calculated to be
\begin{equation}
    \label{eq:leftrighttwistedpprimitiveaction}
    \begin{split}
        &\tilde Y_t^{L}.\delta_{n,x,m}^\pm = A^\pm(n,x,m)\delta_{n,x,m+2}^\pm + \frac{1}{2}q^{-m}x^{-1}t\rho_{\infty, t}.\delta_{n,x,m}^{\pm} +A^\pm(n,x,m-2)\delta_{n,x,m-2}^{\pm},\\
        &\delta_{n,x,m}^{\pm}. \tilde Y_s^R = \pm A^\pm(n,x,m)e.\delta_{n+2, x, m}^{\pm} - \frac{1}{2}q^{-n}x^{-1}s\rho_{s, \infty} . \delta_{n,x,m}^{\pm} \pm A^\pm (n-2, x, m)e.\delta_{n-2, x, m}^{\pm},
    \end{split}
\end{equation}
where $A^\pm(n,x,m) = -\frac{1}{2}\sqrt{1 \pm q^{-n-m-2}x^{-1}}$. In order to diagonalise with respect to, for example, the left action of $\tilde Y_t^L$, we would like to bring the form of the action into that of a three-term recurrence relation, such that we can use the results on special functions as described in Appendix~\ref{sec:specialfunctions}.\\

Therefore, let us begin with the diagonalisation of $\rho_{\infty,t}$. From this point onwards, we will always assume $|s|,|t| > q$. It can be calculated that the spherical elements with $s=\infty$ is given by $t\rho_{\infty, t} = \gamma \alpha + \alpha^*\gamma^* + (t+t^{-1})e\gamma^*\gamma$, such that its action is given by
\begin{equation}
    \begin{split}
        \left(t\rho_{\infty, t}\right).\delta_{n,x,m}^\pm= q^{-1}x\sqrt{1+x^{-1}}\delta_{n, q^{-2}x, m+2}^{\pm}+ (t+t^{-1})x \delta_{n,x,m}^{\pm} +q x\sqrt{1+q^{-2}x^{-1}}\delta_{n, q^{2}x, m-2}^{\pm}.
    \end{split}
\end{equation}
For $x = -q^{-2k}$ with $k\in \Z_{>0}$ one can recognize the Jacobi operator for the Al-Salam-Carlitz II polynomials \eqref{eq:alsalamcarlitzoperator}. Similarly, for $x = q^{-2k}$ with $k\in \Z$ one can recognise the doubly-infinite Jacobi operator of the Ciccoli-Koelink-Koornwinder functions \eqref{eq:qlaguerreoperator}. Therefore, we define
\begin{equation}
    \label{eq:preIwasawaBasis}
    \begin{split}
        g_{n,x,m}^{\pm, t} &\coloneq \sum_{k\in \Z_{> 0}}(-1)^kp_{k-1}^{t^{-2}}(tq^{2}x; q^2) \delta^\pm_{n, -q^{-2k}, m+2k} \quad \text{for } x\in \{-q^{-2k} : k\in \Z_{>0}\},\\
        g_{n,x,m}^{\pm, t} &\coloneq \sum_{k\in \Z}\tilde V_k^t(tx; 1; q^2) \delta_{n, q^{-2k}, m + 2k}^\pm \quad \text{for } x\in \{q^{2k} : k\in \Z\}\cup\{-t^{-2}q^{2k} :k\in \Z_{\geq 0}\},\\
    \end{split}
\end{equation}
where $p_k^{t^{-2}}$ and $\tilde V_k^t$ are the Al-Salam-Carlitz II functions and the Ciccoli-Koelink-Koornwinder functions as defined in Appendix~\ref{sec:AlSalamCarlitzII} and \ref{sec:dualqLaguerre}, respectively. From Propositions~\ref{prop:alsalamcarlitzspectralmeasure} and \ref{thm:spectralmeasureqlaguerre}, it follows that these elements form an orthogonal basis of $L^2_q(\mathrm{SU}(1,1)\rtimes \Z_2)$ with $\rho_{\infty, t}.g_{n,x,m}^{\pm, t} = xg^{\pm, t}_{n,x,m}$ for $x\in \sigma(\rho_{\infty, t})$, where the spectrum of (an appropriate self-adjoint extension\footnote{For $|t| > q^{-1}$, $\rho_{\infty, t}$ is essentially self-adjoint, and thus has a unique self-adjoint extension. Moreover, for $q <|t|< q^{-1}$, the operator has a self-adjoint extension such that the elements $g^{\pm, t}_{n,x,m}$ are contained in the domain of this extension (see Propositions~\ref{prop:alsalamcarlitzspectralmeasure} and \ref{thm:spectralmeasureqlaguerre}).} of) $\rho_{\infty, t}$ is given by
\begin{equation}
    \sigma(\rho_{\infty, t}) = \{-q^{-2k} : k\in \Z_{>0}\}\cup \{-t^{-2} q^{2k} : k \in \Z_{\geq 0}\}\cup \{q^{2k} : k\in \Z\}\cup \{0\}.
\end{equation}
Here, it can be noted that in the limit $t\to \infty$, we retrieve the spectrum of $\rho_{\infty,\infty} = e\gamma^*\gamma$ (see e.g. under \eqref{eq:actionsofSUonGNSspace}), i.e. the coordinates corresponding to the Cartan decomposition. For future use, we define $e \colon \sigma(\rho_{\infty, t}) \to \{\pm 1\}$ via $e.g^{\pm, t}_{n,x,m} = e(x) g^{\pm , t}_{n,x,m}$. Also, note that the conditions on the basis \eqref{eq:Lq2Cartanbasis} imply that $g^{\pm, t}_{n,x,m}$ is only non-zero if $n \equiv m \text{ mod } 2$ and $\pm e(x) q^{-n-m} > -1$.\\

From \eqref{eq:leftrighttwistedpprimitiveaction} it now follows that the left action of $\tilde Y_t^L$ on $g^{\pm,t}_{n,x,m}$ is given by
\begin{equation}
    \label{eq:YtLactionong}
    \tilde Y_t^{L}.g_{n,x,m}^{\pm, t}= A(n,x,m)g_{n,x,m+2}^{\pm, t} + \frac{1}{2}e(x)q^{-m}txg_{n,x,m}^{\pm, t} + A(n,x,m-2)g_{n,x,m-2}^{\pm, t},
\end{equation}
with $A(n,x,m) = -\frac{1}{2}\sqrt{1 \pm e(x)q^{-n-m-2}}$. Here, one can recognise the three-term recurrence relation of the little $q^2$-Jacobi functions \eqref{eq:JacobilittleqJacobi} for $\pm e(x) = 1$, and the recurrence relation for the Al-Salam-Chihara polynomials \eqref{eq:JacobiOperatorAl-Salam-Chihara} for $\pm e(x) = -1$. Therefore, for $n\in \Z$ and $x\in \sigma(\rho_{\infty,t})$ fixed, it follows from Propositions~\ref{thm:spectralmeasurelittleqJacobi} and \ref{thm:spectralmeasureboundedAl-Salam_Chihara} that the spectrum of (the appropriate self-adjoint extension\footnote{Since the operator $\tilde Y_t^{L}$ is essentially self-adjoint, this extension is unique.} of) $\tilde Y_t^{L}$ acting on the subspace $\{g^{\pm e(x), t}_{n,x,m}\}_{m\in \Z}$ is given by
 \begin{equation}
    \label{eq:spectrumLeftPrimitiveConditional}
    \begin{split}
        \sigma(\tilde Y_t^L)_{n,x}^{-e(x)} &= [-1,1]\cup \left\{e(x)\mu(q^{-n}(tx)^{-1}q^{-2k}) : |q^{-n}(tx)^{-1}q^{-2k}| < 1\ \mathrm{for}\ k\in \Z_{> 0}\right\},\\
        \sigma(\tilde Y_t^L)_{n,x}^{e(x)} &= [-1,1]\cup\left\{-e(x)\mu(q^n tx q^{-2k}) : |q^n txq^{-2k}| < 1\ \mathrm{for}\ k\in \Z_{\geq 0}\right\}\\
        &\qquad \qquad \cup \left\{e(x)\mu(q^{-n}(tx)^{-1}q^{-2k}) : |q^{-n}(tx)^{-1}q^{-2k}| < 1\ \mathrm{for}\ k\in \Z\right\},
    \end{split}
\end{equation}
where $\mu(x) = \frac{1}{2}(x+x^{-1})$. For $\mu(\xi_{\pm}) \in \sigma(\tilde Y_t^L)_{n,x}^{\pm}$, the corresponding eigenvectors are given by
\begin{equation}
    \label{eq:IwasawaBasis}
    \begin{split}
        \delta_{n,x,\xi_{-e(x)}}^{-e(x), (\infty, t)} &= \sum_{k\in \Z_{\geq 0}}(-1)^k h_k(\mu(\xi_{-e(x)}); e(x)q^{2+n}tx, 0 ; q^2) g^{-e(x), t}_{n,x, -n-2-2k},\\
        \delta_{n,x, \xi_{e(x)}}^{e(x), (\infty, t)} &= \sum_{k\in \Z}(-1)^k j_k(\mu(\xi_{e(x)}); 0, -e(x)q^{-n} (tx)^{-1}; q^2, -1) g^{e(x), t}_{n,x, -n-2 + 2k},
    \end{split}
\end{equation}
where $j_k$ and $h_k$ are the little $q^2$-Jacobi functions and the Al-Salam-Chihara polynomials as defined in Appendix~\ref{sec:littleqJacobi} and \ref{sec:AlSalamChihara}, respectively. Note that these $\delta^{\pm, (\infty, t)}_{n,x,\xi}$ form a (generalised) orthonormal basis of $L^2_q(\mathrm{SU}(1,1)\rtimes\Z_2)$, that is 
\begin{equation}
    \langle \delta_{n,x,\xi}^{\epsilon, (\infty, t)} , \delta^{\epsilon', (\infty,t)}_{n',x',\xi'}\rangle = \delta_{\epsilon, \epsilon'}\delta_{n,n'}\delta_{x,x'}\delta_\nu(\mu(\xi) - \mu(\xi')),
\end{equation}
where $\epsilon \in \{\pm 1\}$ and $\delta_\nu$ is the delta-function with respect to the measure $\nu$ as defined in \eqref{eq:specialFunctionsMeasure}, which is just the Dirac-delta function for $\mu(\xi), \mu(\xi')$ in the continuous spectrum and the Kronecker delta function in the discrete spectrum.\\

We can now see that these basis elements correspond to the Iwasawa decomposition of the quantum group $\mathrm{SU}_q(1,1)\rtimes \Z_2$, that is
\begin{equation}
        \tilde Y_t^{L}.\delta_{n,x,\xi}^{\pm, (\infty, t)} = \mu(\xi)\delta_{n,x,\xi}^{\pm, (\infty, t)}, \quad \rho_{\infty, t}.\delta_{n,x, \xi}^{\pm, (\infty, t)} = x\delta_{n,x,\xi}^{\pm, (\infty, t)}, \quad \delta_{n,x,\xi}^{\pm, (\infty, t)}.K = q^{-\frac{n}{2}}\delta_{n,x,\xi}^{\pm, (\infty, t)}.
\end{equation}
From \eqref{eq:spectrumLeftPrimitiveConditional} it follows that the total spectrum of $\tilde Y_t^L$ on $L^2_q(\mathrm{SU}(1,1)\rtimes\Z_2)$ is given by
\begin{equation}
    \label{eq:spectrumlefttwisted}
    \sigma(\tilde Y_t^L) = \{-\mu(tq^m) : |tq^m| < 1, m \in \Z\}\cup [-1,1]\cup \{\mu(t^{-1}q^m) : |t^{-1}q^m| < 1, m\in \Z\}.
\end{equation}
However, not all $\delta_{n,x,\xi}^{\pm, (\infty, t)}$ are non-zero for general $n,x,\xi$, as the conditions in \eqref{eq:spectrumLeftPrimitiveConditional} need to be satisfied.\\

The action of the coordinate algebra on the basis in the Iwasawa decomposition has been calculated in Appendix~\ref{sec:actionsIwasawaDecomp} and is given by
\begin{equation}
    \label{eq:coordinateAlgebraActionIwasawa}
    \begin{split}
        \alpha_{\infty, qt} . \delta_{n,x,\xi}^{\pm, (\infty, t)} &= -e(x)\sqrt{e(x)(1+x)}~\delta_{n+1, q^{-2}x, \xi}^{\pm, (\infty, qt)},\\
        \gamma_{\infty, qt}.\delta_{n,x,\xi}^{\pm, (\infty, t)} &= e(x)\sqrt{e(x)(q^{-2}t^{-2} + x)} ~\delta_{n-1, x, \xi}^{\pm, (\infty, qt)},
    \end{split}
\end{equation}
The dynamical structure of the quantum group here can be seen from the fact that the value of $t$ changes when acting on basis elements with the coordinate algebra.
Note that if we take $\xi = t^{-1}q^m$, then the vectors $\delta^{\pm,(\infty,t)}_{n,x,tq^{m}}$ reduce to $\delta^{\pm}_{n,x,m}$ in the limit $t\to \infty$. Similarly, the actions of the coordinate algebra reduce to those of the Cartan decomposition \eqref{eq:actionsofSUonGNSspace} up to a phase difference.\\

Next, we would like to diagonalise this basis with respect to the right action of $\tilde Y_s^{R}$. As calculated in Appendix~\ref{sec:actionsIwasawaDecomp}, the right action of the twisted primitive element is given by
\begin{equation}
    \label{eq:rightactionTwistedPrimitiveIwasawa}
    \begin{gathered}
        \delta^{\pm, (\infty, t)}_{n,x,\xi}.\tilde Y_s^R = A(n,x,\xi) \delta^{\pm, (\infty, t)}_{n+2,x, \xi} -q^{-n-1}(tx)^{-1}s\tilde \rho_{st}.\delta^{\pm, (\infty, t)}_{n,x,\xi} + A(n-2,x,\xi)\delta^{\pm, (\infty, t)}_{n-2, x, \xi},\\
        A(n,x,\xi) = \mp \sign{x}\frac{1}{2}\sqrt{(1\pm \xi^{-1} (tx)^{-1} q^{-2-n})(1\pm \xi(tx)^{-1}q^{-2-n})},
    \end{gathered}
\end{equation}
where\footnote{We introduce this renormalised version of the spherical elements, since its continuous spectrum is simply given by the interval $[-1,1]$. Note, however, that $\tilde \rho_{st}$ does not have well-defined limits $s,t\to \infty$. Instead, for example, $\lim_{s\to \infty} s^{-1}\tilde \rho_{st} = -\frac{1}{2}qt\rho_{\infty, t}$.}
\begin{equation}
    \label{eq:normalizedSphericalElement}
    \tilde \rho_{st} \coloneq -\frac{1}{2}qst \rho_{st} = \frac{1}{2}qte\gamma_{\infty, qt}^*\alpha_{\infty, qt} - \frac{1}{2}qt(s+s^{-1})\rho_{\infty, t} + \frac{1}{2}qte\alpha_{\infty, qt}^*\gamma_{\infty, qt}.
\end{equation}
Therefore, we should again first diagonalise with respect to the spherical element $\tilde \rho_{st}$ to make $\tilde Y_s^R$ into a three-term recurrence relation. Using \eqref{eq:coordinateAlgebraActionIwasawa}, the action of $\tilde \rho_{st}$ can be calculated to be given by
\begin{equation}
    \begin{gathered}
        \tilde \rho_{st}.\delta_{n,x,\xi}^{\pm, (\infty, t)} = A(x) \delta_{n+2, q^{-2}x, \xi}^{\pm, (\infty, t)} + B(x) \delta_{n,x,\xi}^{\pm, (\infty, t)} + A(q^2x)\delta_{n-2, q^2x, \xi}^{\pm, (\infty, t)},\\
        A(x) =-\frac{1}{2}e(x)\sqrt{(1+x)(1+t^2x)}, \quad B(x) = -\frac{1}{2}qt(s+s^{-1})x.
    \end{gathered}
\end{equation}
One can note that $\tilde \rho_{st}$ acts irreducibly on the three distinct subspaces\footnote{Note here that an additional irreducible subspace $V$ appears in the Iwasawa decomposition, which has to do with the richer spectrum of $\sigma(\rho_{\infty, t})$. This subspace vanishes in the limit $t\to \infty$.}
\begin{equation}
    \begin{split}
        U &= \overline{\text{span}}\{\delta_{n,x,\xi}^{\pm, (\infty, t)} : x = -q^{-2k - 2} \text{ and } k\in \Z_{\geq 0}\},\\
        V &= \overline{\text{span}}\{\delta_{n,x,\xi}^{\pm, (\infty, t)} : x = -t^{-2}q^{2k}\text{ and } k\in \Z_{\geq 0}\},\\
        W &= \overline{\text{span}}\{\delta_{n,x,\xi}^{\pm, (\infty, t)} : x = q^{-2k} \text{ and } k\in \Z\},
    \end{split}
\end{equation}
where $\overline{\text{span}}$ denotes a closed linear span. For $x = -q^{-2k - 2}, k \in \Z_{\geq 0}$, one can recognize that the action of the spherical element $\tilde \rho_{st}$ is given by the Jacobi operator for the $q^{-2}$-Al-Salam-Chihara polynomials \eqref{eq:JacobiOperatorAl-Salam-Chihara}. Similarly, for $x = -t^{-2}q^{2k}, k\in \Z_{\geq 0}$, one recognizes the Jacobi operator for the $q^2$-Al-Salam-Chihara polynomials. Lastly, for $x = q^{-2k}, k\in \Z$, we have the doubly infinite Jacobi operator for the little $q^2$-Jacobi functions \eqref{eq:JacobilittleqJacobi}. Therefore, it follows from Propositions~\ref{thm:spectralmeasureunboundedAlSalamChihara}, \ref{thm:spectralmeasureboundedAl-Salam_Chihara} and \ref{thm:spectralmeasurelittleqJacobi} that the spectrum of (an appropriate self-adjoint extension\footnote{For $|s| > q^{-1}$, the operator $\tilde \rho_{st}$ is essentially self-adjoint, and thus has a unique self-adjoint extension. For $q<|s|<q^{-1}$, there exists a self-adjoint extension such that the vectors $h^{\pm, (st)}_{n,y,\xi}$ are elements of its domain (see Propositions~\ref{thm:spectralmeasurelittleqJacobi}, \ref{thm:spectralmeasureboundedAl-Salam_Chihara} and \ref{thm:spectralmeasureunboundedAlSalamChihara}). } of) $\tilde \rho_{st}$ on these subspaces is given by
\begin{equation}
    \begin{split}
        \sigma\left(\tilde \rho_{st}|_{U}\right) &= \left\{\mu(qs^{-1}t^{-1} q^{2k}) : k\in \Z_{\geq 0}\right\},\\
        \sigma\left(\tilde \rho_{st}|_{V}\right) &= [-1,1]\cup \left\{\mu(q^{-1}s^{-1}tq^{-2k}) : |q^{-1}s^{-1}tq^{-2k}| < 1\ \mathrm{for}\ k\in \Z_{\geq 0}\right\},\\
        \sigma\left(\tilde \rho_{st}|_{W}\right) &= [-1,1]\cup \left\{\mu(q^{-1}st^{-1}q^{-2k}) : |q^{-1}st^{-1}q^{-2k}| < 1\ \mathrm{for}\ k\in \Z_{\geq 0}\right\}\\
        &\qquad \qquad \cup \left\{-\mu(q^{-1}s^{-1}t^{-1}q^{2k}) : |q^{-1}s^{-1}t^{-1}q^{2k}| < 1\ \mathrm{for}\ k\in \Z\right\}.
    \end{split}
\end{equation}
The corresponding eigenvectors are given by
\begin{equation}
    \label{eq:preGaussBasis}
    \begin{split}
        h_{n,y_U,\xi}^{\pm, (st)} &\coloneq \sum_{k\in \Z_{\geq 0}} h_k(\mu(y_U); q^{-1}st,q^{-1}s^{-1}t; q^{-2})\delta_{n+2k+2, -q^{-2k - 2}, \xi}^{\pm, (\infty, t)},\\
        h_{n, y_V, \xi}^{\pm, (st)} &\coloneq \sum_{k\in \Z_{\geq 0}}h_k(-\mu(y_V); -qs^{-1}t^{-1}, -qst^{-1} ; q^2) \delta_{n-2k, -t^{-2}q^{2k}, \xi}^{\pm, (\infty, t)},\\
        h_{n, y_W, \xi}^{\pm, (st)} &\coloneq \sum_{k\in \Z}j_k(-\mu(y_W) ; q^2 s^{-2}, -qs^{-1}t^{-1} ; q^2, -1) \delta_{n+2k, q^{-2k}, \xi}^{\pm, (\infty, t)},
    \end{split}
\end{equation}
 for $n\in \Z$, $\mu(y_{U,V,W}) \in \sigma(\tilde \rho_{st}|_{U,V,W})$ and $\mu(\xi) \in \sigma(\tilde Y_t^L)$, where the functions $h_k$ and $j_k$ are the Al-Salam-Chihara (\ref{sec:AlSalamChihara}) and the little $q^2$-Jacobi functions (\ref{sec:littleqJacobi}) respectively.\\
 
 Let us introduce the `indicator' functions for the irreducible subspaces described above. Namely, define $\kappa \colon \sigma(\tilde \rho_{st})\to \{\pm 1\}$, where $\kappa(y) = -1$ for $\mu(y) \in \sigma(\tilde \rho_{st}|_V)$ and $\kappa(y) = 1$ otherwise. Similarly, let $e \colon \sigma(\tilde \rho_{st}) \to \{\pm 1\}$ where $e(y) = -1$ if $\mu(y) \in \sigma(\tilde \rho_{st}|_U)$ and $e(y) = 1$ otherwise, i.e. $e.h^{\pm, (st)}_{n,y,\xi} = e(y)h^{\pm, (st)}_{n,y,\xi}$. Now if we let $n\in \Z$, $\epsilon \in \{\pm 1\}$ and $\xi = \epsilon t^{-\epsilon}q^m$ such that $|\xi|<1$, it follows from \eqref{eq:spectrumLeftPrimitiveConditional} that $h^{\pm, (st)}_{n,y, \epsilon t^{-\epsilon}q^m}$ can only be non-zero if $n\equiv m\text{ mod } 2$ and, moreover,
\begin{alignat}{2}
    \label{eq:hnonzerocases}
    h^{-e(y), (st)}_{n,y,  \epsilon t^{-\epsilon}q^m} \text{ non-zero} \quad \Leftrightarrow \quad &\kappa(y) = \epsilon \text{ and } n \leq -m - 2&\\
    h^{e(y), (st)}_{n,y,  \epsilon t^{-\epsilon}q^m} \text{ non-zero} \quad \Leftrightarrow \quad &\bigg[\kappa(y) =  -\epsilon \text{ and } n \geq m \bigg]\text{ or } \kappa(y) = \epsilon& \nonumber.
\end{alignat}
Note that if $\mu(\xi)$ is in the continuous part of the spectrum, the basis elements are always non-zero.\\

We are now ready to diagonalise with respect to the right action of $\tilde Y_s^R$. From \eqref{eq:rightactionTwistedPrimitiveIwasawa}, it follows that
\begin{equation}
    \begin{gathered}
        h_{n,y, \xi}^{\pm, (st)}.\tilde Y_s^R = A(n,y,\xi) h_{n+2, y, \xi}^{\pm, (st)} +  B(n,y,\xi)h_{n,y,\xi}^{\pm, (st)} + A(n-2,y,\xi) h_{n-2, y, \xi}^{\pm, (st)},\\
        A(n,y,\xi) = \mp e(y)\kappa(y)\frac{1}{2}\sqrt{(1\pm e(y)\kappa(y) \xi^{-1} t^{-\kappa(y)} q^{-2-n})(1\pm e(y)\kappa(y) \xi t^{-\kappa(y)}q^{-2-n})},\\
        B(n,y,\xi) = e(y)\kappa(y)t^{-\kappa(y)}\mu(y)q^{-n-1}.
    \end{gathered}
\end{equation}
In order to `diagnose' this difference operator, we have to go through the different cases in which $h^{\pm, (st)}_{n,y,\xi}$ is non-zero as described in \eqref{eq:hnonzerocases}. For these cases, one will recognise the difference equations for both the Al-Salam-Chihara functions (\ref{sec:AlSalamChihara}) and the little $q^2$-Jacobi functions (\ref{sec:littleqJacobi}). In particular, let us fix $\mu(y) \in \sigma(\tilde \rho_{st})$ and $\mu(\xi) \in \sigma(\tilde Y_t^{L})$. We then get the spectrum of (an appropriate self-adjoint extension\footnote{The operator $\tilde Y_s^{R}$ is essentially self-adjoint for $\mu(y) \in \sigma(\tilde \rho_{st})\setminus [-1,1]$, and thus has a unique self-adjoint extension. For $\mu(y) \in [-1,1]$, there exists a self-adjoint extension such that the vectors $\delta^{\pm, (st)}_{\eta, y, \xi}$ are in its domain (see Propositions~\ref{thm:spectralmeasurelittleqJacobi}, \ref{thm:spectralmeasureboundedAl-Salam_Chihara} and \ref{thm:spectralmeasureunboundedAlSalamChihara}).} of) $\tilde Y_s^R$ and the associated eigenvectors split into the following different cases:
\begin{itemize}[leftmargin=*]
    \item For $\xi = \kappa(y)t^{-\kappa(y)}q^m$ such that $|\xi| < 1$, the right action of $\tilde Y^R_s$ is given by the recurrence relations for the Al-Salam-Chihara polynomials \ref{sec:AlSalamChihara} for $\pm e(x) = -1$ and the little $q^2$-Jacobi functions \ref{sec:littleqJacobi} for $\pm e(x) = +1$, such that its spectrum is given by 
    \begin{equation}
        \label{eq:rightPrimitivespectrum1}
        \begin{split}
            \sigma(\tilde Y_s^{R})_{y, \xi}^{-e(y)} &= [-1,1]\cup \left\{e(y)\mu(q^{-1}\xi^{-1}yq^{-2k}) : q^{-1}\xi^{-1}yq^{-2k}\in (-1,1), k\in \Z_{\geq 0}\right\},\\
            \sigma(\tilde Y_s^{R})_{y,\xi}^{e(y)} &= [-1,1]\cup \left\{-e(y)\mu(q^{-1}\xi y^{-1}q^{-2k}) : q^{-1}\xi y^{-1}q^{-2k} \in (-1,1), k\in \Z_{\geq 0}\right\}\\
            &\qquad \qquad \cup \left\{e(y)\mu(q^{-1}\xi^{-1}yq^{-2k}) : q^{-1}\xi^{-1}yq^{-2k} \in (-1,1), k\in \Z\right\},
        \end{split}
    \end{equation}
    where for each $\mu(\eta) \in \sigma(\tilde Y_s^{R})^{\pm e(y)}_{y,\xi}$, we have the corresponding eigenvectors
    \begin{equation}
        \begin{split}
            \delta^{-e(y), (st)}_{\eta, y, \xi} &\coloneq \sum_{k\in \Z_{\geq 0}}(\kappa(y))^k h_k(\mu(\eta); e(y)q\xi y, e(y)q\xi y^{-1}; q^2) h^{-e(y), (st)}_{-m-2-2k, y, \xi},\\
            \delta^{e(y), (st)}_{\eta, y, \xi} &\coloneq \sum_{k\in \Z}(-\kappa(y))^k j_k(\mu(\eta); q^2y^2, -e(y)qy \xi^{-1}; q^2, -1) h^{e(y), (st)}_{-m-2+2k, y, \xi}.
        \end{split}
    \end{equation}
    \item For $\xi = -\kappa(y)t^{\kappa(y)}q^m$ such that $|\xi| < 1$ and $\mu(y) \in \sigma(\tilde \rho_{st})\setminus[-1,1]$\footnote{For $\mu(y) \in [-1,1]$ and $\xi = -\kappa(y)t^{\kappa(y)}q^m$, the recurrence relation of $\tilde Y_s^R$ corresponds to an indeterminate moment problem of the $q^{-1}$-Al-Salam-Chihara polynomials, which was resolved in \cite{G2019}. In particular, one can combine the action that corresponds to the $q^{-1}$-Al-Salam-Chihara polynomials with the action that corresponds to the little $q$-Jacobi functions to obtain a (combined) self-adjoint extension of $\tilde Y_s^R$, with spectrum given by
    \begin{equation}
        \label{eq:rightPrimitivespectrum4}
        \sigma(\tilde Y_s^{R})^{e(y)}_{y,\xi} = [-1,1]\cup \{\pm \mu(\xi q^{2k}) : |\xi q^{2k}|<1, k\in \Z\},
    \end{equation}
    and corresponding eigenfunctions as given in \cite{G2019}. For this paper, we will not require the explicit eigenfunctions for $\mu(y) \in [-1,1]$, and will therefore not discuss them.} such that $|y| < 1$, the right action of $\tilde Y^R_s$ is given by the recurrence relation for the $q^{-2}$-Al-Salam-Chihara polynomials \ref{sec:AlSalamChihara}, such that is spectrum is given by
    \begin{equation}
        \label{eq:rightPrimitivespectrum2}
        \sigma(\tilde Y_s^{R})^{-e(y)}_{y,\xi} = \emptyset,\quad \sigma(\tilde Y_s^{R})^{e(y)}_{y,\xi} = \{e(y)\mu(q\xi y q^{2k}) : k\in \Z_{\geq 0}\},
    \end{equation}
    where for each $\mu(\eta) \in \sigma(\tilde Y_s^{R})^{e(y)}_{y,\xi}$, we have the corresponding eigenvector
    \begin{equation}
        \delta^{e(y), (st)}_{\eta, y, \xi} \coloneq \sum_{k\in \Z_{\geq 0}}(-\kappa(y))^k h_k(\mu(\eta); -e(y)q^{-1}\xi^{-1} y^{-1}, -e(y)q^{-1}\xi^{-1} y; q^{-2}) h^{e(y), (st)}_{m+2k, y, \xi}.
    \end{equation}
    \item For $\mu(\xi) \in [-1,1]$ and $\varepsilon \in \{0,1\}$, the right action of $\tilde Y^R_s$ is given by the recurrence relation for the little $q^2$-Jacobi functions \ref{sec:littleqJacobi}, such that its spectrum is given by
    \begin{equation}
        \label{eq:rightPrimitivespectrum3}
        \begin{split}
            \sigma(\tilde Y_s^R)^{\pm}_{y,\xi} = [-1,1]\cup \left\{e(y)\kappa(y)\mu(t^{\kappa(y)}yq^{\varepsilon-2k}) : t^{\kappa(y)}yq^{\varepsilon-2k}\in (-1,1), k\in \Z\right\}
        \end{split}
    \end{equation}
    where for each  $\mu(\eta) \in \sigma(\tilde Y_s^{R})^{\pm}_{y,\xi}$, the corresponding eigenvectors are given by
    \begin{equation}
        \delta^{\pm, (st)}_{\eta, y, \xi} \coloneq \sum_{k\in \Z}(\mp e(y)\kappa(y))^k j_k(\mu(\eta); q^2y^2, \mp q\xi^{-1}y; q^2, \mp q^{2+\epsilon}e(y)\kappa(y)\xi t^{\kappa(y)}) h^{\pm, (st)}_{\varepsilon + 2k, y, \xi}.
    \end{equation}
\end{itemize}

By plugging in specific values for $y$ and $\xi$ into the restricted spectra above, one can find that the total spectrum of the right action of $\tilde Y_s^{R}$ on $L_q^2(\mathrm{SU}(1,1)\rtimes\Z_2)$ is given by
\begin{equation}
    \sigma(\tilde Y_s^R) = \{-\mu(s^{-1}q^n) : |s^{-1}q^n| < 1, \ n \in \Z\}\cup [-1,1]\cup \{\mu(sq^n) : |sq^n| < 1, n\in \Z\},
\end{equation}
which is similar to the spectrum of $\tilde Y_t^{L}$ in \eqref{eq:spectrumlefttwisted}. The vectors $\delta^{\pm, (st)}_{\eta, y, \xi}$ now form an orthonormal basis\footnote{Note here that there is a multiplicity in the continuous spectra of the operators. In particular, for basis elements $\delta^{\pm, (st)}_{\eta, y, \xi}$ with $\mu(\eta), \mu(y), \mu(\xi) \in [-1,1]$, one can have either $\kappa(y) = +1$ or $\kappa(y) = -1$. Therefore, there are in fact two distinct orthogonal basis elements with $\mu(\eta), \mu(y), \mu(\xi) \in [-1,1]$, labelled by the sign of $\kappa(y)$. In our notation of the basis elements, we do not make this distinction explicit in order to keep the notation less complicated. For the discrete parts of the spectrum (which are the ones that will interest us most in Section~\ref{sec:reductiondsSYK}), there is no such degeneracy.} of $L_q^{2}(\mathrm{SU}(1,1)\rtimes\Z_2)$, that is
\begin{equation}
    \label{eq:orthonormalityGauss}
    \langle \delta_{\eta,y,\xi}^{\epsilon, (st)}, \delta^{\epsilon', (st)}_{\eta',y',\xi'}\rangle = \delta_{\epsilon, \epsilon'}\delta_\nu(\mu(\eta) - \mu(\eta'))\delta_\nu(\mu(y) - \mu(y'))\delta_\nu(\mu(\xi) - \mu(\xi')),
\end{equation}
where $\epsilon \in \{\pm 1\}$ and $\eta, y$, $\xi$ correspond to the eigenvalues of $\tilde Y_s^{R}$, $\tilde \rho_{st}$ and $\tilde Y_t^{L}$ respectively, that is
\begin{equation}
    \tilde Y_t^{L}.\delta_{\eta,y,\xi}^{\pm, (st)} = \mu(\xi)\delta_{\eta,y,\xi}^{\pm, (st)}, \quad \tilde \rho_{st}.\delta_{\eta,y, \xi}^{\pm, (st)} = \mu(y)\delta_{\eta,y,\xi}^{\pm, (st)}, \quad \delta_{\eta,y,\xi}^{\pm, (st)}.\tilde Y_s^R = \mu(\eta)\delta_{\eta,y,\xi}^{\pm, (st)},
\end{equation}
Importantly, not all eigenvectors $\delta^{\pm, (st)}_{\eta,y,\xi}$ are non-zero for arbitrary $\eta,y,\xi$. In particular, the conditions we found for the spectrum of $\tilde Y_s^{R}$ (\eqref{eq:rightPrimitivespectrum1}, \eqref{eq:rightPrimitivespectrum2}, \eqref{eq:rightPrimitivespectrum3}) need to be satisfied.\\

The action of the coordinate algebra on the basis in the Gauss decomposition is calculated in Appendix~\ref{sec:actionsGaussDecomp}, and is given by
\begin{equation}
    \label{eq:Gaussquantumgroupaction}
    \begin{split}
        \alpha_{s, qt}. \delta_{\eta,y,\xi}^{\pm, (st)} &= \sqrt{e(y)(1+q^{-2}s^{-2}t^{-2} - 2q^{-1}s^{-1}t^{-1}\mu(y))}~\delta_{\eta, y, \xi}^{\pm, (qs, qt)},\\
        \gamma_{s,qt}. \delta_{\eta, y, \xi}^{\pm, (st)} &= e(y)\sqrt{e(y)(s^{-2} + q^{-2}t^{-2} - 2q^{-1}s^{-1}t^{-1}\mu(y))}~\delta_{\eta, y, \xi}^{\pm, (q^{-1}s, qt)}.
    \end{split}
\end{equation}
The dynamical structure of the quantum group in this decomposition with respect to twisted primitive elements becomes apparent in the change of both the dynamical parameters $s$ and $t$ here. Taking $x = -2q^{-1}s^{-1}t^{-1}\mu(y)$ (to undo the normalisation on $\tilde \rho_{st}$ \eqref{eq:normalizedSphericalElement}) and $\eta = -s^{-1}q^{n}$, the basis elements and the above action reduce to the Iwasawa decomposition in the limit $s\to \infty$. If we specialise to the quantum Gauss decomposition, i.e. set $s = -q^{n}$ and $t = q^{m}$ with $n\equiv m \text{ mod }2$, the spectra of $\tilde Y_t^{L}$, $\tilde \rho_{st}$ and $\tilde Y_s^{R}$ are given by
\begin{equation}
    \begin{gathered}
        \sigma\left(\tilde \rho_{st}\right) = -\cosh\left(\left(\Z_{\geq 0} + \frac{1}{2}\right)\ln q^2\right)\cup [-1,1] \cup \cosh\left(\left(\Z_{\geq 0} + \frac{1}{2}\right)\ln q^2\right),\\
        \sigma(\tilde Y_{t}^L) = -\cosh\left(\frac{1}{2}\Z_{> 0}\ln q^2\right)\cup [-1,1] \cup \cosh\left(\frac{1}{2}\Z_{> 0}\ln q^2\right) = \sigma(\tilde Y_{s}^R),
    \end{gathered}
\end{equation}
with the additional constraints as given by \eqref{eq:rightPrimitivespectrum1}, \eqref{eq:rightPrimitivespectrum2} and \eqref{eq:rightPrimitivespectrum3}.\\

Let us summarise these constraints for the discrete spectra. Let us first consider $\mu(y) \in \sigma(\tilde \rho_{st}) \setminus [-1,1]$, then the basis elements $\delta^{\pm, (st)}_{\xi, y, \eta}$ are non-zero in the following cases:
\begin{itemize}[leftmargin=*]
    \item For $\xi = \kappa(y)t^{-\kappa(y)}q^m$ and $\mu(\eta) \in \sigma(\tilde Y_s^R)\setminus [-1,1]$ such that $|\xi|, |\eta|<1$,
    \begin{equation}
        \label{eq:nonzeroconditionsGauss1}
        \begin{alignedat}{6}
            &\delta^{-e(y), (st)}_{\eta, y, \xi} &\text{ non-zero} \quad &\Leftrightarrow \quad &y& = e(y)q\eta\xi q^{2k}\quad  &\text{for }& k\in \Z_{\geq 0}\\
            &\delta^{e(y), (st)}_{\eta,y, \xi} &\text{ non-zero} \quad &\Leftrightarrow \quad &y& = -e(y)q^{-1}\eta^{-1}\xi q^{-2k}\quad  &\text{for }& 0\leq k < \log_{q^2}|\xi| - \log_{q^2}|\eta|\\
            &&&\ \text{or} &y& = e(y)q\eta \xi q^{2k} \quad &\text{for }& k\geq -(\log_{q^2}|\xi| + \log_{q^2}|\eta|).
        \end{alignedat}
    \end{equation}
    \item for $\xi = -\kappa(y)t^{\kappa(y)} q^m$ and $\mu(\eta) \in \sigma(\tilde Y_s^R)\setminus [-1,1]$ such that $|\xi|, |\eta|<1$,
    \begin{equation}
        \label{eq:nonzeroconditionsGauss2}
        \begin{split}
            &\delta^{-e(y), (st)}_{\eta, y, \xi} = 0\\
            &\delta^{e(y), (st)}_{\eta, y, \xi} \text{ non-zero} \quad \Leftrightarrow \quad y = e(y)q^{-1}\eta \xi^{-1}q^{-2k}\quad \text{for } 0\leq k < \log_{q^2}|\eta| - \log_{q^2}|\xi|. 
        \end{split}
    \end{equation}
\end{itemize}
Note that these constraints are only there for the discrete spectra. If one would pick $\mu(y)\in [-1,1]$ and $\mu(\xi)\notin [-1,1]$, then it follows from  \eqref{eq:rightPrimitivespectrum4} that $\mu(\eta)$ can take any value in $\sigma(\tilde Y_s^R)$ for $\delta^{e(y), (st)}_{\eta, y, \xi}$ to be non-zero. Similarly, if $\mu(\xi) \in [-1,1]$ and $\mu(y) \not\in [-1,1]$, then it follows that $\mu(\eta) \in [-1,1]$, or $\eta$ is discrete with $\sign{\eta} = e(y)\kappa(y)\sign{ty}$ for $\delta^{\pm, (st)}_{\eta, y, \xi}$ to be non-zero. Lastly, if $\mu(\xi), \mu(y) \in [-1,1]$, then it is enforced that also $\mu(\eta)\in[-1,1]$.\\

\begin{figure}
    \centering
    \includegraphics[width=0.7\linewidth]{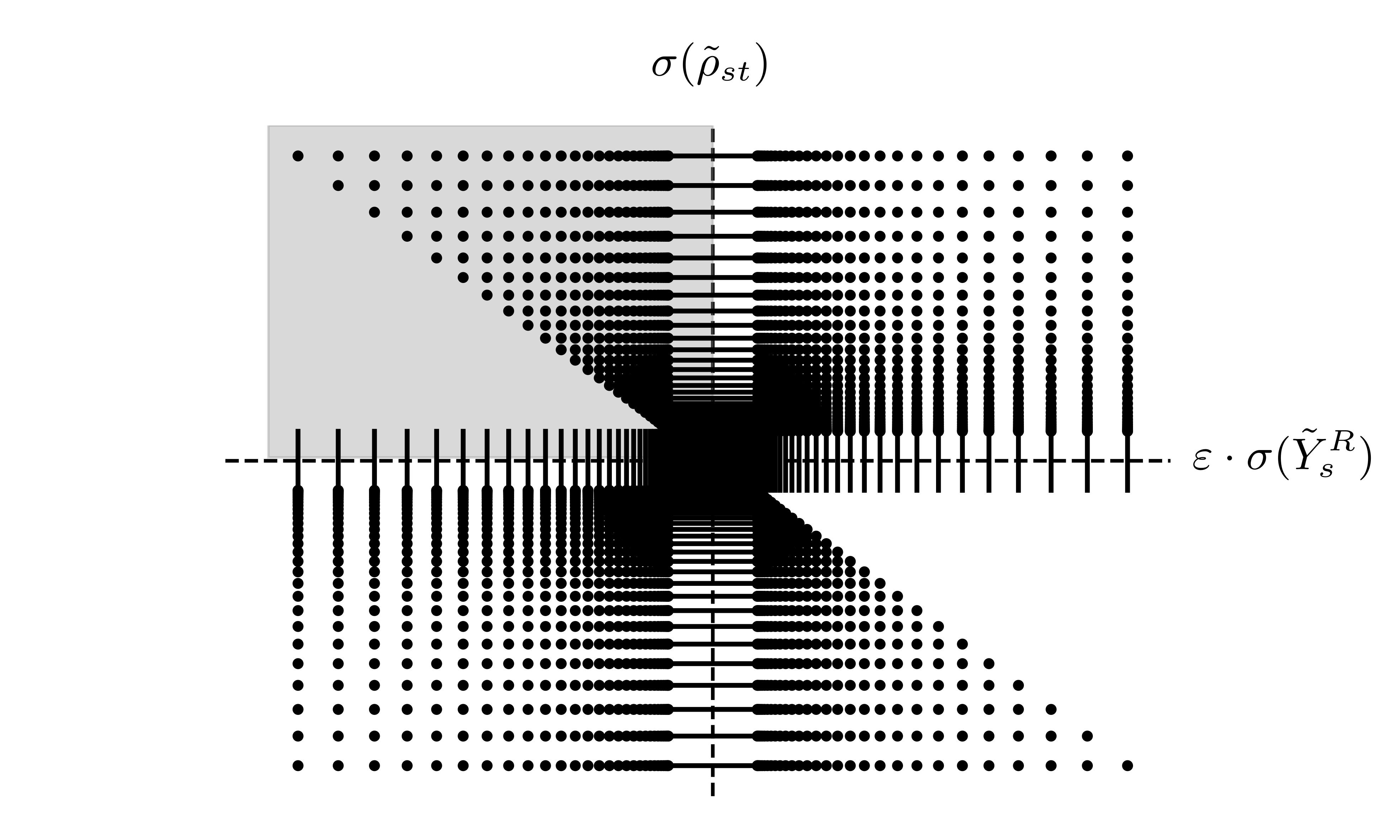}
    \caption{Plot of the allowed spectra $\sigma(\tilde \rho_{st})$ and $\epsilon\cdot \sigma(\tilde Y_s^{R})$ for the subspace of $L^2_q(\mathrm{SU}(1,1)\rtimes \Z_2)$ spanned by $\delta^{\epsilon, (st)}_{\eta, y, \xi}$ with $(s,t) = (-1,1)$ and $\xi = q$, for $q=0.95$. The continuous spectra have a possible multiplicity. Moreover, the discrete spectra can be divided into four quadrants based on the value of $(\epsilon, e(y))$. In particular, the top-left corresponds to $(-1,1)$, the top-right to $(1,1)$, the bottom-left to $(-1,-1)$ and the bottom-right to $(1,-1)$. The top-left quadrant, which is coloured in grey, corresponds to the {\it chord subsector}, as defined in Section~\ref{sec:reductiondsSYK}.}
    \label{fig:nonrescaledcoordinates}
\end{figure}

In Figure~\ref{fig:nonrescaledcoordinates}, we give a plot of the allowed spectra $\sigma(\tilde \rho_{st})$ and $\epsilon\cdot \sigma(\tilde Y_s^{R})$ for the subset of $L^2_q(\mathrm{SU}(1,1)\rtimes \Z_2)$ spanned by $\delta^{\epsilon, (st)}_{\eta, y, \xi}$ with $(s,t) = (-1,1)$ and $\xi = q$. The continuous spectra have a possible multiplicity. Moreover, the discrete points can be divided into four quadrants based on the signs of $\epsilon$ and $e(y)$. Note here that only the top-right quadrant is present in the representation theory of pure (formal) $\mathrm{SU}_q(1,1)$, and the other three quadrants appear due to the extension to the normaliser. We remark that only a bounded region of the spectra is shown: the spectra extend infinitely far in all directions.\\

Let us now also make the connection to the classical coordinates of the Gauss decomposition as in Section~\ref{sec:classicalGaussdecomp}. There we saw that the eigenvalues of $\tilde E$ and $\tilde F$ correspond to the momenta $p_R$ and $p_L$, respectively. As we have $\lim_{q\to 1}K^{-1}Y_{q^n} = 2i\tilde E$ and since $K^{-1}Y_{t}$ is self-adjoint, we infer that the quantised coordinate $p_R$ should correspond to the eigenvalues of the left action of $\frac{1}{2}K^{-1}Y_t$. Similarly, the quantised coordinate $p_L$ corresponds to the eigenvalues of the right action of $\frac{1}{2}Y_s K$. Lastly, in the limit $q\to 1$, $\tilde \rho_{-1,1}$ corresponds to the classical coordinate $-1 + 2e^{2\varphi}$, that is, $\tilde \rho_{-1,1}(g(\gamma_L, \varphi, \gamma_R)) = -1 + 2e^{2\varphi}$. This, therefore, gives the following correspondence between the quantised variables and the classical variables in the Gauss decomposition:
\begin{equation}
    \label{eq:connectionClassicalVariables}
    p_L = -\epsilon \frac{\mu(\eta) + \mu(s)}{q^{-1} - q},\quad  e^{\varphi} = \mu(y^{1/2}), \quad p_R = \frac{\mu(\xi) - \mu(t)}{q^{-1} - q},
\end{equation}
where $\mu(x) = \frac{1}{2}(x+x^{-1})$, $\mu(\eta) \in \sigma(\tilde Y_s^{R})$, $\mu(\xi) \in \sigma(\tilde Y_t^{R})$ and $\mu(y) \in \sigma(\tilde \rho_{st})$ with $\text{Re}(y) > -1$. Note here that we need $\text{Re}(y) > -1$ for $\mu(y^{1/2})$ to be positive, which corresponds to the restriction to $\mathrm{SU}_q^\circ(1,1)$ as in Section~\ref{sec:classicalGaussdecomp}. For $\text{Re}(y) < -1$, we go outside of the subset of Gauss decomposable elements, and there is no relation to $\varphi$. Moreover, we have an additional factor of $-\epsilon$ in the definition of $p_L$, such that $p_Lp_R > 0$ will correspond to elements in $\F_q^{-}$. As we will see in Section~\ref{sec:reductiondsSYK}, this is indeed the correct choice to get the correct classical reduction in the $q\to 1$ limit. The von Neumann algebraic quantum group description now shows explicitly how these coordinates become quantised through the discrete spectra of the associated operators.

\subsection{The Casimir operator}
\label{sec:actionCasimiroperator}
We can now calculate the action of the Casimir element of $\U_q(\mathfrak{su}(1,1))$ on the different basis elements in the decompositions that we studied above. Here, we will consider a more symmetric version of the Casimir element, given by $\tilde \Omega = \frac{1}{2}(q-q^{-1})^2\Omega + 1$, with $\Omega$ the Casimir element as defined in \eqref{eq:CasimirElement}. By the definition of the left action of the universal enveloping algebra on $L^2_q(\mathrm{SU}(1,1)\rtimes\Z_2)$ in \eqref{eq:leftregularaction}, the action of the Casimir operator in the Cartan decomposition can be directly calculated as:\footnote{From now on, we will denote the superscript of the basis elements with $\epsilon \in \{\pm 1\}$ instead of $\pm$, to allow for nicer notation in \eqref{eq:CasimiractionGauss}.}
\begin{equation}
    \label{eq:CasimiractionCartan}
    \begin{gathered}
        \tilde \Omega. \delta_{n,x,m}^{\epsilon} = A(x)\delta_{n, q^{-2}x,m}^{\epsilon} + B(x)\delta_{n,x,m}^{\epsilon} + A(q^2x)\delta_{n, q^2x,m}^{\epsilon},\\
        A(x) = \frac{1}{2}\sqrt{(1+x^{-1})(1+\epsilon q^{-n-m}x^{-1})},\quad B(x) = -\frac{1}{2}(q^{-m}+\epsilon q^{-n})q^{-1}x^{-1}.
    \end{gathered}
\end{equation}
The action of the Casimir operator in the Gauss decomposition can be calculated through the intermediate bases given in the previous section. This is a rather technical derivation, which can be found in our Appendix~\ref{sec:actionsIwasawaDecomp} and \ref{sec:actionsGaussDecomp}. For $\mu(y) \in \sigma(\tilde \rho_{st}) \setminus [-1,1]$, we derived the action of the Casimir operator in the Gauss decomposition to be given by the following $q$-difference operator:
\begin{equation}
    \label{eq:CasimiractionGauss}
    \begin{gathered}
        \tilde \Omega. \delta_{\eta, y, \xi}^{\epsilon, (st)} = A(y)\delta_{\eta, q^{-2}y, \xi}^{\epsilon, (st)} + B(y)\delta_{\eta, y, \xi}^{\epsilon, (st)} + A(y^{-1})\delta_{\eta, q^2y, \xi}^{\epsilon, (st)},\\
        A(y) = \frac{1}{2}\sqrt{\frac{(1-q^{-1}s^\pm t^{\pm}y)(1+q^{-1}(\epsilon\eta)^{\pm}\xi^{\pm}y)}{(1-y^2)(1-q^{-2}y^2)^2(1-q^{-4}y^2)}},\\
        B(y) = \frac{\left[\mu(s)\mu(y)-\mu(q)\mu(t)\right]\mu(\xi) -\left[\mu(t)\mu(y)-\mu(q)\mu(s)\right]\mu(\epsilon\eta)}{(\mu(y) + \mu(q))(\mu(y) - \mu(q))},
    \end{gathered}
\end{equation}
where 
\begin{equation}
    (1 - q^{-1}s^{\pm}t^{\pm}y) \coloneq (1 - q^{-1}sty)(1 - q^{-1}s^{-1}t y)(1 - q^{-1}st^{-1} y)(1 - q^{-1}s^{-1}t^{-1} y).
\end{equation}

Note here that this action can, in principle, be naturally extended to all $\mu(y) \in \sigma(\tilde \rho_{st})$. However, one has to be careful with the fact that $\mu(q^2y)$ and $\mu(q^{-2}y)$ are not always in the spectrum of $\tilde{\rho}_{st}$, in which case the resulting vector on the right hand side becomes a more complicated linear combination of eigenvectors. Of course, this does not mean that the Casimir action is not defined on the whole spectrum, but just that for the boundary spectral points, one cannot uniformly write down this action as a three-term recurrence relation in terms of our orthonormal basis. If we take $x = -2q^{-1}s^{-1}t^{-1}\mu(y)$ and $\eta = -s^{-1}q^n$, then the action \eqref{eq:CasimiractionGauss} reduces to the action in the Iwasawa decomposition \eqref{eq:CasimiractionIwasawa} in the limit $s\to\infty$. If we then also take $\xi = t^{-1}q^m$, then the action reduces to the action in the Cartan decomposition \eqref{eq:CasimiractionCartan} in the limit $t\to \infty$.\\

The Casimir operator action \eqref{eq:CasimiractionGauss} corresponds to the recurrence relation for the Askey-Wilson functions/polynomials. To see this, consider the rescaled (non-normalised) eigenbasis given by the elements
\begin{equation}
    \tilde \delta^{\epsilon, (st)}_{\eta, y, \xi} = \sqrt{\frac{(qsty, qsty^{-1}, -q\epsilon \eta\xi y, -q\epsilon\eta\xi y^{-1}; q^2)_\infty}{|y^{-1} - y|(qs^{-1}ty, qs^{-1}ty^{-1}, -q(\epsilon \eta)^{-1}\xi y, -q(\epsilon \eta)^{-1}\xi y^{-1}; q^2)_\infty}}\delta^{\pm, (st)}_{\eta, y, \xi},
\end{equation}
and define the shift operator $\mathcal{T}_{q^2}.\tilde \delta^{\epsilon, (st)}_{\eta, y, \xi} = \tilde \delta^{\epsilon, (st)}_{\eta, q^2y, \xi}$. The action of the (non-symmetric) Casimir element $\Omega$ on $\tilde \delta^{\epsilon, (st)}_{\eta, y, \xi}$ can then be calculated to yield:
\begin{equation}
    -qs\epsilon \eta(q-q^{-1})^2\Omega = \psi(y)(\mathcal{T}_{q^{2}} - 1) + \psi(y^{-1})(\mathcal{T}_{q^{-2}} - 1)+ (1+qs\epsilon\eta)^2
\end{equation}
where
\begin{equation}
     \psi(y) \coloneq \frac{(1-qst^{-1}y)(1-q sty)(1+q\epsilon \eta\xi^{-1}y)(1+ q\epsilon \eta\xi y)}{(1-y^2)(1-q^2y^2)}.
\end{equation}
Here, one can recognise the $q$-difference equation for the Askey-Wilson functions and polynomials as given in \cite{KS2000, KS2001_3}. Moreover, this expression is well-defined for all $\mu(y)\in \sigma(\tilde \rho_{st})$ over the space of functions on the support of the measure corresponding to the Askey-Wilson functions. This action reduces to the action given by Koelink and Stokman in \cite{KS2001} for $\epsilon = +1$, $\eta = -s^{-1}q^{2i}$ and $\xi = t^{-1}q^{2j}$. Koelink and Stokman only considered this action in the representation theory of the algebraic quantum group $\A_q(\mathrm{SU}(1,1))$, as by this time the locally compact quantum group $\mathrm{SU}_q(1,1)\rtimes \Z_2$ had not yet been constructed. Therefore, the above action is a generalisation of the Gauss decomposition to the locally compact (dynamical) quantum group $\mathrm{SU}_q(1,1)\rtimes \Z_2$.

\section{Reduction to the double-scaled SYK model}
\label{sec:reductiondsSYK}
In the previous section, we have described all the necessary ingredients to consider a particle-on-a-quantum-group-manifold theory that corresponds to the double-scaled SYK model. Namely, as discussed in \cite{BINT2019, BINN2023}, the bulk theory corresponding to the DSSYK model is realised on a quantum deformation of AdS$_2$ space, which in turn can be realised as a $q$-homogeneous space of the quantum group $\mathrm{SU}_q(1,1)\rtimes \Z_2$. Similar to Section~\ref{sec:classicalNormaliserandLiouvilleTheory}, the quantum homogeneous spaces relevant for the double-scaled SYK model can be constructed by restricting to subspaces with fixed left and/or right actions of the parabolic elements on the function space, the eigenvalues of which impose constraints on the homogeneous spaces, and are therefore additional parameters in the reduced theory. In particular, consider the {\it quantum Gauss decomposition}, i.e. $(s,t) \in \left(-q^{-\Z_{\geq 0}}\right)\times q^{-\Z_{\geq 0}}$ such that $|s|, |t| > q$, and restrict to the subsector of vectors given by $\delta^{\epsilon, (st)}_{\eta, y, \xi}$ with $\epsilon = -1$, $\mu(y)\in \sigma(\rho_{st}|_W)\setminus [-1,1]$ with $0<y<1$, $\xi = t^{-1}q^m$ for $m\in \Z$ such that $0<\xi < 1$, and $\mu(\eta) \in \sigma(\tilde Y_s^R)^{-}_{y,\xi}$ such that $0<\eta<1$. Since $\kappa(y) = +1$, $\epsilon = -1$ and $e(y) = +1$, it follows from the \eqref{eq:nonzeroconditionsGauss1} that
\begin{equation}
    \label{eq:reductionCoordinatesrelation}
    y = q\eta \xi q^{2k}, \quad\eta,\xi< 1 \quad \text{and}\quad  k\in \Z_{\geq 0},
\end{equation}
such that in particular $y \leq q\eta\xi$. We will call the subspace of $L^2_q(\mathrm{SU}(1,1)\rtimes \Z_2)$ spanned by these vectors the {\it chord subsector}. This subspace corresponds to the top-left quadrant\footnote{Note that we could have instead chosen the chord subsector to correspond to the bottom-right quadrant in Figure~\ref{fig:nonrescaledcoordinates}, which would have given the same result. In that case, the states do not correspond to the `Gauss decomposable subset' $\mathrm{SU}_q^\circ(1,1)$, but rather to $\left(-\mathrm{SU}_q^\circ(1,1)\right)$.} of Figure~\ref{fig:nonrescaledcoordinates}.\\

It might be unexpected that the chord subsector lies in $\F_q^-$, i.e. corresponds to $\epsilon = -1$. However, we do argue that this is indeed the correct subsector to consider. Namely, the eigenstates of the transfer matrix in the double-scaled SYK model are given by the continuous $q$-Hermite polynomials \cite{BNS2018, BINT2019}. One does not expect these polynomials to appear in the representation theory of purely (formal) $\mathrm{SU}_q(1,1)$, as the eigenfunctions of the Casimir operator in its regular representation correspond to the Askey-Wilson functions \cite{KS2001}. However, in the extension to the normaliser $\mathrm{SU}_q(1,1)\rtimes \Z_2$, the action of the Casimir operator \eqref{eq:CasimiractionGauss} restricted to the chord subsector is instead precisely given by the recurrence relation for the Askey-Wilson polynomials\footnote{Note here the important distinction between the Askey-Wilson \textit{functions} and the Askey-Wilson \textit{polynomials}. In particular, the Askey-Wilson functions diagonalise the corresponding doubly-infinite Jacobi operator on $\ell^2(\Z)$, while the Askey-Wilson polynomials diagonalise a Jacobi operator on $\ell^2(\Z_{\geq 0})$. It should be noted that for specific values of the spectral parameters, the Askey-Wilson functions may reduce to the Askey-Wilson polynomials \cite{KS2001_3}, but that such a reduction does not appear in the allowed spectra for $\mathrm{SU}_q(1,1)$ \cite{KS2001}.} \cite{AW1985} (see also \cite{KLS2010}) with $a = qs\eta, b = qt^{-1}\xi, c = qt\xi$ and $d = qs^{-1}\eta$. The continuous $q$-Hermite polynomials, and hence the transfer matrix for the double-scaled SYK model, can then be obtained from the Askey-Wilson polynomials by taking the limit $a,b,c,d\to 0$ \cite{KLS2010}. Therefore, we consider the rescaling of variables
\begin{equation}
    \label{eq:rescalingVariables}
    \eta \to q^r\eta, \quad \xi\to q^r \xi, \quad y \to q^{2r}y,
\end{equation}
and, upon \eqref{eq:connectionClassicalVariables}, also the rescaling of the classical variables
\begin{equation}
    p_L \to q^{-r} p_L, \quad p_R \to q^{-r} p_R, \quad \varphi \to \varphi -r\ln q,
\end{equation}
and take the limit $r \to \infty$. Note that this rescaling is compatible with the chosen coordinates \eqref{eq:reductionCoordinatesrelation}, and can physically be interpreted as zooming into the boundary of quantum AdS$_2$. Moreover, the classical Casimir action \eqref{eq:classicalCasimirAction} is invariant under such a rescaling, such that classically the above transformation is just a symmetry.
\begin{figure}
    \centering
    \includegraphics[width=0.7\linewidth]{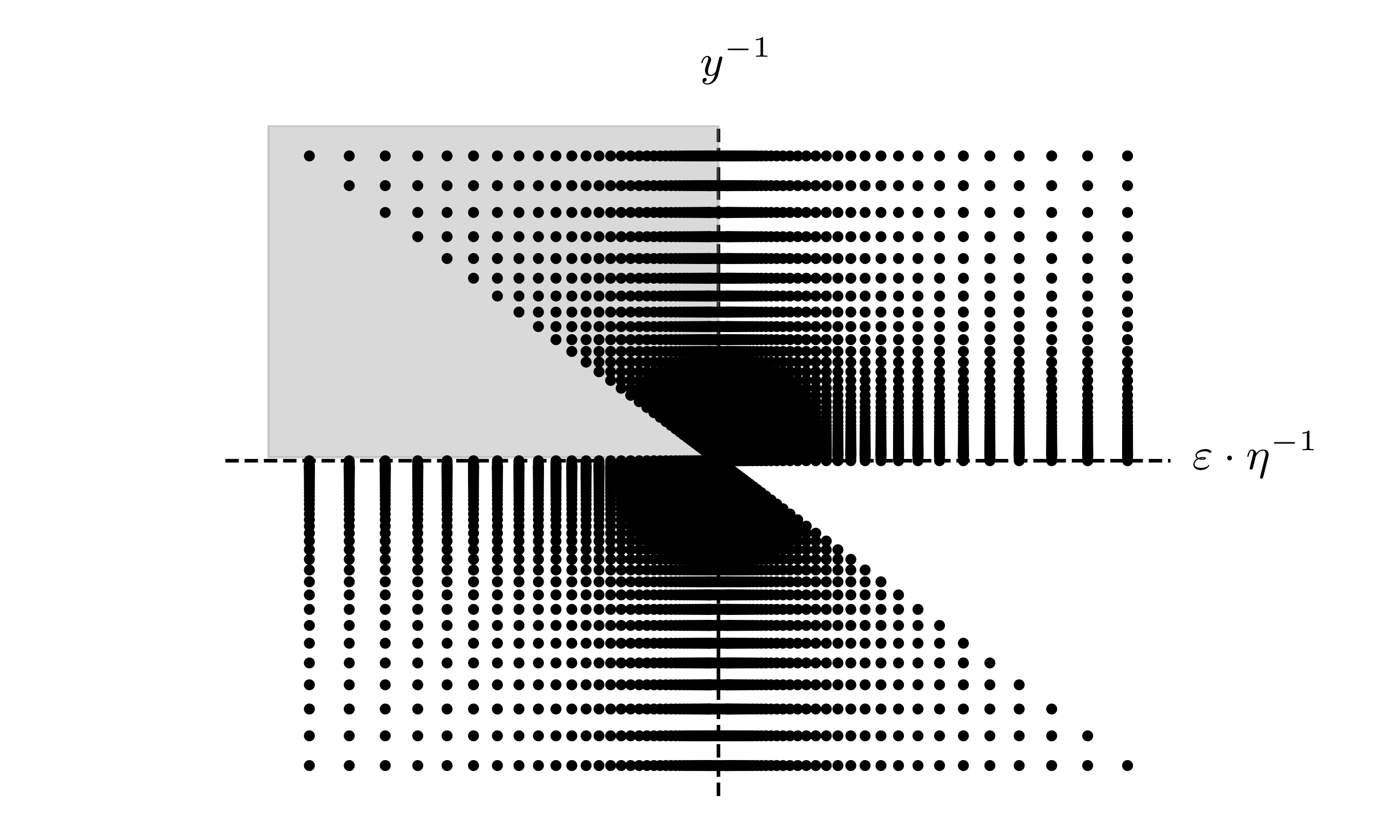}
    \caption{Plot of the allowed rescaled coordinates $y^{-1}$ and $\eta^{-1}$ for the subspace of $L^2_q(\mathrm{SU}(1,1)\rtimes \Z_2)$ spanned by the rescaled states $\lim_{r\to \infty}\delta^{\epsilon, (st)}_{q^r\eta, q^{2r}y, q^r\xi}$ with $(s,t) = (-1,1)$ and $\xi = q$, for $q=0.95$. Note that this plot now only contains discrete points. The chord subsector is coloured in grey.}
    \label{fig:rescaledcoordinates}
\end{figure}
In Figure~\ref{fig:rescaledcoordinates}, we plotted the allowed rescaled coordinates $y^{-1}$ and $\eta^{-1}$  for the subspace of $L^2_q(\mathrm{SU}(1,1)\rtimes \Z_2)$ spanned by the rescaled $\lim_{r\to \infty}\delta^{\epsilon, (st)}_{q^r\eta, q^{2r}y, q^r\xi}$ with $(s,t) = (-1,1)$ and $\xi = q$. Note that this plot is obtained by zooming `infinitely far' out of the plot in Figure~\ref{fig:nonrescaledcoordinates}, such that the continuous spectra decouple, and the discrete points can now get arbitrarily close to the two axes. Also here, only a bounded region of the coordinates is shown, and they extend infinitely far in all directions.\\

In these rescaled coordinates, the relation to the classical coordinates \eqref{eq:connectionClassicalVariables} becomes
\begin{equation}
    \label{eq:rescaledClassicalrelation}
    p_L = \frac{1}{2}\frac{\eta^{-1}}{q^{-1} - q}, \quad e^{\varphi} = \frac{1}{2}y^{-\frac{1}{2}}, \quad p_R = \frac{1}{2}\frac{\xi^{-1}}{q^{-1}-q},
\end{equation}
where $\eta, \xi \in q^{\Z}$ with $\log_q\eta \equiv \log_q \xi \text{ mod } 2$ and $y \in q^{2\Z + 1}$ such that $y \leq q\eta \xi$. Note here that $\epsilon = -1$ and  $p_Lp_R > 0$. It follows from \eqref{eq:reductionCoordinatesrelation} that the value of $y$ is determined by the non-negative integer $k \in \Z_{\geq 0}$. Therefore, for $\eta, \xi \in q^{\Z}$ with $\log_q\eta \equiv \log_q \xi \text{ mod } 2$, we denote the vectors in the chord subsector in these rescaled coordinates by 
\begin{equation}
    \label{eq:chordstates}
    |k\rangle_{\eta,\xi} \coloneq \lim_{r \to \infty}\delta^{-, (st)}_{q^r\eta, q^{2r}y, q^{r}\xi} \quad \text{with } y = q\eta\xi q^{2k} \text{ and } k\in \Z_{\geq 0},
\end{equation} 
which we will call the \textit{chord states}.

\subsection{Length positivity}
\label{sec:lengthpositivity}
Similarly to Section~\ref{sec:LiouvillePartitionFunction}, the chord states $|k\rangle_{\eta, \xi}$ correspond to wormhole states with boundary conditions $\eta$ and $\xi$, but now separated by $k$ chords. These states have the natural interpretation of counting the chord number in the two-sided picture of the double-scaled SYK model \cite{L2022}. Pictorially, a chord state is given by
\begin{equation}
    \begin{tikzpicture}[baseline=(current  bounding  box.center)]
        \begin{scope}
            \draw[thick, black] (1,-0.5) to (1,0.5);
            \draw[thick, black] (4,-0.5) to (4,0.5);
            \draw[thick, black, dotted] (1,0) to (4,0);

           \draw[thick, black] (1.5,-0.375) to (1.5,0.375);
           \draw[thick, black] (1.75,-0.375) to (1.75,0.375);
           \draw[thick, black] (2,-0.375) to (2,0.375);

           \draw[thick, black] (3,-0.375) to (3,0.375);
           \draw[thick, black] (3.25,-0.375) to (3.25,0.375);
           \draw[thick, black] (3.5,-0.375) to (3.5,0.375);

           \draw[draw=black, fill=black] (2.25, 0) circle (0.05);
           \draw[draw=black, fill=black] (2.5, 0) circle (0.05);
           \draw[draw=black, fill=black] (2.75, 0) circle (0.05);
            
            \draw(0,0) node {$|k\rangle_{\eta, \xi} = $};
            \draw(1,-0.8) node {\footnotesize $\eta$};
            \draw(4,-0.8) node {\footnotesize$\xi$};
            \draw(2.5,-0.6) node {\footnotesize$k$};
        \end{scope}
    \end{tikzpicture}
\end{equation}
where the geodesic length $L$ in \eqref{eq:wormholestate} has been replaced by the chord number $k$. Remember from Section~\ref{sec:LiouvillePartitionFunction} that the coordinate $\varphi$ corresponds to the renormalised geodesic length, i.e. $2\varphi = 2L + \ln \epsilon$, and we had to impose length positivity $L\geq 0$ by hand. Here, equation \eqref{eq:rescaledClassicalrelation} gives the relation between the quantised coordinate $\varphi$ and the non-negative chord number $k$ (via \eqref{eq:reductionCoordinatesrelation}). It can now be noted that the constraint $y\leq q\eta\xi$ also gives a constraint on $\varphi$. In particular, in terms of the classical coordinates, we have
\begin{equation}
     2\varphi \geq \ln\left(p_Lp_R\lambda^2\right) \quad \text{for } \lambda \to 0,
\end{equation}
where $q = e^{-\lambda/2}$ and $\lambda$ is the renormalization parameter. If one imposes the asymptotic \text{AdS}$_2$ boundary conditions $p_L, p_R \sim \lambda^{-1}$, the coordinate $\varphi$ is forced to be positive, leading to length positivity of the bare geodesic length. This gives a clean, first principle derivation of the length positivity, which we regard as an improvement over \cite{BMT2025}, where the derivation of the gravitational matrix elements was based on a representation of a so-called \textit{twisted Hopf $*$-algebra}, corresponding to $\U_q(\mathfrak{sl}(2,\R))$ with unusual unitarity properties (see also our discussion in Section~\ref{sec:quantumgroupactiononAdS}). Notice that in other earlier discussions (see e.g. \cite{BM2024, BINN2023}), length positivity had to be imposed by hand. Here, length positivity is a consequence of the constraint on the allowed eigenvalues for $\eta$, $\xi$ and $y$ in the von Neumann algebraic description of the normaliser $\mathrm{SU}_q(1,1)\rtimes \Z_2$. Note that the normaliser plays a significant role in the length positivity, as the chord subsector corresponds to states with $\epsilon = -1$.

\subsection{The partition function and $n$-point functions}
\label{sec:sykpartitionfunction} 
We can now repeat the considerations of Section~\ref{sec:LiouvillePartitionFunction} for the $q$-deformed case. While the resulting expressions are well known, the goal of this subsection is to set straight the conventions of our current way to reduce to the DSSYK model correlators from the von Neumann algebraic quantum group.\\

Starting with the partition function, we notice that, similarly to Section~\ref{sec:LiouvillePartitionFunction}, and as discussed in \cite{L2022}, it can be defined by starting with a state of 0 chords, evolving this state over Euclidean time $\beta$, and closing the circle again by taking the inner product with a state of 0 chords. Also, to reduce to the double-scaled SYK mode, we choose the same boundary conditions on the left and right-hand side, in particular $\eta = \xi = 1$. We will not always explicitly denote these boundary conditions from now on. Pictorially, the partition function is therefore given by
\begin{equation}
    \mathcal{Z}(\beta) \coloneq \langle 0 | e^{-\beta T} |0\rangle=
    \begin{tikzpicture}[baseline={([yshift=-.5ex]current bounding box.center)}]
        \draw[black, thick] (0:0.75) arc (0:360:0.75);
        \draw[draw=black, fill=black] (-90:0.75) circle (0.07);
        \draw[draw=black, fill=black] (90:0.75) circle (0.07);

        \draw[gray] (0:0.75) to [bend right = -30]  (150:0.75);
        \draw[gray] (-80:0.75) to [bend right = -20]  (50:0.75);
        \draw[gray] (-110:0.75) to [bend right = 20]  (90:0.75);
        \draw[gray] (-140:0.75) to [bend right = -20]  (-30:0.75);
        \draw[gray] (-60:0.75) to [bend right = 10]  (120:0.75);
        \draw[gray] (190:0.75) to [bend right = 10]  (30:0.75);

        \draw[black, ->] (1.1,-0.5) -- (1.1,0.5);

        \draw (-90:1) node {\footnotesize $k = 0$};
        \draw (90:1) node {\footnotesize $k = 0$};
        \draw (1.7,0) node {$e^{-\beta T}$};
    \end{tikzpicture},
\end{equation}
where the DSSYK transfer-operator is related to the Casimir as $\frac{1}{2}\sqrt{1-q^2} T = \tilde \Omega$ and the action of the Casimir operator \eqref{eq:CasimiractionGauss} in the rescaled coordinates is given, as discussed earlier, by the recurrence relations for the continuous $q^2$-Hermite polynomials, i.e.
\begin{equation}
    \tilde \Omega .|k\rangle = \frac{1}{2}\sqrt{1-q^{2k+2}}|k+1\rangle + \frac{1}{2}\sqrt{1-q^{2k}}|k-1\rangle.
\end{equation}
The eigenvectors of the Casimir operator are given by 
\begin{equation}
    \label{eq":quantumSYKeigenfunction}
    |\P^\theta\rangle_{\eta, \xi} = \sum_{k\in \Z_{\geq 0}} \frac{H_k(\cos \theta ; q^2)}{\sqrt{(q^2;q^2)_k}}|k\rangle_{\eta, \xi},
\end{equation}
for $\theta \in [0,\pi]$, where we use standard notations $(\cdot; q)_\infty$ for the $q$-Pochhammer symbol \eqref{eq:qpochhammer} and $H_k(\cdot; q)$ for the continuous $q$-Hermite polynomial (see for example \cite{KLS2010}), which can be obtained as $a,b\to 0$ limit of the Al-Salam-Chihara polynomial \eqref{eq:al-salam-chiharapolynomials}. 
The vector $|\P^\theta\rangle_{\eta, \xi}$ has eigenvalue $\cos(\theta)$, and therefore corresponds to a matrix element of the principal unitary series \cite{BK1993}. The inner product of two such states is now calculated as \cite{KLS2010}\footnote{Alternatively, this also follows from Proposition~\ref{thm:spectralmeasureboundedAl-Salam_Chihara} after taking the limit $a,b\to 0$.}
\begin{equation}
    \langle \P^{\theta}| \P^{\theta'}\rangle_{\eta, \xi} = \sum_{k\in \Z_{\geq 0}} \frac{H_k(\cos \theta ; q^2) H_{k}(\cos \theta' ; q^2)}{(q^2;q^2)_k} = \frac{2\pi}{(q^2, e^{\pm 2i\theta};q^2)_{\infty}}\delta(\theta - \theta'),
\end{equation}
where the corresponding Plancherel measure can be read off on the right side via \eqref{eq:Plancherelmeasureformula}. It can also be noted that $\langle 0 | \P\rangle_{\eta, \xi} = 1$. The partition function is therefore given by
\begin{equation}
    \begin{split}
        \mathcal{Z}(\beta) &= \int d\mu(\P^\theta)d\mu(\P^{\theta'})\langle 0| \P^{\theta}\rangle\langle\P^{\theta}| e^{-\beta T} |\P^{\theta'}\rangle \langle \P^{\theta'}|0\rangle = \int_{0}^{\pi}\frac{d\theta}{2\pi}(q^2, e^{\pm 2i\theta};q^2)_{\infty}e^{-\beta E(\theta)},
    \end{split}
\end{equation}
where $E(\theta) = 2\cos(\theta) / \sqrt{1-q^2}$. This is, of course, exactly the partition function for the double-scaled SYK model \cite{BNS2018, BINT2019} (under the redefinition $q^2 \to q$).\\

We now continue to calculate the $2$-point functions corresponding to the discrete series. Similarly to Section~\ref{sec:LiouvillePartitionFunction}, the operators corresponding to the discrete series should not change the boundary conditions, and we therefore impose that the twisted primitive elements act trivially by their left and right action. In other words, these operators correspond to having quantised momenta equal to zero, i.e. $p_L = p_R = 0$. Using \eqref{eq:connectionClassicalVariables}, this amounts to $\eta = -s^{-1}$ and $\xi = t^{-1}$, such that any operator spanned by $\Lambda^{-1}(\delta^{\epsilon, (st)}_{-s^{-1},y,t^{-1}})$\footnote{Recall that $\Lambda^{-1}$ is well-defined on the compactly supported functions.} leaves the boundary conditions intact. Note that these operators do not have to be elements of the chord subsector. However, it is difficult to explicitly calculate the action of these operators on the chord vectors, and thus, at the moment, we found it impractical to use these for the bilocal operators. Luckily, we already know of another operator on which the twisted primitive elements act trivially, namely the $(s,t)$-spherical element $\rho_{st}$ \eqref{eq:sphericalelement}. Therefore, any function $f(\rho_{st})$ (within a reasonable class of functions we consider) defines an operator that does not change the boundary conditions. An important subtlety here is that, in taking the rescaling limit, we also have to rescale this operator, for it to have a well-defined action on the chord states \eqref{eq:chordstates}. That is, we set $\hat \rho_{st} = q^{2r}\tilde \rho_{st}$ (with $\tilde \rho_{st}$ the normalised spherical element \eqref{eq:normalizedSphericalElement}) and take the rescaling limit $r \to \infty$ at the same rate as for the chord states, i.e.
\begin{equation}
    \label{eq:actionsphericalelement}
    \hat \rho_{st}.|k\rangle \coloneq \lim_{r\to\infty}q^{2r}\tilde \rho_{st}.\delta^{-,(st)}_{q^r, q^{2r}y, q^r} = \lim_{r\to\infty}q^{2r}\mu(q^{2r}y)\delta^{-,(st)}_{q^r, q^{2r}y, q^r} = \frac{1}{2}q^{-1}q^{-2k}|k\rangle ,
\end{equation}
where we assumed $\eta = \xi = 1$ and $y = q^{2k+1}$.\\

Next, we have to figure out which functions $f(\hat \rho_{st})$ correspond to the discrete series. It is known that the generators $\alpha, \gamma$ of the coordinate algebra $\A_q(\mathrm{SU}(1,1))$ correspond to matrix elements of the (non-unitary) spin-$\frac{1}{2}$ representation (see for example \cite{K1993}). Therefore, a product of $2\ell$ generators corresponds to a sum of matrix elements of spin-$\frac{n}{2}$ representations with $n\leq 2\ell$. Therefore, for $\ell \in \frac{1}{2}\Z_{\geq 0}$, the action of the Casimir operator on powers of the (rescaled) spherical element is given by $\Omega \triangleright (\hat \rho_{st})^\ell = \left[\ell + 1/2\right]_q^2\cdot (\hat \rho_{st})^\ell$,\footnote{More precisely, for every $\ell \in \Z$ there exists a polynomial $p_\ell(\tilde \rho_{st})$ of degree $\ell$ such that  $\Omega \triangleright p_\ell(\tilde \rho_{st}) = \left[\ell + 1/2\right]_q^2\cdot p_\ell(\tilde \rho_{st})$ \cite{K1993}. After rescaling $\tilde \rho_{st} = q^{-2r}\hat \rho_{st}$ and taking the limit $r\to \infty$, only the highest order term in the polynomial survives. A similar argument can be made for half-integer values of $\ell$, where some additional care has to be taken with the square root.} where $[\cdot]_q$ is the $q$-number \eqref{eq:qnumber}. By taking its inverse, it can be directly calculated\footnote{Let $X \in \U_q$ with $\Delta(X) = \sum X_{(1)}\otimes X_{(2)}$, then $X\triangleright (ab) = \sum (X_{(1)}\triangleright a)(X_{(2)}\triangleright b)$. In particular, one can use this `Leibniz rule' to consistently calculate the action of the universal enveloping algebra on the inverses of elements in the coordinate algebra, via $X\triangleright(a\cdot a^{-1}) = \sum (X_{(1)}\triangleright a)(X_{(2)}\triangleright a^{-1}) = \epsilon(X)\cdot1$. Applying this to the action of the Casimir element on the spherical element, although somewhat tedious, gives the result.} that $\Omega \triangleright (\hat \rho_{st})^{-\ell} = \left[1/2 - \ell\right]_q^2\cdot(\hat \rho_{st})^{-\ell}$, where one can now recognise the Casimir eigenvalue for the discrete series \cite{BK1993}, so that the negative powers of the spherical element correspond to matrix elements of the discrete series. Also, it follows from \eqref{eq:actionsphericalelement} that the action of $\hat \rho_{st}^{-\ell}$ on the chord subsector is given by a bounded operator. We therefore define the bilocal operator for the discrete series with $\ell \in \frac{1}{2}\Z_{> 0}$\footnote{Similarly to Section~\ref{sec:LiouvillePartitionFunction}, the `conformal weights' of the matter representations can only take integer/half-integer values here. For a full continuous spectrum of positive conformal weights, one would instead have to consider the von Neumann algebraic quantum group for the universal cover of the normaliser, i.e. $\widetilde{\mathrm{SU}_q(1,1)\rtimes \Z_2}$, which at this point has not yet been constructed. However, just as in the classical case, most of the results are expected to follow in close analogy.} at Euclidean time $\tau_{12} = \tau_1 - \tau_2$  to be given by
\begin{equation}
    \O^\ell(\tau_{12}) \coloneq e^{\tau_{12} T}\left[(2q\hat \rho_{st})^{-\ell}\right] e^{-\tau_{12} T}.
\end{equation}
We remark that, although $\rho_{st}$ generates the $(s,t)$-spherical elements of the coordinate algebra $\A_q$, it is not necessarily guaranteed that this operator generates all the `spinless' operators in the von Neumann algebra. To ensure a complete description of the set of operators that are invariant under the actions of the twisted primitive elements, one would therefore have to calculate the action of $\Lambda^{-1}(\delta^{\epsilon, (st)}_{-s^{-1},y,t^{-1}})$.\\

Similarly to Section~\ref{sec:LiouvillePartitionFunction}, the $2$-point function is given in terms of the 3j-symbol, which in turn is calculated as the overlap between the eigenfunctions \eqref{eq":quantumSYKeigenfunction}. From \eqref{eq:actionsphericalelement}, this overlap is given by \cite{BNS2018, BINT2019}
\begin{equation}
    \langle \P^{\theta_1}|\O^\ell(0)|\P^{\theta_2}\rangle = \sum_{k \in \Z_{\geq 0}}\frac{q^{2 k \ell}}{(q^2; q^2)_k} H_k(\cos \theta_1 ; q^2)H_k (\cos \theta_2;q^2) = \frac{(q^{4\ell} ; q^2)_{\infty}}{(q^{2\ell}e^{i(\pm \theta_1 \pm \theta_2)};q^2)_{\infty}}.
\end{equation}
Thus, the $2$-point function for the discrete series is given by
\begin{equation}
    \label{eq:2pointdiscreteseries}
    \langle 0| e^{-\beta T}\O^\ell(\tau_{12})| 0\rangle=\begin{tikzpicture}[baseline={([yshift=-.5ex]current bounding box.center)}]
    
        \draw[gray] (0:0.75) to [bend right = -30]  (150:0.75);
        \draw[gray] (-80:0.75) to [bend right = -20]  (50:0.75);
        \draw[gray] (-140:0.75) to [bend right = -20]  (-30:0.75);
        \draw[gray] (190:0.75) to [bend right = 20]  (30:0.75);
        \draw[gray] (60:0.75) to [bend right = -10] (160:0.75);
        
        \draw[black, thick] (0:0.75) arc (0:360:0.75);
        \draw[draw=black, fill=black] (0:0.75) circle (0.07);
        \draw[draw=black, fill=black] (180:0.75) circle (0.07);
        \draw[black, dashed, thick] (180:0.75) -- (0:0.75);
        
        \draw (0, 0.2) node {\footnotesize $\ell$};
    \end{tikzpicture}= \int \prod_{n=1}^{2}\left\{d\mu(\P^{\theta_n})e^{-\beta_nE(\theta_n)}\right\}\frac{(q^{4\ell} ; q^2)_{\infty}}{(q^{2\ell}e^{i(\pm\theta_1 \pm \theta_2)};q^2)_{\infty}},
\end{equation}
where $\beta_1 = \tau_{12}$ and $\beta_2 = \beta - \tau_{12}$, giving the 2-point function in the double-scaled SYK model \cite{BNS2018, BINT2019}.\\

To compare with what we will consider in the following subsection, let us also recall the Schwarzian limit of this amplitude \cite{MTV2017}. That is, let $\theta_i = \pi - |\ln(q^2)|k_i$ with $k_i \geq 0$ and take the limit $q\to 1$. Using \eqref{eq:qgamma}, we can write the overlap between eigenstates as a ratio of $q$-gamma functions, such that
\begin{equation}
    \label{eq:Schwarzianlimitdiscreteseries}
    \frac{(q^{4\ell} ; q^2)_{\infty}}{(q^{2\ell}e^{i(\pm \theta_1 \pm \theta_2)};q^2)_{\infty}} \sim \frac{\Gamma_{q^2}(\ell\pm ik_1 \pm ik_2)}{\Gamma_{q^2}(2\ell)} \xrightarrow{q\to 1} \frac{\Gamma(\ell\pm ik_1 \pm ik_2)}{\Gamma(2\ell)}.
\end{equation}
As was observed for the $n$-point functions of the double-scaled SYK model \cite{BNS2018, BINT2019}, higher time-ordered (non-crossing) correlators can be calculated by simply inserting extra operators $\O^\ell$ into the inner product, which gives extra 3j-symbols to integrate over. Similarly (and as in the non-deformed case of Section~\ref{sec:LiouvillePartitionFunction}), any out-of-time ordered $n$-point function, i.e. one that has a crossing, will give rise to additional 6j-symbols.

\subsection{Strange matter}
\label{sec:strangematter}
As discussed at the end of Section~\ref{sec:quantumregularrepresentations}, the principal unitary series and the discrete series are not the only type of irreducible representations that appear in the decomposition of the regular representations of the von Neumann algebraic quantum group $\mathrm{SU}_q(1,1)\rtimes \Z_2$. In particular, a new, non-classical type of irreducible representations that appear in the $q$-deformed Plancherel theorem are the {\it strange series}. Their appearance can be explained by the fact that for every eigenstate of the Casimir operator $\tilde \Omega$ with positive eigenvalue, there also is an eigenstate with negative eigenvalue \cite{GKK2010}, whereby the strange series correspond to the `negative-energy counterparts' of the discrete series. The matrix elements of the strange series can be obtained by simply replacing $\ell \to \ell + \frac{i\pi}{2\ln q}$ \cite{BK1993}.\footnote{Indeed, the eigenvalue of $\tilde \Omega$ for the discrete series is given by $\mu(q^{2\ell-1})$. Under the substitution $\ell \to \ell + \frac{i\pi}{2\ln q}$, this eigenvalue becomes $-\mu(q^{2\ell - 1})$, which therefore corresponds to the strange series.} That is, for $\ell \in \frac{1}{2}\Z_{>0}$, the bilocal operator corresponding to the strange series is given by
\begin{equation}
    \mathcal{S}^\ell(\tau_{12}) \coloneq e^{\tau_{12} T}\left[(2q\hat \rho_{st})^{-\ell - \frac{i\pi}{2\ln q}} \right]e^{-\tau_{12} T}.
\end{equation}
The two-point function for the strange series now follows from \eqref{eq:2pointdiscreteseries} with the substitution $\ell \to \ell + \frac{i\pi}{2\ln q}$.\\

Importantly, the overlap between eigenfunctions for the strange series vanishes in the Schwarzian limit. Indeed, consider $\theta_i = \pi -|\ln(q^2)|k_i$ with $k_i \geq 0$, then
\begin{equation}
      \langle \P^{\theta_1}|\mathcal{S}^\ell(0)|\P^{\theta_2}\rangle = \frac{(q^{4\ell} ; q^2)_{\infty}}{(q^{2\ell}e^{i(\pm \theta_1 \pm \theta_2) + i\pi};q^2)_{\infty}} \sim \frac{\Gamma_{q^2}\left(\ell\pm ik_1 \pm ik_2 + i\pi/\ln q\right)}{\Gamma_{q^2}(2\ell)} \xrightarrow{q\to 1} 0.
\end{equation}
This vanishing of the two-point function in this limit is, of course, expected as the strange series does not have a classical analogue. However, let us consider another limit, where we expand one of the angles around the lower part of the spectrum, while the other around the upper part. That is, we take $\theta_1 = \pi - |\ln(q^2)|k_1$ and $\theta_2 = |\ln(q^2)|k_2$ with $k_1, k_2 \geq 0$. It then follows that the $q\to 1$ limit gives
\begin{equation}
      \langle \P^{\theta_1}|\mathcal{S}^\ell(0)|\P^{\theta_2}\rangle \sim \frac{\Gamma_{q^2}\left(\ell\pm ik_1 \pm ik_2\right)}{\Gamma_{q^2}(2\ell)} \xrightarrow{q\to 1} \frac{\Gamma(\ell\pm ik_1 \pm ik_2)}{\Gamma(2\ell)},
\end{equation}
such that the strange series has a non-zero amplitude between the lower part and upper part of the spectrum . It was proposed in \cite{BLMPP2025} and \cite{O2025} that zooming into the upper part of the spectrum, i.e. $\theta = |\ln(q^2)|k$, corresponds to de Sitter JT-gravity.\footnote{Notice that different identifications with de Sitter space have been proposed up to date, in particular, by Susskind et al. \cite{S2021, S2022, S2025, Susskind2023sitterspacechordsconfined, Sekino2025doublescaledsykqcdflat} and by Verlinde et al. \cite{NV2025, V2025, VZ2025, Tietto2025microscopicmodelsitterspacetime}. The latter proposes that de Sitter space is instead identified with an expansion around $E(\theta) = 0$, or $\theta = \pi/2$.} If this proposal were proven to be correct, the strange series would therefore have a non-zero amplitude between the anti-de Sitter and de Sitter regions.

\subsection{The one-sided Hilbert space}
\label{sec:onesidedHilbertspace}
Let us now discuss the one-sided Hilbert space, i.e. the space of functions with only one fixed boundary condition. Here, we will actually consider a bit more general set of vectors with $e(y) = +1$.\footnote{These will in fact be all the vectors with $e(y) = +1$ that do not vanish upon rescaling.} In particular, in this section we will consider the subset of vectors $\delta^{\epsilon, (st)}_{\eta, y, \xi}$ where we do not fix $\epsilon$, but still pick $\mu(y)\in \sigma(\rho_{st}|_W)\setminus [-1,1]$ with $0<y<1$, $\xi = t^{-1}q^m$ for $m\in \Z$ such that $0<\xi < 1$, and $\mu(\eta) \in \sigma(\tilde Y_s^R)^{\epsilon}_{y,\xi}\setminus [-1,1]$ with $0<\eta<1$. We then do the same rescaling of variables as in \eqref{eq:rescalingVariables}. Since $\kappa(y) = +1$ and $e(y) = +1$, it follows from \eqref{eq:nonzeroconditionsGauss1} (after rescaling of $y$, $\eta$ and $\xi$) that
\begin{equation}
    y = q\eta \xi q^{2k},  \quad \eta,\xi< 1\quad  \text{and}\quad k\in \begin{cases}\Z_{\geq 0} & \text{for }\epsilon = -1\\
    \Z &\text{for } \epsilon = +1\end{cases}.
\end{equation}
Therefore, for $\epsilon = +1$, there is no constraint on $y$. Note that we get the same relation to classical coordinates as in \eqref{eq:rescaledClassicalrelation}, but now with the constraint $y \leq q\eta \xi$ for $\epsilon = -1$ and without any constraints for $\epsilon = +1$. These states correspond to the top two quadrants in Figure~\ref{fig:rescaledcoordinates}.\\

Let us fix $\xi\in q^{2\Z}$ as our right boundary condition and consider the \textit{restricted $q$-lattice}
\begin{equation}
    \label{eq:restricedqlattice}
    \R_{q^2}^{2}(\xi) \coloneq \{(-q^{2n}, q^{2m+1}) : n,m\in \Z\}\cup \{(q^{2n}, q^{2m+1}) : n,m\in \Z \text{ and } q^{2(n+m)} \leq\xi\},
\end{equation}
and we denote its elements by $(\chi, y)$, where we identify $\chi = -\epsilon \eta^{-1}$. In particular, positive values of $\chi$ correspond to vectors with $\epsilon = -1$, and the constraint $y \leq q\eta\xi$ for $\epsilon = -1$ is included in the definition of the lattice. Remember that we stated in Section~\ref{sec:classicalNormaliserandLiouvilleTheory} that matrix elements with $p_Lp_R > 0$ correspond to matrix elements for the normaliser. For the quantum group case, we also see that $\chi\xi > 0$ corresponds to $\epsilon =-1$. Next, we turn our vectors into functions on the lattice $\R_{q^2}^{2}(\xi)$ by considering the basis elements as the functions
\begin{equation}
    \delta^{\epsilon, (st)}_{\eta,y,\xi} \colon \R_{q^2}^{2}(\xi)\to  \C, \quad \delta^{\epsilon, (st)}_{\eta, y', \xi}(\chi, y) \coloneq (q\xi^{-1}\chi y;q^{2})_{\infty}^{-\frac{1}{2}}\delta_{\eta}(-\epsilon\chi^{-1})\delta_{y'}(y),
\end{equation}
where $\delta_p(x)$ is the $q^2$-delta function \eqref{eq:qdeltafunction}. The inner product on the space of square-integrable functions $L^2(\R^2_{q^2}(\xi))$ is now induced by the orthonormality relations \eqref{eq:orthonormalityGauss} of the (rescaled) basis elements $\delta^{\epsilon, (st)}_{\eta,y,\xi}$, and is given by
\begin{equation}
    \label{eq:innerproductAdS2}
    \langle f, g\rangle_{\xi} =  \int_{-\infty}^{\infty} d_{q^2}\chi \int_{0}^{\infty}d_{q^2}y\ (q\xi^{-1}\chi y;q^2)_{\infty}\overline{f(\chi, y)}g(\chi, y).
\end{equation}
It can be noted that $(q\xi^{-1}\chi y;q^2)_{\infty}=0$ for $\chi > 0$ and $|\chi| y \geq q^{-1}\xi$. Therefore, the constraint $y\leq q\eta\xi$ for $\epsilon = -1$ is encoded into the measure of the inner product. If we define the shift operator $(\mathcal{T}_q.f)(\chi, y) = f(\chi, qy)$,\footnote{Although this shift operator itself is not well-defined on the lattice $\R^2_q(\xi)$, the square of this operator is.} then the action of the (non-symmetric) Casimir element \eqref{eq:CasimirElement} on this space of functions can be calculated from \eqref{eq:CasimiractionGauss} (in the rescaled coordinates) to be equal to 
\begin{equation}
    \label{eq:CasimiractionreducedAdS}
    \Omega = \frac{q^{-1}\mathcal{T}_q^{-2} + q\mathcal{T}_q^{2} - 2}{(q-q^{-1})^2} - \mu y\frac{\chi}{1 - q^2}\mathcal{T}_q^{-2},
\end{equation}
where $\mu \coloneq \xi^{-1}/(1-q^2)$.\\

Here, we can recognise a similar action for the Casimir operator as suggested in \cite[(4.10)]{BINN2023} for the hands-on lattice realisation of $q$-deformed AdS$_2$ space. In that paper, the issue of the truncation of the Casimir operator for general realisations of the quantum AdS$_2$ space was discussed. Namely, the only physically reasonable actions for the Casimir operator are those that truncate, which ensures that there are no additional localised bound states on the Poincaré horizon \cite{BINN2023}.  In the von Neumann algebraic description, this truncation is manifest due to the restriction in the $q$-lattice $\R_{q^2}^{2}(\xi)$. In the limit $q\to 1$, this action reduces to the classical Casimir action \eqref{eq:classicalCasimirAction}, where $p_Lp_R < 0$ corresponds to an action on vectors with $\epsilon = +1$ and $p_Lp_R > 0$ corresponds to $\epsilon = -1$. This, therefore, explicitly shows that the Liouville eigenstates \eqref{eq:principalunitaryeigenfunctions} appear as matrix elements of the normaliser with $\epsilon = -1$.\\

Notice that our functions and Casimir operator are currently written in the momentum space, with $\eta$ corresponding to momentum. However, at this moment we have not identified an appropriate $q$-Fourier transform that is compatible with the restricted $q$-lattice and the inner product $\langle \cdot ,\cdot  \rangle_\xi$ \eqref{eq:innerproductAdS2}, and would allow to go back to the coordinate $q$-lattice. Instead, it can be noted that in the limit $\xi \to \infty$, the $q$-lattice $\R_{q^2}^2(\infty)$ becomes unconstrained and the inner product $\langle \cdot ,\cdot \rangle_{\infty}$ is just the ordinary inner product on the $q$-lattice. Moreover, since $\R_{q^2}^2(\xi) \subseteq \R^2_{q^2}(\infty)$, we can consider the space of square-integrable functions $L^2(\R^2_{q^2}(\xi))$ as the quotient space
\begin{equation}
    \label{eq:AdSsqaureintegrablefunctions}
    L^2(\R_{q^2}^2(\xi)) = \frac{\{f \colon \R_{q^2}^2(\infty) \to \C : ||f||_{\xi}^2 < \infty\} }{ \{f \colon \R_{q^2}^2(\infty) \to \C : ||f||_{\xi}^2 =0\}}.
\end{equation}
We can now proceed as follows: perform a Fourier transform on the space of functions on $\R_{q^2}^2(\infty)$ (forgetting about the inner product), and then define the inner product in the position space by imposing Parseval's formula by hand for arbitrary $\xi$. For the purposes of this paper, we will not write down the explicit form of such a $q$-Fourier transform; various explicit versions of the $q$-Fourier transform can be found in \cite{KS1992, R2007, BB2007, OR1997, BINN2023, Feinsilver1989qHarmonicAnalysis}. In other words, we assume there exists an invertible $q$-Fourier transform
\begin{equation}
    \F_{q^2} \colon \{\R_{q^2}^{2}(\infty) \to \C\} \to \{\R_{q^2}^{2}(\infty) \to \C\},
\end{equation}
such that $\D_{q^2}^\gamma\left(\F_{q^2}f\right)(\gamma, y) = \F_{q^2}\left(i\frac{\chi}{1-q^2}f\right)(\gamma, y)$, where $\D^\gamma_{q^2}$ is the $q^2$-derivative in the $\gamma$ coordinate as defined under \eqref{eq:poincareboundaryconditions}. We then define the inner product on these functions in position space, imposing Parseval's formula by hand, that is
\begin{equation}
    \langle \F_{q^2} f, \F_{q^2}g\rangle_{\xi} \coloneq \langle f, g\rangle_{\xi},
\end{equation}
as an appropriate explicit expression for the inner product in the position basis, which is compatible with the constraint and the inner product in the momentum space, is currently not available to us.\\

The above construction then defines a space of square-integrable functions on the $q$-lattice $\R^2_{q^2}(\infty)$ in the position space $(\gamma, y)$, on which the action of the Casimir operator is given by
\begin{equation}
    \Omega = \frac{q^{-1}\mathcal{T}_q^{-2} + q\mathcal{T}_q^{2} - 2}{(q-q^{-1})^2} + i\mu y\mathcal{T}_q^{-2}\D_{q^2}^\gamma,
\end{equation}
where $\mu \coloneq \xi^{-1}/(1-q^2)$. Here, we can now recognise the same formula for the Casimir operator on $q$-deformed AdS$_2$ as found in \cite[(4.8)]{BINN2023}, which gives the Hamiltonian on lattice AdS$_{2,q}$ space. The problems with the Fourier transform that were faced in that paper have now also become apparent in the incompatibility between the Fourier transform and the inner product. In particular, we therefore show that the $q$-deformed AdS$_2$ lattice is most naturally described in the momentum space, and somewhat ill-behaved in the position space.

\subsection{Quantum group action on reduced AdS$_{2,q}$}
\label{sec:quantumgroupactiononAdS}
We can now identify the lattice $\R^{2}_{q^2}(\xi)$ as defined in \eqref{eq:restricedqlattice} for a given $\xi \in q^{2\Z}$ with the \textit{reduced quantum AdS$_{2,q^2}$ space}. We refer to this space as reduced, as it corresponds to the space after the rescaling of the coordinates as given in \eqref{eq:rescalingVariables}. The full quantum AdS$_2$ space would be given in terms of the coordinates before the rescaling. We emphasise here that this $q$-lattice is written in the momentum basis, which is expected to be dual to the coordinate $q$-lattice. For points $(\chi, y)\in \R_{q^2}^{2}(\xi)$, the corresponding classical coordinates are given by
\begin{equation}
    \label{eq:quantumAdScoordinates}
    p = \frac{1}{2}\frac{\chi}{q^{-1} - q}, \quad e^{\varphi} = \frac{1}{2}y^{-\frac{1}{2}}.
\end{equation}

Recall that the space of square-integrable functions $L^2(\R_{q^2}^2(\xi))$ is given by the inner product as defined in \eqref{eq:innerproductAdS2}. Here, one can make the observation that
\begin{equation}
    (q\xi^{-1}\chi y;q^{2})_{\infty} = E_{q^2}(-2p y\xi^{-1} (1-q^2)),
\end{equation}
where $E_{q^2}(z) = (-z;q^2)_\infty$ is the $q^2$-exponential \cite{BHS}. In particular, in the limit $q\to 1$ where $\xi$ would remain fixed, i.e. asymptotic boundary conditions with $p_R \sim \lambda^{-1}$, the term in the measure of \eqref{eq:innerproductAdS2} becomes an ordinary exponential, i.e.  $E_{q^2}(-2p y\xi^{-1} (1-q^2)) \xrightarrow{q\to 1} \exp(-2py\xi^{-1})$. This therefore shows that the space of square-integrable functions over AdS$_2$ with asymptotic boundary conditions acquires a non-trivial measure, given in terms of the asymptotic boundary condition $\xi$. In the limit $q\to 1$ where $p_R$ is fixed, the exponential vanishes, i.e. $E_{q^2}(-2p y\xi^{-1} (1-q^2)) \xrightarrow{q\to 1} 1$, and one retrieves the ordinary measure.\\

Let us now define a different action of the quantum universal enveloping algebra $\U_q(\mathfrak{sl}_2)$ on this space. Firstly, it can be noted that the twisted-primitive element $Y_sK$ acts diagonally on $\delta^{\epsilon, (st)}_{\eta,y,\xi}$ from the right. Under the rescaling of variables \eqref{eq:rescalingVariables}, we should also rescale the operator to give finite-valued eigenvectors. In particular, we define $\tilde E  = \frac{1}{2i}q^{r}Y_sK$ and send $r \to \infty$ at the same rate as the rescaled variables. Moreover, we can turn the right action into a left action by taking the adjoint. That is, for a function $f\in L^2(\R_{q^2}^2(\xi))$, we define $\tilde E\triangleright f \coloneq f.\tilde E^*$. In particular, this gives $(\tilde E \triangleright f)(\chi, y)= ipf(\chi, y)$, where $p$ is defined in \eqref{eq:quantumAdScoordinates}. Inspired by the hands-on lattice realisations of the quantum group action in \cite{BINN2023}, one can now define the full quantum group action on the reduced AdS$_{2,q}$ space. Consider the operators $(\mathcal{T}_qf)(\chi, y) = f(\chi, qy)$, $(\mathcal{R}_q f)(\chi, y) = f(q\chi, y)$ and $(\D_{q^2}^{\chi}f)(\chi, y) = \frac{f(\chi, y)-f(q^2\chi, y)}{\chi(1-q^2)}$, then we define the left action of $\U_q(\mathfrak{sl}_2)$ on $L^2(\R_{q^2}^2(\xi))$ to be given by
\begin{equation}
    \label{eq:Cayleytransformedquantumgroupactions}
        \tilde K = \mathcal{T}_q \mathcal{R}_q^{-1}, \quad \tilde E = \frac{i}{2}\frac{\chi}{q^{-1}-q}, \quad \tilde F = 2i(\D^\chi_{q^{-2}}\mathcal{T}_q^2 - \D^\chi_{q^2}\mathcal{T}_q^{-2}) + 2iq^{-1}\mu y \mathcal{T}_q^{-2},
\end{equation}
where $\mu = \xi^{-1}/(1-q^2)$. Note here that $\tilde K$ itself does not map elements from $L^2(\R_{q^2}^2(\xi))$ to $L^2(\R_{q^2}^2(\xi))$, due to the single shift in $q$, and is therefore not actually a well-defined operator. However, $\tilde K^2$ and $\tilde K^{-2}$ are well-defined operators. It can be easily verified that these operators satisfy the relations for $\U_q(\mathfrak{sl}_2)$ as given in \eqref{eq:quantums2lrelations}.\footnote{More precisely, $\tilde K^2$, $\tilde E$ and $\tilde F$ are generators of the Drinfeld-Jimbo presentation of $\U_q(\mathfrak{sl}_2)$ \cite{QGAR}.} Moreover, the action of the Casimir element \eqref{eq:CasimirElement} for these generators coincides with the Casimir action as given in \eqref{eq:CasimiractionreducedAdS}. The adjoints with respect to the inner product \eqref{eq:innerproductAdS2} are given by $(\tilde K^2)^* = \tilde K^{-2}, \tilde E^* = -\tilde E$ and $\tilde F^* = -\tilde F$. In particular, these elements have a $*$-structure of $\mathrm{\U}_q(\mathfrak{sl}(2,\R))$. It means that, using the above construction, we have performed a Cayley-like transformation for the quantum group, i.e. redefined the generators to have the $*$-structure of $\mathrm{\U}_q(\mathfrak{sl}(2,\R))$ instead of $\U_q(\mathfrak{su}(1,1))$. The generators $\tilde K$, $\tilde E$ and $\tilde F$ can therefore be thought of as the quantum group equivalents of the generators in \eqref{eq:classicalParabolicElements}. However, we emphasise here that, unlike $\tilde E$, the actions of the introduced operators $\tilde K$ and $\tilde F$ are not guaranteed to be expressible in terms of the generators $E,F,K$ of $\U_q(\mathfrak{su}(1,1))$. In fact, it is unknown whether an explicit expression for the Cayley transformation, like \eqref{eq:classicalParabolicElements}, for the quantum group even exists, and even if it does, whether it is unique.\\

Importantly, the above representation is a representation of $\U_q(\mathfrak{sl}_2)$ as an algebra, but not as a Hopf algebra. In particular, the coproduct is not closed among these elements. Indeed, we defined $\tilde E$ as a rescaling of the twisted primitive element $Y_sK$, the coproduct of which is given under \eqref{eq:deftwistedprimitiveelement}. This implies that $\tilde E$ has the coproduct $\Delta(\tilde E) = K^2\otimes \tilde E + \tilde E\otimes 1$. Note here that the Cartan element that appears in this coproduct is the original Cartan element $K$ for $\U_q(\mathfrak{su}(1,1))$ as defined above \eqref{eq:quantums2lrelations} -- it is not $\tilde K$. Of course, it cannot be $\tilde K$, as then the coproduct would not be a $*$-homomorphism. We remark here that in \cite{BMT2025}, the action of $\U_q(\mathfrak{sl}_2)$ with the $*$-structure of $\mathfrak{sl}(2,\R)$ was used to compute gravitational matrix elements using Whittaker vectors. However, in that paper, the coproduct of the quantum group was assumed to be {\it closed} under the generators, giving rise to a \textit{twisted Hopf $*$-algebra} \cite{Coquereaux2000TwistedHopf}. We therefore see that this appearance of the twisted Hopf $*$-algebra here can be understood merely as an artefact due to the non-existence of a proper Cayley-transformation of the quantum group on the level of the Hopf algebra. This also implies that one can not easily take tensor products of such representations \eqref{eq:Cayleytransformedquantumgroupactions}, as one has to consider the action of the original Cartan element $K$ as well, which might be complicated.

\section{Discussion and outlook}
In this paper, we reviewed the construction of the von Neumann algebraic quantum group of $\mathrm{SU}_q(1,1)\rtimes \Z_2$, computed its quantum Gauss decomposition and showed that the dynamics of a particle on the von Neumann algebraic quantum homogeneous space AdS$_{2,q}$ reduces to the double-scaled SYK model. In Section~\ref{sec:classicalSUandLiouville}, when discussing the classical Lie group $\mathrm{SU}(1,1)$ and Liouville quantum mechanics, we gave a handful of arguments of why the extension of the global symmetry group to the normaliser $\mathrm{SU}(1,1)\rtimes \Z_2$ is natural. Moreover, when $q$-deforming the symmetry group, it was discussed in Section~\ref{sec:quantumcoordinatealgebra} that the locally compact quantum group can in fact only be consistently defined for this normaliser, which can be equivalently regarded as a problem with defining multiplication on the naive (non-extended) deformed group manifold. The reduction to the double-scaled SYK model then appears by restricting to the chord states, which correspond to mixed matrix elements for the normaliser. Considering the normaliser here is an important point. Namely, the eigenstates in the double-scaled SYK model correspond to the continuous $q$-Hermite polynomials, which do not appear as ordinary matrix elements in the representation theory of $\mathrm{SU}_q(1,1)$, but rather in a particular subsector of the normaliser $\mathrm{SU}_q(1,1)\rtimes \Z_2$. This also elegantly resolves the issue of length positivity, which in other discussions either had to be imposed by hand \cite{BM2024, BINN2023}, or required the introduction of a twisted Hopf $*$-algebra with unusual unitarity properties \cite{BMT2025}. For the chord subsector in the von Neumann algebraic quantum group $\mathrm{SU}_q(1,1)\rtimes \Z_2$, the chord number is automatically non-negative due to restrictions in the spectra of the twisted primitive elements.\\

\noindent There are several interesting future research directions one can take from here:
\begin{itemize}[leftmargin=*]
    \item We have argued in Section~\ref{sec:classicalNormaliserandLiouvilleTheory} that for Liouville quantum mechanics, the more natural global symmetry group to consider is the normaliser $\mathrm{SU}(1,1)\rtimes \Z_2$. Indeed, we demonstrated that for the quantum group, the reduction to the double-scaled SYK model works exclusively on the level of the normaliser. However, besides having a natural interpretation for the two-sided wormhole states, the exact role of the normaliser in the quantum theory is not fully understood. For example, as is the case for the quantum group, the non-existence of a global dressing action for the non-compact Poisson-Lie group might also lead to inconsistencies in the classical Poisson-geometric setup
    which, to our knowledge, has not been studied. Moreover, using the quantum group derivation, we argued that the wavefunctions for Liouville quantum mechanics can be derived from matrix elements of the normaliser. A corresponding explicit derivation for the classical case, using harmonic analysis on the extended group, still needs to be done cleanly.
    
    \item It is known that the $q\to 1$ limit of the double-scaled SYK model, while zooming into the lower edge of the spectrum, reduces to JT gravity on AdS space. Moreover, it has been argued that zooming into other parts of the spectrum, for example, the upper part (\cite{BLMPP2025, O2025, Aguilar2025CosmologicalEnt}) or the middle part (\cite{NV2025, V2025, VZ2025}), reduces to gravity on 2D or 3D de Sitter space, respectively. Similarly, there are claims that the middle part of the spectrum corresponds to the flat space regime \cite{BLMPP2025}. This would therefore suggest that the quantum homogeneous space of the quantum group $\mathrm{SU}_q(1,1)\rtimes \Z_2$ somehow at least contains information about both two-dimensional AdS and dS spaces. Although it is known how to reduce to AdS$_2$ space from its $q$-deformed version, as constructed in Section~\ref{sec:onesidedHilbertspace}, by taking a $q\to 1$ limit, it is not fully understood, to our knowledge, how de Sitter space would appear here. Moreover, supposing that dS space corresponds to the upper part of the spectrum, it was shown in Section~\ref{sec:strangematter} that the strange series gives non-zero amplitudes between AdS and dS regions, and thus in some sense interpolates between them. It is currently unknown what the correct physics interpretation of such an amplitude would be.

    \item To reduce to the double-scaled SYK model, we had to do a rescaling of the coordinates \eqref{eq:rescalingVariables}. Although classically, this rescaling corresponds to a symmetry that preserves the classical Casimir action, the quantum Casimir action does change. In particular, this rescaling was necessary to reduce to the continuous $q$-Hermite polynomials in the $q$-Askey scheme. This suggests that the double-scaled SYK model is in some sense `too degenerate', and it would be natural to expect that a more general bulk theory exists -- appropriately to be called the `{\it Askey-Wilson DSSYK}' (AW-DSSYK) model -- which reduces to the double-scaled SYK model in a corresponding scaling limit. In fact, one could readily compute the partition function and the $n$-point functions that such a theory should have by taking the same steps as in Section~\ref{sec:sykpartitionfunction}, but starting with the action of the Casimir element before the rescaling, i.e. the one that corresponds to the Askey-Wilson polynomials as opposed to the continuous $q$-Hermite polynomials. On the side of special functions giving eigenstates of the transfer-matrix, this corresponds to going all the way up the $q$-Askey scheme to its highest level. For example, the quantum disk\footnote{In our (representation-theoretic) picture, the quantum disk represents another (distinct from the DSSYK coset considered in this paper) quantum homogeneous space of the von Neumann algebraic quantum group, corresponding to its quotient by a maximal compact subgroup.} model \cite{VaksmanQuantumBoundedDomains} reviewed in \cite{Almheiri2025Holographyquantumdisk}, as well as the End-Of-the-World brane \cite{Okuyama2023EOWbrane, Aguilar2025EOW}, have transfer-matrix eigenstates corresponding to the Al-Salam-Chihara polynomials, which are already two levels up in the $q$-Askey scheme. Another possible way to explore this direction is via the connection to the asymmetric exclusion process, see \cite{OkuyamaHolographicTensor2025} and references therein for some recent work.
    
    \item The quantum Gauss decomposition corresponds to the diagonalisation of the square-integrable function with respect to the twisted primitive elements. It was shown at the algebraic level in \cite{S2003} that such a diagonalisation corresponds to a dynamical twist of the quantum group algebra.  Indeed, our states obtained dynamical labels $s$ and $t$, which change upon action of the quantum group \eqref{eq:Gaussquantumgroupaction}. In essence, it means that in this paper we have constructed an example of a `{\it von Neumann algebraic dynamical quantum group}', of which there is currently no clear general mathematical notion.  After the rescaling of variables \eqref{eq:rescalingVariables}, this dynamical structure got obscured. However, in, for example, the hypothetical AW-DSSYK model, the dynamical quantum group symmetry should become more apparent. The precise physics implications of such a dynamical symmetry group are currently unknown.

    \item We argued in this paper that the harmonic analysis understanding can elegantly clarify subtleties such as length positivity in the DSSYK model. It is well known that harmonic analysis often goes hand in hand with integrability due to the simple fact that having a `big enough' family of commuting self-adjoint operators definitely helps in finding spectral decomposition of the Hilbert space into `small enough' pieces.
    The integrable structure of the DSSYK model is rooted in the chord diagrams and corresponding algebras of observables. A more in-depth study of these algebras was initiated in \cite{LinStanfordSymmetryAlgDSSYK}, see also \cite{HVX2025QuantumSymmetry} for recent related work. It would be important to understand the integrable structure behind these harmonic analysis problems in full detail. This should also make a direct contact with the SUSY gauge theory point of view on the DSSYK model suggested in \cite{GaiottoVerline2025SYKSchur, GaiottoTeschner2024SchurQuant}, see also \cite{LewisMezei2025Schurconnectionschordcounting}.

    \item This paper was devoted to understanding the implications of a specific analytical setting of von Neumann algebraic quantum groups for the physics of the DSSYK model. As mentioned in the introduction, there are further, potentially more refined, setups in non-commutative geometry, such as Connes' spectral triples \cite{ConnesNoncommutativeGeometry1990}. Exploring these latter can be pictured as (a non-commutative analogue of) going from measure theoretical considerations towards a more precise understanding of the structure of smooth functions on the corresponding space, and is expected to be directly related to a full-fledged (non-commutative-)geometric description of such discretised spacetimes and quantum gravity on those. Among other things, this approach is expected to clarify the geometric nature and dynamics of `fat points' corresponding to infinite-dimensional irreducible representations of the coordinate algebra (see Section~\ref{sec:quantumcoordinatealgebra}), whereas the natural metric structure provided by a spectral triple should shed light on the intriguing overlap of correlation functions between  AdS and dS regions of the $q$-deformed homogeneous space, which we observed in Section~\ref{sec:strangematter}. Just like in the von Neumann algebraic quantum group setting, however, the non-compactness of the physics problem in question brings in a lot of technical difficulties. Implementing and exploring this vision in the present context is a work in progress \cite{WPJort}.

    \item Finally, in this paper, we analysed the von Neumann algebraic quantum group structure related to the DSSYK, viewed as a toy model of quantum gravity. A natural question is whether one can `globalise' this construction in a physically meaningful way, to get $q$-deformed models with a similar level of analytical control, but richer dynamics. Namely, one can imagine a structure similar to the usual net structure of algebraic QFT \cite{HaagLocalQuantumPhysics}, but having a `net' of $q$-deformed spaces, with symmetries respecting the net structure. A natural way to approach that is by first producing more interesting toy examples from appropriately described glueing of DSSYK-like models, see e.g. \cite{MertensSimonWong2023Proposal3dQG} for some initial work in this direction. Fleshing out this idea in the operator-algebraic framework discussed in the present paper remains an interesting question for the future.
\end{itemize}

\section*{Acknowledgements}
We thank E. Opdam for support and collaboration in the initial stages of this project. We also thank A. Belaey, J. de Groot, E. Koelink, T. Mertens and T. Tappeiner for comments on the draft of the paper. We would like to thank S. Aguilar, M. Berkooz, J. de Groot, E. Koelink, K. Krawczyk, T. Mertens, E. Opdam, H. Posthuma, J. Stokman, J. Teschner and E. Verlinde for useful discussions and comments. This work is partially based on the Master's thesis of the first author, supervised by the second author and E. Opdam \cite{S2024}.

\begin{appendices}
    \section{Special functions}
    \label{sec:specialfunctions}
    In both the derivation of the quantum Gauss decomposition (Section~\ref{sec:quantumGaussDecomp}) and in the calculation of the partition function and $n$-point functions of the double-scaled SYK model (Section~\ref{sec:reductiondsSYK} and \cite{BNS2018, BINT2019}), one makes heavy use of $q$-orthogonal polynomials/functions. Indeed, their recurrence relations appear in the action of operators in the representation theory of the quantum group $\mathrm{SU}_q(1,1)\rtimes \Z_2$. In this section, we will recall and/or derive some properties of the special functions and give the spectra and eigenvectors of appropriate self-adjoint extensions of some {\it Jacobi operators} (specific tridiagonal operators) that we encounter in this paper. We refer to \cite{K2001, Koelink2018, CK2010} for some overviews on Jacobi operators and special functions and to \cite{KLS2010} for an overview on $q$-orthogonal polynomials.

    \subsection{$q$-Calculus and the basic hypergeometric series}
    Let us begin by recalling some definitions that appear in $q$-calculus and the definition of the basic hypergeometric series. We refer to \cite{QGAR, BHS} for complete discussions. During this section, we will keep the deformation parameter in the range $q\in (0,1)$. For any $z\in \C$, one can define a {\it $q$-number} as
    \begin{equation}
        \label{eq:qnumber}
        [z]_q \coloneq \frac{q^z - q^{-z}}{q-q^{-1}} = \frac{\sinh z\ln q}{\sinh \ln q}.
    \end{equation}
    It is clear that $[z]_q \to z$ in the limit $q\to 1$, such that one can indeed interpret this as the $q$-analogue of a number. Such $q$-numbers make a natural appearance in the representation theory of quantum groups, where, for example, upon quantisation, the Casimir eigenvalues are replaced by the corresponding $q$-numbers. Another quantity that makes a natural appearance in representation theory, $q$-combinatorics, and the theory of the basic hypergeometric series is the {\it $q$-Pochhammer symbol}, which is defined by
    \begin{equation}
        \label{eq:qpochhammer}
        (z;q)_n \coloneq \prod_{k=1}^{n}(1-zq^{k-1}), \quad (z;q)_0 \coloneq 1, \quad (z;q)_{-n} \coloneq (zq^{-n};q)_n^{-1},
    \end{equation}
    for $z \in \C$ and $n \in \Z_{\geq 0}$. We also define the limit $(z;q)_\infty \coloneq \lim_{n\to \infty}(z;q)_n$ and the product $(z_1,...,z_r;q)_n \coloneq (z_1;q)_n\cdots(z_r,q)_n$. We will also occasionally make use of the notation $(z^{\pm}; q)_n \coloneq (z,z^{-1};q)_{n}$. The $q$-Pochhammer symbol is closely related to the factorial, and therefore also to the gamma function. Indeed, the $q$-Pochhammer symbols allows us to define a $q$-analogue of the gamma function, called the {\it $q$-gamma function}, which is given by
    \begin{equation}
        \label{eq:qgamma}
        \Gamma_q(z) \coloneq (1-q)^{1-z}\frac{(q,q)_{\infty}}{(q^z;q)_{\infty}}.
    \end{equation}
    It can be shown that the $q$-gamma function reduces to the ordinary gamma function in the limit $q\to 1$, that is $\Gamma_q(z)\to \Gamma(z)$ as $q\to 1$ \cite{QGAR}. Some of the properties that the $q$-Pochhammer symbol and the $q$-gamma functions have can be found in \cite{QGAR, BHS}.\\

    The {\it basic hypergeometric series} are defined by \cite{BHS}
    \begin{equation}
        \label{eq:basichypergeometricfunctions}
        {}_r\phi_s \left(\nobarfrac{a_1,...,a_r}{b_1,...,b_s} ; q,z\right) := \sum_{n=0}^{\infty} \frac{(a_1,..., a_r ; q)_n}{(b_1,..., b_s ; q)_n}\left((-1)^n q^{n(n-1)/2}\right)^{1+s-r}\frac{z^n}{(q;q)_n},
    \end{equation}
    for $r,s \geq 0$. These are the natural $q$-analogue of the hypergeometric series, and can therefore be used to solve $q$-differential equations (i.e. certain recurrence relations). Here, it can be verified using the definition of the $q$-Pochhammer symbol that the above series terminates if one of the $a_i$'s equals $q^{-N}$ for some $N\geq 1$. Moreover, the series converges for all $z$ if $r\leq s$ and for $|z| < 1$ if $r=s+1$. The basic hypergeometric series (and their analytical continuations), together with the $q$-Pochhammer symbols, will be our main objects in defining eigenfunctions of some Jacobi operators.

    \subsection{Little $q$-Jacobi functions}
    \label{sec:littleqJacobi}
    The {\it little $q$-Jacobi functions} were originally defined to describe a kind of Fourier transform that appears in the representation theory of the quantum group $\mathrm{SU}_q(1,1)$ \cite{Kakehi1995, KS2001}. We refer to these papers, as well as \cite{K2001} and \cite{GKK2010}, for discussions on these functions. Here, we will give some of the basic results. Consider the unbounded doubly-infinite Jacobi operator $L$ on $\ell^2(\Z)$ given by
    \begin{equation}
        \label{eq:JacobilittleqJacobi}
        \begin{gathered}
            Le_k \coloneq a_k e_{k+1} + b_ke_k + a_{k-1}e_{k-1}\\
            a_k = \frac{1}{2}\sqrt{\left(1 - \frac{q^{-k}} {r}\right)\left(1 - \frac{c q^{-k}}{d^2 r}\right)}, \quad b_k = \frac{q^{-k}(c+q)}{2dr},
        \end{gathered}
    \end{equation}
    where $c,d,r \in \R$ such that $a_k > 0$\footnote{More generally, the coefficients can also be complex-valued (see, for example, \cite{KS2001}) under the condition that $a_k > 0$ and $b_k \in \R$. All the same results will follow. However, some points in the discrete spectrum might vanish if the values in \eqref{eq:discretespectrumlittleqJacobi} fail to remain real-valued.}. Let $y\in \C$ and $\mu(y) \coloneq \frac{1}{2}(y+y^{-1})$. Here, one can recognise the recurrence relation for the little $q$-Jacobi functions, which are defined by
    \begin{equation}
        \label{eq:nonnormalizedjacobifunctions}
        f_k(y) \coloneq f_k(\mu(y);c,d;q,r) = d^k\sqrt{\frac{(cq^{1-k}/d^2r;q)_\infty}{(q^{1-k}/r ; q)_{\infty}}} {}_2\phi_1\left(\nobarfrac{dy, d/y}{c}; q, rq^k\right).
    \end{equation}
    The vector $f(y) \coloneq \sum_{k}f_k(y) e_k$ is then an eigenvector of $L$ with eigenvalue $\mu(y)$. Moreover, we can normalise this vector by rescaling the little $q$-Jacobi functions and defining
    \begin{equation}
        j(x;c,d;q,r) = \|f(x;c,d;q,r)\|^{-1}f(x;c,d;q,r),
    \end{equation}
    where this norm is explicitly given by \cite[\S B.5]{GKK2010}. Of course, not all possible values $\mu(y)$ will appear in the spectrum of the self-adjoint extension of $L$. Instead, we define the set
    \begin{equation}
        \label{eq:discretespectrumlittleqJacobi}
        D = \left[\left\{d^{-1}q^{-k} : k\in \Z_{\geq 0}\right\}\cup \left\{\frac{d}{c}q^{-k} : k\in \Z_{\geq 0}\right\}\cup \left\{rdq^k : k\in \Z\right\}\right]\cap\{y \in \R : |y| < 1\}.
    \end{equation}
    and the measure on $\R$ given by
    \begin{equation}
        \label{eq:specialFunctionsMeasure}
        \int_{\R} f(x) d\nu(x) \coloneq \int_{-1}^{1}f(x) dx + \sum_{x\in \mu(D)} f(x),
    \end{equation}
    which contains both a continuous and a discrete support. In particular, the support of this measure is exactly given by the spectrum of the Jacobi operator $L$. In fact, we now have the following result:
    \begin{proposition}[\textit{\cite[\S B.5]{GKK2010}}]
        \label{thm:spectralmeasurelittleqJacobi}
        The operator $L$ is essentially self-adjoint for $0 < c \leq q^2$. Moreover, for $0 < c < 1$, $L$ has a self-adjoint extension with spectrum given by $\sigma(L) = [-1,1] \cup \mu(D)$ and spectral measure given by
        \begin{equation*}
            \langle u| E_L(\Delta)| v\rangle = \int_\Delta \overline{(\F_L u)(x)} (\F_L v)(s) d\nu(x), \quad (\F_L u)(x) = \sum_{k\in \Z}u_kj_k(x)
        \end{equation*}
        for any Borel subset $\Delta \subset \sigma(L)$ and $u,v \in \ell^{2}(\Z)$.
    \end{proposition}

    The above proposition, therefore, gives a Fourier transform that corresponds to the operator $L$. Let us now discuss some additional properties of the little $q$-Jacobi functions. Firstly, the little $q$-Jacobi functions satisfy the symmetry relation \cite[(B.46)]{GKK2010}
    \begin{equation}
        \label{eq:symmetryqJacobi}
        j_k(x;c,d;q,r) = j_k(x; c, c/d; q, rd^2/c).
    \end{equation}
    Moreover, it satisfies the shift condition \cite[(B.46)]{GKK2010}
    \begin{equation}
        \label{eq:qprrelationsJacobi}
        j_k(x;c,d; q, q^{p}r) = \sign{d}^p j_{k+p}(x;c,d;q,r),
    \end{equation}
    where it should be noted that we use a slightly different convention than \cite{GKK2010}. Under the redefinition $j_k \to j_{-k}$, we get the same expression. Next, we will state some useful $q$-difference relations for the little $q$-Jacobi functions. Firstly, we recall the result from \cite{GKK2010}:
    \begin{lemma}[\textit{\cite[Lemma B.16]{GKK2010}}]
        \label{lemma:relationsqJacobi}
        The functions $j_k(x; c,d; q,r)$ satisfy
        \begin{equation*}
            \begin{split}
                &\sqrt{1- \frac{1}{r}q^{-k}}j_{k+1}(x;c,d;q,r) - d\sqrt{1-\frac{c}{d^2r}q^{-k}}j_{k}(x;c,d;q,r)\\
                &= \sign{d}\sqrt{1-2xd+d^2}j_k(x;qc,qd; q,r).
            \end{split}
        \end{equation*}
    \end{lemma}
    Next, we give some new $q$-difference relations:
    \begin{lemma}
        \label{lemma:firstqdifferenceqJacobi}
        For $\mu(y) \in \sigma(L)\setminus [-1,1]$, the functions $j_k(\mu(y); c,d; q,r)$ satisfy the $q$-difference relations
        \begin{equation*}
            \begin{split}
                &q^{\frac{k}{2}}\sqrt{1 - \frac{c}{d^2r}q^{-k}}j_k(\mu(y); c, d; q,r)\\
                &= A(y)j_k(\mu(q^{1/2}y); c, q^{1/2}d; q,r) +  A(y^{-1})j_k(\mu(q^{-1/2}y); c, q^{1/2}d; q,r)\\
            \end{split}
        \end{equation*}
        where $A(y) = \sqrt{-\frac{c}{d^2r}\frac{(1-dy)(1-qdy/c)}{(1-y^2)(1-qy^2)}}$.
        \begin{proof}
            Firstly, by using the definition of the basic hypergeometric series \eqref{eq:basichypergeometricfunctions}, it can be shown that
            \begin{equation*}
                {}_2\phi_1\left(\nobarfrac{dy, d/y}{c}; q, rq^{k}\right) = \frac{1-dy}{1-y^2}{}_2\phi_1\left(\nobarfrac{qdy, d/y}{c}; q, rq^{k}\right) + \frac{1-d/y}{1-1/y^2}{}_2\phi_1\left(\nobarfrac{dy, qd/y}{c}; q, rq^{k}\right).
            \end{equation*}
            Therefore, by definition of $f_k$ as in \eqref{eq:nonnormalizedjacobifunctions}, we have
            \begin{equation*}
                \begin{split}
                    &q^{\frac{k}{2}}\sqrt{1-\frac{c}{d^2r}q^{-k}}f_k(\mu(y); c,d; q,r)\\
                    &= \frac{1-dy}{1-y^2}f_k(\mu(q^{1/2}y); c, q^{1/2}d; q,r) + \frac{1-d/y}{1-1/y^2}f_k(\mu(q^{-1/2}y); c, q^{1/2}d; q,r).
                \end{split}
            \end{equation*}
            To obtain the result, one uses the normalisation of the little $q$-Jacobi functions as given in \cite[(B.37)]{GKK2010}. For example, for $y = q^{1-\ell}/dr$, we have
            \begin{equation}
                \label{eq:normalisationfactorJacobi}
                \|f(\mu(y); c,d; q,r)\|^{2} = \frac{(q,q,cy/d, dy, c/dy, d/y; q)_\infty}{-(1-y^2)(dr)^{2(1-\ell)}q^{-(\ell-2)(\ell - 1)}}.
            \end{equation}
            Therefore also
            \begin{equation*}
                \begin{split}
                    \frac{\|f(\mu(q^{1/2}y); c, q^{1/2}d; q,r)\|^{2}}{\|f(\mu(y); c,d; q,r)\|^{2}} &= -\frac{c}{d^2r}\frac{(1-y^2)(1-qdy/c)}{(1-qy^2)(1-dy)},\\
                    \frac{\|f(\mu(q^{-1/2}y); c, q^{1/2}d; q,r)\|^{2}}{\|f(\mu(y); c,d; q,r)\|^{2}} &= -\frac{c}{d^2r}\frac{(1-1/y^2)(1-qd/cy)}{(1-q/y^2)(1-d/y)}.
                \end{split}
            \end{equation*}
            One can find the same expressions for other values of $\mu(y) \in \sigma(L) \setminus [-1,1]$. Now, using the fact that $j_k = \frac{1}{\|f_k\|} f_k$, the result follows.
        \end{proof}
    \end{lemma}

    By using the symmetry relation \eqref{eq:symmetryqJacobi} and \eqref{eq:qprrelationsJacobi}, one also gets the second relation
    \begin{corollary}
        \label{cor:firstqdifferenceqJacobicor}
        For $\mu(y) \in \sigma(L)\setminus [-1,1]$, the functions $j_k(\mu(y); c,d; q,r)$ satisfy the $q$-difference relations
        \begin{equation*}
            \begin{split}
                &q^{\frac{k}{2}}\sqrt{1 - \frac{1}{r}q^{-k}}j_k(\mu(y); c, d; q,r)\\
                &= A(y)j_{k+1}(\mu(q^{1/2}y); c, q^{-1/2}d; q,r) +  A(y^{-1})j_{k+1}(\mu(q^{-1/2}y); c, q^{-1/2}d; q,r)\\
            \end{split}
        \end{equation*}
        where $A(y) = \sign{d}\sqrt{-\frac{1}{r}\frac{(1-cy/d)(1-qy/d)}{(1-y^2)(1-qy^2)}}$.
    \end{corollary}
    
    Lastly, there is a third $q$-difference equation that will be useful.
    
    \begin{lemma}
        \label{lemma:secondqdifferenceqJacobi}
        For $\mu(y) \in \sigma(L)\setminus [-1,1]$, the functions $j_k(\mu(y); c,d; q,r)$ satisfy the $q$-difference relations
        \begin{equation*}
            \begin{split}
                &q^{\frac{k}{2}}j_k(\mu(y); c,d;q,r)\\
                &= A(y)j_k(\mu(q^{1/2}y); qc, q^{1/2}d; q,r) + A(y^{-1})j_k(\mu(q^{-1/2}y); qc, q^{1/2}d; q,r),
            \end{split}
        \end{equation*}
        where $A(y) = \sqrt{\frac{qy}{dr}\frac{(1-dy)(1-cy/d)}{(1-y^2)(1-qy^2)}}$.
        \begin{proof}
            The proof is analogous to the proof for \textit{Lemma~\ref{lemma:firstqdifferenceqJacobi}}. However, here one starts with the relation
            \begin{equation*}
                \begin{split}
                    &(1-qdy/c){}_2\phi_1\left(\nobarfrac{dy, d/y}{c}; q, rq^k\right)\\
                    &= (1-dy){}_2\phi_1\left(\nobarfrac{qdy, d/y}{c}; q, rq^k\right) -\frac{qdy}{c}(1-q^{-1}c){}_2\phi_1\left(\nobarfrac{dy, d/y}{q^{-1}c}; q, rq^k\right),
                \end{split}
            \end{equation*}
            which can be derived from the definition of the basic hypergeometric series \eqref{eq:basichypergeometricfunctions}. Applying these relations to the functions $f_k$ as defined in \eqref{eq:nonnormalizedjacobifunctions} gives
            \begin{equation*}
                \begin{split}
                    q^{\frac{k}{2}}f_k(\mu(y); c,d; q,r) &= \frac{(1-dy)(1-cy/d)}{1-y^2}f_k(\mu(q^{1/2}y); qc, q^{1/2}d; q,r)\\
                    &+ \frac{(1-d/y)(1-c/dy)}{1-1/y^2}f_k(\mu(q^{-1/2}y); qc, q^{1/2}d; q,r).
                \end{split}
            \end{equation*}
            We now again use the normalisation of the little $q$-Jacobi functions as given in \cite[(B.37)]{GKK2010}. For example, for $y = q^{1-\ell}/dr$, this normalisation is given in \eqref{eq:normalisationfactorJacobi}, and therefore
            \begin{equation*}
                \begin{split}
                    \frac{\|f(\mu(q^{1/2}y); qc, q^{1/2}d; q,r)\|^{2}}{ \|f(\mu(y); c,d; q,r)\|^{2}} &= \frac{qy}{dr}\frac{(1-y^2)}{(1-qy^2)(1-dy)(1-cy/d)}\\
                    \frac{\|f(\mu(q^{-1/2}y); qc, q^{1/2}d; q,r)\|^{2}}{\|f(\mu(y); c,d; q,r)\|^{2}} &= \frac{q}{dry}\frac{(1-1/y^2)}{(1-q/y^2)(1-d/y)(1-c/dy)}.
                \end{split}
            \end{equation*}
            One can find the same expressions for other values of $\mu(y) \in \sigma(L) \setminus [-1,1]$. Now using, the fact that $j_k = \frac{1}{\|f\|} f_k$, the result follows.
        \end{proof}
    \end{lemma}

    All the above lemmas can be extended to hold for all $\mu(y)\in \sigma(L)$, i.e. also for $\mu(y) \in [-1,1]$. One has to be careful here with the fact that now $\mu(q^{1/2}y)$ and $\mu(q^{-1/2}y)$ do not necessarily lie in the spectrum of $L$. However, the functions $j_k$ can be naturally extended to these values.

    \subsection{Al-Salam-Chihara polynomials}
    \label{sec:AlSalamChihara}
    In this section, we consider the {\it Al-Salam-Chihara polynomials}. These are actually very closely related to the little $q$-Jacobi functions, and one might view them as some bounded variant. Indeed, as can be seen in Section~\ref{sec:quantumGaussDecomp}, the appearance of the little $q$-Jacobi function is often accompanied by the appearance of the Al-Salam-Chihara polynomials. For $0 < q < 1$, consider the bounded Jacobi operator $J_q$ on $\ell^2(\Z_{\geq 0})$ given by
    \begin{equation}
        \label{eq:JacobiOperatorAl-Salam-Chihara}
        \begin{gathered}
            J_q e_k \coloneq a_k e_{k+1} + b_ke_k + a_{k-1}e_{k-1},\\
            a_k = \frac{1}{2}\sqrt{(1-q^{k+1})(1-abq^k)}, \quad b_k = \frac{1}{2}q^k(a+b),
        \end{gathered}
    \end{equation}
    where $a,b\in \R$\footnote{More generally, $a$ and $b$ can also be complex conjugates such that $\text{max}(|a|, |b|) < 1$. In this case, the same results follow, but the discrete spectrum will vanish.} such that $ab < 1$. Again, one can observe the similarity to \eqref{eq:JacobilittleqJacobi}. This operator is given by the (symmetric) recurrence relation for the Al-Salam-Chihara polynomials (see \cite[\S 14.8]{KLS2010} or \cite[\S B.4]{GKK2010}), which are given by
    \begin{equation}
        \label{eq:al-salam-chiharapolynomials}
        \begin{split}
            P_k(\mu(y); a,b ; q) &= a^{-k}(ab; q)_k~{}_3\phi_2\left(\nobarfrac{q^{-k}, ay, a/y}{ab, 0} ; q, q\right)\\
            &= (a/y ; q)_k y^k {}_2\phi_1\left(\nobarfrac{q^{-k}, by}{q^{1-k}y/a} ; q , q/ya\right),
        \end{split}
    \end{equation}
    for $y\in \C\setminus \{0\}$ and $\mu(y) = \frac{1}{2}(y + y^{-1})$. These polynomials only solve for the non-symmetric recurrence relation. The polynomials that solve the symmetric recurrence relation are given by
    \begin{equation}
        \label{eq:smallalsalamchiharapolynomials}
        p_k(x) \coloneq p_k(x; a,b; q) = (q, ab; q)_k^{-\frac{1}{2}}P_k(x; a,b ;q),
    \end{equation}
    such that $p(x) = \sum_k p_k(x)e_k$ are eigenvectors of $L$ with eigenvalue $x$. These vectors can be normalised by defining
    \begin{equation}
        \label{eq:normalizedal-salam-chiharapolynomials}
        h(x; a,b; q) \coloneq \|p(x;a,b;q)\|^{-1}p(x;a.b;q),
    \end{equation}
    where the norm is given in \cite[\S B.4]{GKK2010}. Not all values of $x$ appear in the self-adjoint extension of $J_q$. Instead, we define the set
    \begin{equation}
        D_q \coloneq \left[\{a^{-1}q^{-k} : k\in \Z_{\geq 0}\}\cup \{b^{-1}q^{-k} : k\in \Z_{\geq 0}\}\right]\cap \{y\in \R : |y| < 1\},
    \end{equation}
    and the measure on $\R$ given by
    \begin{equation}
        \int_{\R} f(x) d\nu(x) \coloneq \int_{-1}^{1}f(x) dx + \sum_{x\in \mu(D_q)} f(x).
    \end{equation}
    The support of this measure now corresponds to the spectrum of the operator $J_q$. Indeed, we have the following result:
    \begin{proposition}[\textit{\cite[\S B.4]{GKK2010}}]
        \label{thm:spectralmeasureboundedAl-Salam_Chihara}
        $J_q$ is a bounded self-adjoint operator with spectrum $\sigma(J_q) = [-1,1]\cup \mu(D_q)$. Moreover, its spectral measure is given by
        \begin{equation*}
            \langle u| E_{J_q}(\Delta)| v\rangle = \int_{\Delta} \overline{(\F_{J_q} u )(x)} (\F_{J_q} v)(x) d\nu(x), \quad (\F_{J_q} u)(x) = \sum_{k\in \Z_{\geq 0}} u_k h_k(x),
        \end{equation*}
        for any Borel subset $\Delta \subset \sigma(J_q)$ and $u,v\in \ell^{2}(\Z_{\geq 0})$.
    \end{proposition}

    Next, we are also interested in the unbounded case. That is, we replace $q\to q^{-1}$ and therefore consider the unbounded Jacobi operator $J_{q^{-1}}$. Here, we assume $ab > 1$. The eigenfunctions of this Jacobi operator are then given by ${q}^{-1}$-Al-Salam-Chihara polynomials (i.e. replace $q\to q^{-1}$ in \eqref{eq:smallalsalamchiharapolynomials}), which are actually dual to the little $q$-Jacobi polynomials (not functions), see \cite[Remark 3.1]{G2004} or \cite{K2004}. The spectrum for the self-adjoint extension of $J_{q^{-1}}$ will not simply be given by replacing $q \to q^{-1}$ in the spectrum of $J_q$. Instead, we consider the discrete set
    \begin{equation}
        D_{q^{-1}} \coloneq \{aq^{-k} : k\in \Z_{\geq 0}\}.
    \end{equation}
    On this discrete set, we define the functions
    \begin{equation}
        h_k(\mu(aq^{-n}); a,b;q^{-1}) = \sqrt{\frac{q^{k^2}w_{n}(a,b;q)}{(ab)^k(q, (ab)^{-1} ; q)_k}}P_k(\mu(aq^{-n}) ; a,b ; q^{-1}),
    \end{equation}
    where
    \begin{equation}
        w_n(a,b;q) = \left(\frac{b}{a}\right)^nq^{n^2}\frac{1-a^{-2}q^{2n}}{1-a^{-2}}\frac{(a^{-2}, (ab)^{-1} ; q)_n}{(q, a^{-1}bq ; q)_n}\frac{(a^{-1}bq ; q)_{\infty}}{(qa^{-2} ; q)_{\infty}}.
    \end{equation}
    Since $J_{q^{-1}}$ is symmetric in $a$ and $b$, we can assume w.l.o.g that $a/b > q$. In this case, the functions $h_k$ form an orthonormal set and satisfy \cite{K2004}
    \begin{equation}
        \sum_{n = 0}^{\infty} (h_k h_\ell)(\mu(aq^{-n}); a,b; q^{-1}) = \delta_{k\ell}.
    \end{equation}
    This then gives the following result
    \begin{proposition}[\textit{\cite{K2004}}]
        \label{thm:spectralmeasureunboundedAlSalamChihara}
        $J_{q^{-1}}$ is essentially self-adjoint for $a/b > q^{-1}$. Moreover, for $a/b > q$, there exists a self-adjoint extension of $J_{q^{-1}}$ with spectrum $J_{q^{-1}} = \{\mu(a^{-1}q^{k}) : k\in \Z_{\geq 0}\}$ and spectral measure
        \begin{equation*}
            \langle u| E_{J_{q^{-1}}}(\Delta) | v\rangle = \sum_{x\in \Delta} \overline{(\F_{J_{q^{-1}}} u)(x)}\F_{J_{q^{-1}}}v(x), \quad (\F_{J_{q^{-1}}} u)(x) = \sum_{k\in \Z_{\geq 0}} u_k h_k(x; a,b; q^{-1}),
        \end{equation*}
        for any Borel subset $\Delta \subseteq \sigma(J_{q^{-1}})$ and $u,v \in \ell^{2}(\Z_{\geq 0})$.
    \end{proposition}
    Let us now discuss some relations on the functions $h_k$ corresponding to the $q$-Al-Salam-Chihara polynomials \eqref{eq:normalizedal-salam-chiharapolynomials}. Similar relations for the $q^{-1}$-Al-Salam-Chihara polynomials can be found by simply replacing $q \to q^{-1}$. Firstly, we have
    \begin{lemma}[\textit{\cite[Lemma B.14]{GKK2010}}]
        \label{lemma:relationsAlSalamChihara}
        The functions $h_k(x; a,b; q)$ satisfy
        \begin{equation*}
            \sqrt{1-abq^k}h_k(x;a,qb; q) - b\sqrt{1-q^k}h_{k-1}(x; a,bq; q) = \sqrt{1-2bx + b^2}h_k(x; a,b; q),
        \end{equation*}
        for $x\in \sigma(J_q)$.
    \end{lemma}
    We now derive some additional $q$-difference equations.
    \begin{lemma}
        \label{lemma:firstqdifferenceAlSalamChihara}
        For $\mu(y) \in \sigma(J_q)\setminus [-1,1]$, the functions $h_k(\mu(y); a,b; q)$ satisfy the $q$-difference relations
        \begin{equation*}
            \begin{split}
                &q^{-\frac{k}{2}}\sqrt{1 - abq^k}~h_k(\mu(y); a, b; q)\\
                &= A(y)h_k(\mu(q^{1/2}y); q^{1/2}a, q^{1/2}b; q) +  A(y^{-1})h_k(\mu(q^{-1/2}y); q^{1/2}a, q^{1/2}b; q)\\
            \end{split}
        \end{equation*}
        where $A(y) = \sqrt{-qy^2\frac{(1-ay)(1-by)}{(1-y^2)(1-qy^2)}}$.
        \begin{proof}
            Firstly, by the definition of the basic hypergeometric series \eqref{eq:basichypergeometricfunctions}, we have
            \begin{equation*}
                \begin{split}
                    &(1-qy/b){}_3\phi_2\left(\nobarfrac{q^{-k}, ay, a/y}{ab, 0}; q,q\right)\\
                    &= (1-ay){}_3\phi_2\left(\nobarfrac{q^{-k}, qay, a/y}{ab, 0}; q,q\right) - \frac{qy}{b}(1-q^{-1}ab){}_3\phi_2\left(\nobarfrac{q^{-k}, ay, a/y}{q^{-1}ab, 0}; q,q\right).
                \end{split}
            \end{equation*}
            From the definition of the Al-Salam-Chihara polynomials in \eqref{eq:smallalsalamchiharapolynomials} this now gives
            \begin{equation*}
                \begin{split}
                    q^{-\frac{k}{2}}\sqrt{1-abq^{k}}p_k(\mu(y); a,b;q)&= \frac{(1-ay)(1-by)}{(1-y^2)\sqrt{1-ab}}p_k(\mu(q^{1/2}y); q^{1/2}a, q^{1/2}b; q) \\
                    &+ \frac{(1-a/y)(1-b/y)}{(1-1/y^2)\sqrt{1-ab}}p_k(\mu(q^{-1/2}y); q^{1/2}a, q^{1/2}b; q).
                \end{split}
            \end{equation*}
            We can now use the normalisation of the Al-Salam-Chihara polynomials as given in \cite[(B.18)]{GKK2010}. In particular, for $y = aq^{\ell}$, we have
            \begin{equation}
                \label{eq:normalisationAlSalamChihara}
                \|p(\mu(y); a,b; q)\|^2 = \frac{(b/a; q)_\infty(q, aq/b; q)_\ell(1-a^2)}{(a^{-2}; q)_\infty (a^2, ab; q)_\ell (1-a^2 q^{2\ell})}q^{\ell^2}a^{3\ell} b^\ell.
            \end{equation}
            This gives
            \begin{equation*}
                \begin{split}
                    \frac{\|p(\mu(q^{1/2}y); q^{1/2}a, q^{1/2}b ; q)\|^2}{\|p(\mu(y); a,b; q)\|^2} &= -qy^2\frac{(1-y^2)(1-ab)}{(1-ay)(1-by)(1-qy^2)}\\
                    \frac{\|p(\mu(q^{-1/2}y); q^{1/2}a, q^{1/2}b ; q)\|^2}{\|p(\mu(y); a,b; q)\|^2} &= -qy^{-2}\frac{(1-1/y^2)(1-ab)}{(1-a/y)(1-b/y)(1-q/y^2)}.
                \end{split}
            \end{equation*}
            By using that $\|h_k\| = \frac{1}{\|p\|}p_k$, this gives the result.
        \end{proof}
    \end{lemma}

    \begin{lemma}
    \label{lemma:secondqdifferenceAlSalamChihara}
        For $\mu(y) \in \sigma(J_q)\setminus [-1,1]$, the functions $h_k(\mu(y); a,b; q)$ satisfy the $q$-difference relations
        \begin{equation*}
            \begin{split}
                &q^{-\frac{k}{2}}\sqrt{1 - q^k}~h_{k-1}(\mu(y); a, b; q)\\
                &= A(y)h_k(\mu(q^{1/2}y); q^{-1/2}a, q^{-1/2}b; q) +  A(y^{-1})h_k(\mu(q^{-1/2}y); q^{-1/2}a, q^{-1/2}b; q)\\
            \end{split}
        \end{equation*}
        where $A(y) = \sqrt{-q^{-1}aby^2\frac{(1-qy/a)(1-qy/b)}{(1-y^2)(1-qy^2)}}$.
        \begin{proof}
            This proof is analogous to the proof of \textit{Lemma~\ref{lemma:firstqdifferenceAlSalamChihara}}. By the definition of the basic hypergeometric series \eqref{eq:basichypergeometricfunctions}, we have
            \begin{equation*}
                q^{-k}{}_2\phi_1\left(\nobarfrac{q^{-k}, by}{q^{1-k}y/a}; q,q/ya\right) = {}_2\phi_1\left(\nobarfrac{q^{-k}, by}{q^{1-k}y/a}; q,1/ya\right) - (1-q^{-k}){}_2\phi_1\left(\nobarfrac{q^{1-k}, by}{q^{1-k}y/a}; q,q/ya\right)
            \end{equation*}
            From the definition of the Al-Salam-Chihara polynomials in \eqref{eq:smallalsalamchiharapolynomials}, this now gives
            \begin{equation*}
                \begin{split}
                    q^{-\frac{k}{2}}\sqrt{1-q^{k}}p_{k-1}(\mu(y); a,b;q)&= \frac{y\sqrt{1-q^{-1}ab}}{1-y^2}p_k(\mu(q^{1/2}y); q^{-1/2}a, q^{-1/2}b; q) \\
                    &+ \frac{y^{-1}\sqrt{1-q^{-1}ab}}{1-1/y^2}p_k(\mu(q^{-1/2}y); q^{-1/2}a, q^{-1/2}b; q).
                \end{split}
            \end{equation*}
            We can now again use the normalisation of the Al-Salam-Chihara polynomials as given in \cite[(B.18)]{GKK2010} and \eqref{eq:normalisationAlSalamChihara}. This then gives
            \begin{equation*}
                \begin{split}
                    \frac{\|p(\mu(q^{1/2}y); q^{-1/2}a, q^{-1/2}b ; q)\|^2}{\|p(\mu(y); a,b; q)\|^2} &= -q^{-1}ab\frac{(1-y^2)(1-qy/a)(1-qy/b)}{(1-qy^2)(1-q^{-1}ab)}\\
                    \frac{\|p(\mu(q^{-1/2}y); q^{-1/2}a, q^{-1/2}b ; q)\|^2}{\|p(\mu(y); a,b; q)\|^2} &= -q^{-1}ab\frac{(1-1/y^2)(1-q/ay)(1-q/by)}{(1-q/y^2)(1-q^{-1}ab)}.
                \end{split}
            \end{equation*}
            By using the fact that $\|h_k\| = \frac{1}{\|p\|}p_k$, this gives the result.
        \end{proof}
    \end{lemma}

    \begin{lemma}
        \label{lemma:thirdqdifferenceAlSalamChihara}
        For $\mu(y) \in \sigma(J_q)\setminus [-1,1]$, the functions $h_k(\mu(y); a,b; q)$ satisfy the $q$-difference relations
        \begin{equation*}
            \begin{split}
                &q^{-\frac{k}{2}}h_{k}(\mu(y); a, b; q)\\
                &= A(y)h_k(\mu(q^{1/2}y); q^{1/2}a, q^{-1/2}b; q) +  A(y^{-1})h_k(\mu(q^{-1/2}y); q^{1/2}a, q^{-1/2}b; q)\\
            \end{split}
        \end{equation*}
        where $A(y) = \sqrt{by\frac{(1-ay)(1-qy/b)}{(1-y^2)(1-qy^2)}}$.
        \begin{proof}
            This proof is again analogous to the previous proofs. By the definition of the basic hypergeometric series \eqref{eq:normalisationAlSalamChihara}, we have
            \begin{equation*}
                \begin{split}
                    &{}_3\phi_2\left(\nobarfrac{q^{-k}, ay, a/y}{ab, 0}; q,q\right)\\
                    &= \frac{1-ay}{1-y^2}{}_3\phi_2\left(\nobarfrac{q^{-k}, qay, a/y}{ab, 0}; q,q\right) - \frac{1-a/y}{1-1/y^2}{}_3\phi_2\left(\nobarfrac{q^{-k}, ay, qa/y}{ab, 0}; q,q\right)
                \end{split}
            \end{equation*}
            From the definition of the Al-Salam-Chihara polynomials in \eqref{eq:smallalsalamchiharapolynomials}, this now gives
            \begin{equation*}
                \begin{split}
                    q^{-\frac{k}{2}}p_{k}(\mu(y); a,b;q)&= \frac{1-ay}{1-y^2}p_k(\mu(q^{1/2}y); q^{1/2}a, q^{-1/2}b; q) \\
                    &+ \frac{1-a/y}{1-1/y^2}p_k(\mu(q^{-1/2}y); q^{1/2}a, q^{-1/2}b; q).
                \end{split}
            \end{equation*}
            Using the normalisation of the Al-Salam-Chihara polynomials as given in \cite[(B.18)]{GKK2010} and \eqref{eq:normalisationAlSalamChihara}, this gives
            \begin{equation*}
                \begin{split}
                    \frac{\|p(\mu(q^{1/2}y); q^{1/2}a, q^{-1/2}b ; q)\|^2}{\|p(\mu(y); a,b; q)\|^2} &= by\frac{(1-y^2)(1-qy/b)}{(1-qy^2)(1-ay)}\\
                    \frac{\|p(\mu(q^{-1/2}y); q^{1/2}a, q^{-1/2}b ; q)\|^2}{\|p(\mu(y); a,b; q)\|^2} &= by^{-1}\frac{(1-1/y^2)(1-q/by)}{(1-q/y^2)(1-a/y)}.
                \end{split}
            \end{equation*}
            By using the fact that $\|h_k\| = \frac{1}{\|p\|}p_k$, this gives the result.
        \end{proof}
    \end{lemma}
    Again, by replacing $q \to q^{-1}$, the above relations give the $q$-difference equations for the $q^{-1}$-Al-Salam-Chihara polynomials. Also, the above lemmas can be extended to hold for all $\mu(y)\in \sigma(J_q)$, so also for $\mu(y)\in [-1,1]$. However, one has to be careful with the fact that $\mu(q^{1/2}y)$ and $\mu(q^{-1/2}y)$ are not part of the spectrum of $J_q$. The functions $h_k$ need to be extended to these values for the relations to make sense. 
    
    \subsection{Al-Salam-Carlitz II polynomials}
    \label{sec:AlSalamCarlitzII}
    Another polynomial that shows up in the representation theory of $\mathrm{SU}_q(1,1)\rtimes \Z_2$, and actually only in the normaliser, is the {\it Al-Salam-Carlitz II polynomial} \cite{AlSalamCarlitz1965}. Consider the Jacobi operator $J$ on $\ell^2(\Z_{\geq 0})$ given by
    \begin{equation}
    \label{eq:alsalamcarlitzoperator}
        \begin{gathered}
            Je_k = a_ke_{k+1} + b_k e_k + a_{k-1}e_k,\\
            a_k = q^{-k - \frac{1}{2}}\sqrt{1-q^{k+1}}, \quad b_k = \epsilon (a^{\frac{1}{2}} + a^{-\frac{1}{2}})q^{-k},
        \end{gathered}
    \end{equation}
    where $0 < a < q^{-1}$ and $\epsilon \in \{-1,1\}$. This Jacobi operator is indeed exactly given by the (symmetric) recurrence relation for the Al-Salam-Carlitz polynomials of type II \cite[\S 14.25]{KLS2010}, which are given by
    \begin{equation}
        \label{eq:definitionAlSalamCarlitzII}
        P_k^{(a)}(x;q) \coloneq (-a)^kq^{-\frac{1}{2}k(k-1)}{}_2\phi_0\left(\nobarfrac{q^{-k}, x}{-}; q, q^k / a\right).
    \end{equation}
    In order for these to define an eigenvector for $J$, the polynomial should be symmetrised. In fact, we will also immediately normalise these functions. The symmetrised and normalised Al-Salam-Carlitz II polynomials are then given by
    \begin{equation}
        \label{eq:normalizedAlSalamCarlitz}
        p_k^{a}(\epsilon q^{-n}/\sqrt{a} ; q) \coloneq \sqrt{\frac{q^{n^2 + k^2} a^{n-k} (aq; q)_{\infty}}{(q;q)_n (q;q)_k (aq; q)_n}}\epsilon^k P_k^{(a)}(q^{-n} ; q),
    \end{equation}
    for $n \in \Z_{\geq 0}$. The vector $p^a(x) = \sum_{k} p^a_k(x)e_k$ is now an eigenfunction of $J$ with eigenvalue $x$ for $x = \epsilon q^{-n}/\sqrt{a}$. Moreover, they satisfy the orthonormality relation \cite[\S 14.25]{KLS2010}
    \begin{equation}
        \sum_{n \in \Z_{\geq 0}} p_k^{a}(\epsilon q^{-n} /\sqrt{a} ; q) p_\ell^{a}(\epsilon q^{-n}/\sqrt{a} ; q) = \delta_{k\ell}
    \end{equation}
    for $0 < a < q^{-1}$. We now get the following result:
    \begin{proposition}[\cite{Chihara1986}]
        \label{prop:alsalamcarlitzspectralmeasure}
        $J$ is essentially self-adjoint for $0 < a < q$. For $0 < a < q^{-1}$, there exists a self-adjoint extension of $J$ with spectrum $\sigma(J) = \{\epsilon q^{-k}/\sqrt{a} : k \in \Z_{\geq 0}\}$ and spectral measure
        \begin{equation*}
            \langle u| E_J(\Delta)|v\rangle = \sum_{x\in \Delta}\overline{(\F_J u)(x)} (\F_J v) (x), \quad (\F_J u)(x) = \sum_{k\in \Z_{\geq 0}} u_k p_k^{a}(x; q)
        \end{equation*}
        for any Borel set $\Delta \subseteq \sigma(J)$ and $u,v \in \ell^2(\Z_{\geq 0})$.
    \end{proposition}
    These polynomials satisfy the following useful $q$-difference equation:
    \begin{lemma}
        \label{lemma:qdifferenceAlSalamCarlitz}
        For $x\in \sigma(J)$, the functions $p_k^a$ satisfy the $q$-difference equation
        \begin{equation*}
            q^{\frac{k}{2}}p_k^a(x) = q^\frac{1}{2}|x|^{-1}\sqrt{q^{-1}|x|/\sqrt{a} - 1} p_k^{qa}(q^{-1/2}x) + q^{-\frac{1}{2}}|x|^{-1}\sqrt{|x|\sqrt{a} - 1} p_k^{qa}(q^{1/2}x).
        \end{equation*}
        \begin{proof}
            Define ${}_2\phi_0^a(x) = {}_2\phi_0\left(q^{-k}, x; - : q, q^{k}/a\right)$. Then, using the definition of the basic hypergeometric series \eqref{eq:basichypergeometricfunctions}, this function satisfies the $q$-difference relation
            \begin{equation*}
                (1-x){}_2\phi_0^a(qx) + x{}_2\phi_0^{q^{-1}a}(x) = {}_2\phi_0^a(x).
            \end{equation*}
            Therefore, from the definition of the Al-Salam-Carlitz II polynomials \eqref{eq:definitionAlSalamCarlitzII}, these satisfy
            \begin{equation*}
                (1-q^{-n}) P_k^{(a)}(q^{1-n}) + q^{-n}q^k P_k^{(q^{-1}a)}(q^{-n}) = P_k^{(a)}(q^{-n}).
            \end{equation*}
            If we write $x = \epsilon q^{-n}/\sqrt{a}$, then
            \begin{equation*}
                \begin{split}
                    \|P^{(q^{-1}a)}(q^{-n})\|^2 &= -q^{-k}\frac{a^{-1}}{1-\epsilon x/\sqrt{a}}\|P^{(a)}(q^{-n})\|.
                \end{split}
            \end{equation*}
            Also note that $P_k^{(q^{-1}a)}(q^{-n})\sim p_k^{q^{-1}a}(q^{1/2}x)$. Combining these, one can write the $q$-difference equation for $p$, and after some rewriting, one gets the result.
        \end{proof}
    \end{lemma}
    
    \subsection{Ciccoli-Koelink-Koornwinder functions}
    \label{sec:dualqLaguerre}
    We will now consider a set of functions of which the recurrence relation may be regarded as a non-terminating version of the recurrence for the Al-Salam-Carlitz II polynomials \eqref{eq:alsalamcarlitzoperator}, which were first discussed in the paper by Ciccoli, Koelink and Koornwinder \cite{CKK1999}. Consider the doubly-infinite Jacobi operator $L$ on $\ell^2(\Z)$ given by
    \begin{equation}
        \label{eq:qlaguerreoperator}
        \begin{gathered}
            L e_k \coloneq a_k e_{k+1} + b_k e_k + a_{k-1}e_{k-1},\\
            a_k = q^{-\frac{1}{2}(k+1)}\sqrt{1 + c^{-1}q^{-k}}, \quad b_k = \sqrt{c^{-1}}(t + t^{-1})q^{-k},
        \end{gathered}
    \end{equation}
    where $c > 0$ and $t\in \R\setminus \{0\}$ such that $a_k > 0$ and $b_k \in \R$. For $z\in \C$, we consider the functions
    \begin{equation}
        \label{eq:dualqLaguerre}
        \begin{split}
            V_k^{t}(z) &= q^{\frac{1}{2}k} (-t)^{-k}(-cq^k ; q)_\infty^{\frac{1}{2}}{}_2\phi_1\left(\nobarfrac{-qz / \sqrt{c} t, 0}{qt^{-2}} ; q, -cq^k\right)\\
            &= \frac{(-qz/\sqrt{c} t ; q)_\infty}{(qt^{-2} ; q)_\infty} q^{\frac{1}{2}k}(-t)^{-k}(-cq^k; q)_\infty^{-\frac{1}{2}} {}_2\phi_1\left(\nobarfrac{-\sqrt{c} / tz, -cq^k}{0} ; q, -qz / \sqrt{c}t\right),
        \end{split}
    \end{equation}
    then $V^t(z) = \sum_k V_k^t(z)e_k$ is an eigenvector of $L$ with eigenvalue $z$ with norm given by \cite[(3.2), (3.4)]{CKK1999}
    \begin{equation}
        \label{eq:normofqLaguerreFunction}
        \| V^t(z)\|^2 = \frac{t}{\sqrt{c}|z|}\frac{(-qz/\sqrt{c}t ; q)_{\infty}}{(-qtz/\sqrt{z} ; q)_\infty}\frac{(q,q,-c/t^2, -qt^2/c; q)_\infty}{(-c, -q/c, qt^{-2}, qt^{-2}; q)_\infty}.
    \end{equation}
    Therefore, we will define the normalised functions to be equal to $\tilde V^t_k(z) = \|V^t(z)\|^{-1}V^t_k(z)$. For $t^{-2} = q^\alpha$ with $\alpha > -1$ and $z = -\sqrt{c}t^{-1}q^n$ with $n\geq 0$, these functions are dual to the $q$-Laguerre functions (see \cite{CKK1999}). We now have the following result for the Jacobi operator $L$:

    \begin{proposition}[\cite{CKK1999}]
        \label{thm:spectralmeasureqlaguerre} $L$ is essentially self-adjoint for $|t| > q^{-\frac{1}{2}}$. For $|t| > q^{\frac{1}{2}}$, $L$ has a self-adjoint extension with spectrum
        \begin{equation*}
            \sigma(L) = \{-\sqrt{c}t^{-1} q^k : k \in \Z_{\geq 0}\}\cup \{\sqrt{c^{-1}} t q^k : k\in \Z\}\cup\{0\}
        \end{equation*}
        and the spectral measure given by
        \begin{equation*}
            \langle u| E_L(\Delta)| v\rangle = \sum_{x\in \Delta} \overline{(\F_L u)(x)}(\F_L v)(x), \quad (\F_L u)(x) = \sum_{k\in \Z} u_k\tilde V_k ^t(x; c; q)
        \end{equation*}
        for any Borel set $\Delta \subseteq \sigma(L)$ and $u,v \in \ell^{2}(\Z)$.
    \end{proposition}
    
    The functions $\tilde V_k^{t}(x)$ now satisfy a few useful relations:

    \begin{lemma}
        \label{lemma:qLaguerreContiguousRelations}
        Let $x\in \sigma(L)$, then the functions $\tilde V_k^t$ satisfy the relations
        \begin{equation*}
            \sqrt{1+c^{-1}q^{1-k}}\tilde V_{k-1}^{t}(x) + \sqrt{c^{-1}}~tq^{-\frac{1}{2}k} \tilde V_k^{t}(x) = \sqrt{1 + xt/\sqrt{c}}~\tilde V_k^{q^{-1/2}t}(q^{-1/2}x)
        \end{equation*}
        and
        \begin{equation*}
            \sqrt{c}\sqrt{1+c^{-1}q^{-k}}~\tilde V_{k+1}^t(x) + t^{-1}q^{\frac{1}{2} - \frac{1}{2}k}\tilde V_k^t(x) = -\sqrt{c}\sqrt{1 + qx / \sqrt{c}t}~\tilde V_k^{q^{-1/2}t}(q^{1/2}x).
        \end{equation*}
        \begin{proof}
            Firstly, it can be shown from the definition of the basic hypergeometric series \eqref{eq:basichypergeometricfunctions} that (see also \cite[ex. 1.9]{BHS})
            \begin{equation*}
                {}_2\phi_1\left(\nobarfrac{aq, b}{0}; q,z\right) - {}_2\phi_1\left(\nobarfrac{a, b}{0}; q,z\right)  = az(1-b) {}_2\phi_1\left(\nobarfrac{aq, bq}{0}; q,z\right).
            \end{equation*}
             From the definition \eqref{eq:dualqLaguerre}, it then follows that $V_k^t(z)$ satisfies the relation
            \begin{equation*}
                \sqrt{1+c^{-1}q^{1-k}}V_{k-1}^t(z) + \sqrt{c^{-1}}tq^{-\frac{1}{2}k}V_k^t(z) = \frac{\sqrt{c} t^{-1} + z}{1 - qt^{-2}}V_k^{q^{-1/2}t}(q^{-1/2}z).
            \end{equation*}
            Moreover, using the normalisation  \eqref{eq:normofqLaguerreFunction} of the functions, we have
            \begin{equation*}
                \|V^{q^{-1/2}t}(q^{-1/2}z)\|^2 = \frac{(1-qt^{-2})^2}{(1+tz/\sqrt{c})}\frac{t^2}{c}\|V^t(z)\|^2,
            \end{equation*}
            which gives the first relation. The second relation follows analogously, but now starting from the contiguous relation
            \begin{equation*}
                {}_2\phi_1\left(\nobarfrac{a,b}{0} ; q,z\right) - (1-b)\ {}_2\phi_1\left(\nobarfrac{a,bq}{0} ; q,z\right) = b\ {}_2\phi_1\left(\nobarfrac{a,b}{0} ; q,qz\right)
            \end{equation*}
            and then performing the same steps as before.
        \end{proof}
    \end{lemma}

    \begin{lemma}
        \label{lemma:qdifferenceqLaguerre}
        For $x\in \sigma(L)$, the functions $\tilde V_k^t$ satisfy the $q$-difference equation
        \begin{equation*}
            q^{\frac{k}{2}}\tilde V_k^t(x) = x^{-1}\sqrt{1 + tx/\sqrt{c}}~\tilde V_k^{q^{-1/2}t}(q^{-1/2}x) - x^{-1}\sqrt{q^{-1} + x/\sqrt{c}t}\tilde V_k^{q^{-1/2}t}(q^{1/2}x).
        \end{equation*}
        \begin{proof}
            Define ${}_2\phi_1^t(z) = {}_2\phi_1\left(-qz/\sqrt{c}t, 0 ; qt^{-2} : q, -cq^k\right)$.
            From the definition of the basic hypergeometric series \eqref{eq:basichypergeometricfunctions}, it then follows that
            \begin{equation*}
                (1+qz/\sqrt{c} t){}_2\phi_1^t(qz) + qzt/\sqrt{c}(1-t^{-2}){}_2\phi_1^{q^{1/2}t}(q^{1/2}z) = (1+qzt/\sqrt{c}){}_2\phi_1^t(z).
            \end{equation*}
            From \eqref{eq:dualqLaguerre} and after some rewriting, it then follows that
            \begin{equation*}
                (1+qz/\sqrt{c}t)V_k^{q^{-1/2}t}(q^{1/2}z) + zt/\sqrt{c}(1-qt^{-2})q^{\frac{1}{2}k} V_k^t(z) = (1+zt/\sqrt{c}) V_k^{q^{-1/2}t}(q^{-1/2}z).
            \end{equation*}
            Now use \eqref{eq:normofqLaguerreFunction} to get
            \begin{equation*}
                \begin{split}
                    \|V^{q^{-1/2}t}(q^{1/2}x)\|^2 &= q^{-1}\frac{t^2}{c}\frac{(1-qt^{-2})^2}{1+qx/\sqrt{c}t}\|V^t(x)\|^2,\\
                    \|V^{q^{-1/2}t}(q^{-1/2}x)\|^2 &= \frac{t^2}{c}\frac{(1-qt^{-2})^2}{1+tx/\sqrt{c}}\|V^t(x)\|^2.
                \end{split}
            \end{equation*}
            The result now follows by substituting in $\tilde V_k^t(x) = \|V^t(x)\|^{-1}V_k^t(x)$ for $x\in \sigma(L)$.
        \end{proof}
    \end{lemma}

    \section{Calculations for the operator-algebraic quantum group action}
    \label{sec:calcsqgactions}
    In this section, we calculate the action of some of the operators of the coordinate algebra $\A_q(\mathrm{SU}(1,1)\rtimes \Z_2)$ and the quantum universal enveloping algebra $\U_q(\mathfrak{su}(1,1))$. Most of these calculations are somewhat tedious, and make use of the $q$-difference relations of the special functions as given/derived in Appendix~\ref{sec:specialfunctions}. 
    \subsection{Actions in the quantum Iwasawa decomposition}
    \label{sec:actionsIwasawaDecomp}
     Let us first calculate the action of the coordinate algebra on the basis elements $g^{\pm, t}_{n,x,m}$ of $L_q^2(\mathrm{SU}(1,1)\rtimes\Z_2)$ as given in \eqref{eq:preIwasawaBasis}. We follow a similar derivation as given in \cite{KS2001}. First, let $x = q^{2k}$ or $x = -t^{-2}q^{k2}$, i.e. $e(x) = 1$, and consider the element $\alpha_{\infty,t}^{*} = \alpha^* + qt^{-1}\gamma$ \eqref{eq:coordinateAlgebraGauss}. By Lemma~\ref{lemma:qLaguerreContiguousRelations}, we have
    \begin{equation}
        \begin{split}
            &\alpha_{\infty, t}^*.g_{n,x,m}^{\pm ,t}\\
            &= \sum_{k\in \Z} \tilde V_k^{t}(tx; 1; q^2)(\alpha^* + qt^{-1}\gamma). \delta_{n, q^{-2k}, m+2k}^{\pm}\\
            &= \sum_{k\in \Z}\left(\sqrt{1+ q^{-2k}}~\tilde V_{k+1}^{t}(tx; 1; q^2) + t^{-1}q^{1-k}\tilde V_k^t(tx; 1; q^2)\right)\delta_{n-1, q^{-2k}, m+1+2k}^{\pm}\\
            &= -\sqrt{1 + q^2x}~g^{\pm, q^{-1}t}_{n-1, q^2x, m+1}.
        \end{split}
    \end{equation}
    This also gives the action of $\alpha_{\infty, qt}$ by taking the adjoint.
    For $x = -q^{-2k}$, i.e. $e(x) = -1$, one can assume that $\alpha_{\infty, qt}.g_{n,x,m}^{\pm, t} = C\cdot g_{n+1, q^{-2}x, m-1}^{\pm, qt}$, which can be motivated by the commutation relation of $\alpha_{st}$ with the spherical element $\rho_{st}$ \eqref{eq:sphericalElementcommutation}. We get
    \begin{equation}
        \label{eq:SUactiononPreIwasawaBasis}
        \begin{split}
            -C p_0^{(qt)^{-2}}(qtx ; q^2) &= \langle \alpha_{\infty, qt} .g^{\pm, t}_{n,x,m}, \delta^{\pm}_{n+1, q^{-2}, m+1}\rangle = \langle g^{\pm, t}_{n,x,m}, \alpha_{\infty, qt}^* .\delta^{\pm}_{n+1, q^{-2}, m+1}\rangle\\
            &= -q^{-1}t^{-1}p_0^{t^{-2}}(tq^2x ; q^2).
        \end{split}
    \end{equation}
    Therefore, we have $C = q^{-1}t^{-1}p_0^{t^{-2}}(tq^2x ; q^2) / p_0^{(qt)^{-2}}(qtx ; q^2)$, which can be calculated directly using the definition of the Al-Salam-Carlitz II polynomial \eqref{eq:normalizedAlSalamCarlitz}. Analogously, the above steps can be done for $\gamma_{\infty, t}$. The actions on the basis elements $\delta^{\pm, (\infty, t)}_{n,x,\xi}$ \eqref{eq:IwasawaBasis} can now be calculated directly and are given by \eqref{eq:coordinateAlgebraActionIwasawa}.\\

    Let us now calculate the left action of $K$ on the basis $g_{n,x,m}^{\pm, t}$ \eqref{eq:preIwasawaBasis}. For $x = q^{2k}$ or $x = -t^{-2}q^{-2k}$ such that $e(x) = 1$, one finds
    \begin{equation}
        \begin{split}
            K. g^{\pm, t}_{n,x,m} &= q^{\frac{m}{2}}\sum_{k\in \Z}q^{k}\tilde V_k^{t}(tx; q; q^2)\delta^{\pm}_{n, q^{-2k}, m+2k}\\
            &= q^{\frac{m}{2}}(tx)^{-1}\left[\sqrt{1+t^2x}~g_{n,x,m}^{\pm, q^{-1}t} -\sqrt{q^{-2}+x}~g_{n, q^2x, m}^{\pm, q^{-1}t}\right],
        \end{split}
    \end{equation}
    where we used Lemma~\ref{lemma:qdifferenceqLaguerre} in the second equality. For $x = -q^{-2k}$ such that $e(x) = -1$ one can use Lemma~\ref{lemma:qdifferenceAlSalamCarlitz} to find a similar expression. Also, since $K^* = K$, one can obtain a second expression for the action of $K$ by taking the adjoint of the above result. In particular, the action of $K$ is given by
    \begin{equation}
        \label{eq:actionKIwasawa}
        \begin{split}
            K.g_{n,x,m}^{\pm, t} &= q^{\frac{m}{2}}(tx)^{-1}\left[\sqrt{e(x)(1+t^2x)} e.g_{n,x,m}^{\pm, q^{-1}t} -\sqrt{e(x)(q^{-2}+x)}g_{n, q^2x, m}^{\pm, q^{-1}t}\right]\\
            &= q^{\frac{m}{2}}(tx)^{-1}\left[\sqrt{e(x)(q^{-2} + t^2x)} e. g_{n,x,m}^{\pm, qt} - \sqrt{e(x)(1+x)}g_{n, q^{-2}x, m}^{\pm, qt}\right].
        \end{split}
    \end{equation}
    Combining these two expressions then gives the $q$-difference equation for $K^2$, which is given by
    \begin{equation}
    \label{eq:IwasawaK2action}
        \begin{gathered}
            \frac{1}{2}e(x)q^{-m-1}K^2.g_{n,x,m}^{\pm, t} = A(x)g_{n,q^{-2}x, m}^{\pm, t} + B(x)g_{n,x,m}^{\pm, t} + A(q^2x)g_{n, q^2x, m}^{\pm, t},\\
            A(x) = -\frac{1}{2}(tx)^{-2}\sqrt{(1+t^2x)(1+x)},\quad B(x) = \frac{1}{2}q^{-1}(tx)^{-2}(1+q^{-2} + (1+t^2)x),
        \end{gathered}
    \end{equation}
    where one can see that the value of $t$ remains unchanged. Next, let us calculate the right action of $\tilde Y_s^R$ on $g^{\pm, t}_{n,x,m}$. From the definition of $\tilde Y_s^{R}$ \eqref{eq:normalizedTwistedPrimitives}, it follows that
    \begin{equation}
        2\tilde Y_s^R = q^{1/2}(q-q^{-1})EK - q^{-1/2}(q-q^{-1})FK - (s+s^{-1})K^2.
    \end{equation}
    It can be directly calculated from \eqref{eq:rightregularaction} that
    \begin{equation}
        \begin{split}
            &q^{1/2}(q-q^{-1})\delta^\pm_{n,x,m}.EK\\
            &= \mp e(x)\sqrt{1\pm q^{-n-m-2}x^{-1}}\delta^\pm_{n+2,x,m} + q^{-n-1}e(x)\sqrt{1+x^{-1}}\delta^\pm_{n+2, q^{-2}x, m}.
        \end{split}
    \end{equation}
    For $x = q^{2k}$ or $x = -t^{-2}q^{-2k}$ such that $e(x) = 1$, we can act with $EK$ on the basis $g_{n,x,m}^{\pm,t}$ \eqref{eq:preIwasawaBasis}, which gives
    \begin{equation}
        \begin{split}
            &q^{1/2}(q-q^{-1})g^{\pm, t}_{n,x,m}.EK\\
            &= \mp\sqrt{1 + q^{-n-m-2}}g^{\pm, t}_{n+2,x,m} \\
            &\quad + q^{-n-1}\sum_k q^{k-1}\sqrt{1+q^{2-2k}}\tilde V_{k-1}^{t}(tx; 1; q^2)\delta^{\pm}_{n+1, q^{-2k}, m-2 + 2k},\\
            &= \mp\sqrt{1 + q^{-n-m-2}}g^{\pm, t}_{n+2,x,m}\\
            &\quad - tq^{-n-2}g^{\pm, t}_{n+2, x, m-2} + q^{-n-2}q^{-\frac{m}{2}}\sqrt{1+t^2x} K. g^{\pm, q^{-1}t}_{n+2, x, m-2},
        \end{split}
    \end{equation}
    where we used Lemma~\ref{lemma:qLaguerreContiguousRelations} in the last step. A similar derivation can be done when $e(x) = -1$. Here, it can be noted that the actions of the coordinate algebra \eqref{eq:coordinateAlgebraActionIwasawa} define new relations on the Al-Salam-Carlitz II polynomials, which also appear in the right action of $EK$. Note here that the above equation still has a remaining left action of $K$. Writing out this action using \eqref{eq:actionKIwasawa}, we get
    \begin{equation}
        \begin{split}
            &q^{1/2}(q-q^{-1})g^{\pm, t}_{n,x,m}.EK\\
            &= \mp e(x)\sqrt{1 \pm e(x) q^{-n-m-2}}g^{\pm, t}_{n+2,x,m} + e(x)q^{-n-2}t^{-1}x^{-1}g^{\pm, t}_{n+2, x, m+2} \\
            &+ q^{-n}x^{-1}e\gamma_{\infty, qt}^*\alpha_{\infty, qt}.g^{\pm, t}_{n,x,m}.
        \end{split}
    \end{equation}
    A similar expression can be found for the right action of $q^{-1/2}(q-q^{-1})FK$ by taking the adjoint. Now, we can substitute in $\tilde \rho_{st}$ as given in \eqref{eq:normalizedSphericalElement} to find that the right action of $\tilde Y_s^{R}$ on basis element $g^{\pm, t}_{n,x,m}$ is given by
    \begin{equation}
        \begin{split}
            g^{\pm, t}_{n,x,m}.\tilde Y_s^{R} &= \left(\mp \frac{1}{2}e(x)\sqrt{1\pm e(x)q^{-n-m-2}}g^{\pm, t}_{n+2, x, m} + \frac{1}{2}q^{-n-2}t^{-1}x^{-1}e(x)g^{\pm, t}_{n+2, x, m-2}\right)\\
            &+\left(\mp \frac{1}{2}e(x)\sqrt{1\pm e(x)q^{-n-m}}g^{\pm, t}_{n-2, x, m} + \frac{1}{2}q^{-n}t^{-1}x^{-1}e(x)g^{\pm, t}_{n-2, x, m+2}\right)\\
            &- q^{-n-1}(tx)^{-1}\tilde \rho_{st}.g^{\pm, t}_{n,x,m}.
        \end{split}
    \end{equation}
    Lastly, we can calculate the right action of $\tilde Y_s^R$ on the basis elements $\delta^{\pm, (\infty, t)}_{n,x,\xi}$ \eqref{eq:IwasawaBasis}. Upon this action, the terms in the brackets can be simplified using the $q$-difference relation as given in Lemma~\ref{lemma:relationsqJacobi} and Lemma~\ref{lemma:relationsAlSalamChihara} for the $q^2$-Jacobi functions (if $e(x) = 1$) or the Al-Salam-Chihara polynomials (if $e(x) = -1$), respectively. Applying these relations then gives \eqref{eq:rightactionTwistedPrimitiveIwasawa}.\\

    We can also calculate the action of the Casimir element in the Iwasawa decomposition. Firstly, let us split up the action of the Casimir element \eqref{eq:CasimiractionCartan} as
    \begin{equation}
        \tilde \Omega.\delta^\pm_{n,x,m} = \tilde \Omega^L.\delta^\pm_{n,x,m} + \tilde \Omega^C.\delta^\pm_{n,x,m} + \tilde \Omega^R.\delta^\pm_{n,x,m},
    \end{equation}
    where
    \begin{equation}
        \begin{split}
            \tilde \Omega^L.\delta^{\pm}_{n,x,m} &= \frac{1}{2}\sqrt{(1+x^{-1})(1\pm q^{-n-m}x^{-1})}\delta_{n, q^{-2}x, m}^{\pm},\\
            \tilde \Omega^C.\delta^{\pm}_{n,x,m} &= -\frac{1}{2}(q^{-m} \pm q^{-n})q^{-1}x^{-1}\delta^{\pm}_{n,x,m}
        \end{split}
    \end{equation}
    and $\tilde \Omega^R = (\tilde \Omega^L)^*$. Also, we will write the basis elements $g^{\pm, t}_{n,x,m}$ \eqref{eq:preIwasawaBasis} as
    \begin{equation}
        g^{\pm,t}_{n,x,m} = \sum_{k} g_k^t(x)\delta^\pm_{n,e(x) q^{-2k}, m+2k},
    \end{equation}
    where $g_k^t$ corresponds to the Al-Salam-Carlitz II polynomials or the Ciccoli-Koelink-Koornwinder functions for $e(x) =-1$ or $e(x)=1$, respectively. The action of $\tilde \Omega^L$ on this basis can be calculated to be equal to
    \begin{equation}
        \begin{split}
            &\tilde \Omega^L.g^{\pm, t}_{n,x,m}\\
             &= \sum_k \frac{1}{2}\sqrt{(1+e(x)q^{2k})(1\pm e(x) q^{-n-m})}g_k^t(x)\delta^\pm_{n, e(x) q^{-2(k+1)}, m-2 + 2(k+1)}\\
            &= \frac{1}{2}\sqrt{1\pm e(x)q^{-n-m}}\sum_k\bigg(t q^{k-1}\sqrt{e(x)(t^{-1}+x)} g_k^{q^{-1}t}(x)\\
            &\qquad\qquad\qquad\qquad\qquad\qquad- tq^{-1}g_k^t(x)\bigg) \delta^\pm_{n,e(x)q^{-2k}, m-2+2k}\\
            &= \frac{1}{2}\sqrt{1\pm e(x)q^{-n-m}}\left(-tq^{-1}g^{\pm, t}_{n,x, m-2} + q^{-\frac{m}{2}}t\sqrt{e(x)(t^{-2} + x)}K.g^{\pm, t}_{n,x,m-2}(q^{-1}t)\right),
        \end{split}
    \end{equation}
    where in the second step we used an identity for $g_k^t(x)$ that can be derived from the action of $\beta_{\infty, q^{-1}t}$ on $g^{\pm, t}_{n,x,m}$. By evaluating the left action of $K$ as given in \eqref{eq:actionKIwasawa}, and also by direct calculation of the action of $\tilde \Omega^C$, we have
    \begin{equation*}
        \begin{split}
        \tilde \Omega^L.g^{\pm, t}_{n,x,m} &= \frac{1}{2}t^{-1}x^{-1}\sqrt{1\pm e(x)q^{-n-m}}\bigg(q^{-1}g^{\pm, t}_{n,x,m-2} \\
        &\qquad\qquad\qquad\qquad\qquad\qquad\qquad- \sqrt{(1+t^2x)(1+x)}g^{\pm, t}_{n, q^{-2}x, m-2}\bigg),\\
        \tilde \Omega^C. g^{\pm, t}_{n,x,m} &= -\frac{1}{2}e(x)q^{-1}q^{-m} g^{\pm, t}_{n,x,m} \mp \frac{1}{2}e(x)q^{-1}q^{-n-m} K^2.g^{\pm, t}_{n,x,m}.
        \end{split}
    \end{equation*}
    One can now write out the action of $K^2$ using \eqref{eq:IwasawaK2action} and also substitute in the action of $\tilde Y^L_t$ as given in \eqref{eq:YtLactionong}. This now allows us to recombine terms to rewrite the Casimir action with
    \begin{equation}
        \begin{split}
            &\tilde \Omega^C.g^{\pm, t}_{n,x,m} = \left(\mp\frac{1}{2}q^{-1-n}(tx)^{-1}(1+q^{-2} + (1+t^2)x) - q^{-1}t^{-1}x^{-1}\tilde Y_t^{L}\right).g^{\pm, t}_{n,x,m}\\
            &\tilde \Omega^L. g^{\pm, t}_{n,x,m} \\
            &= -\frac{1}{2}(tx)^{-1}\sqrt{(1+t^2x)(1+x)}\bigg(\sqrt{1\pm e(x)q^{-n-m}}g^{\pm, t}_{n, q^{-2}x, m-2} \\
            &\qquad\qquad\qquad\qquad\qquad\qquad\qquad\qquad\qquad\mp q^{-n}t^{-1}x^{-1}g^{\pm, t}_{n, q^{-2}x, m}\bigg).
        \end{split}
    \end{equation}
    This action can now be evaluated on the basis $\delta^{\pm, (\infty, t)}_{n,x,\xi}$ as given in \eqref{eq:IwasawaBasis}. Since $\tilde Y_t^L$ acts diagonally on this basis, the action of $\tilde \Omega^C$ is immediately clear. Moreover, when one acts with $\tilde \Omega^L$ on $\delta^{\pm, (\infty,t)}_{n,x,\xi}$, the term in the brackets an be simplified using the relation of Lemma~\ref{lemma:relationsqJacobi} and Lemma~\ref{lemma:relationsAlSalamChihara} for the little $q^2$-Jacobi functions (if $e(x) = 1$) or the Al-Salam-Chihara polynomials (if $e(x) = -1$), respectively. After combining all the terms, one finds that the action of the Casimir element in the Iwasawa decomposition is given by
    \begin{equation}
    \label{eq:CasimiractionIwasawa}
        \begin{gathered}
            \tilde \Omega.\delta^{\pm, (\infty, t)}_{n,x,\xi} = A(x)\delta^{\pm, (\infty, t)}_{n, q^{-2}x, \xi} + B(x)\delta^{\pm, (\infty, t)}_{n,x,\xi} + A(q^2x)\delta^{\pm, (\infty, t)}_{n, q^2x, \xi},\\
            A(x) = -\frac{1}{2}e(x)\sqrt{(1+x^{-1})(1+t^{-2}x^{-1})(1 \pm \xi^{-1}(tx)^{-1} q^{-n})(1 \pm \xi (tx)^{-1}q^{-n})},\\
            B(x) = -\frac{1}{2}(2t^{-1}\mu(\xi) \pm t^{-2}x^{-1}(1 + q^{-2} + (1+t^2)x)q^{-n})q^{-1}x^{-1}.
        \end{gathered}
    \end{equation}
\subsection{Actions in the quantum Gauss decomposition}
\label{sec:actionsGaussDecomp}
Let us now calculate some of the actions of operators in the quantum Gauss decomposition. The methods used here are completely analogous to the ones used in Section~\ref{sec:actionsIwasawaDecomp}, and we will therefore be a bit more brief about the derivations.\\

Firstly, we comment on the action of the coordinate algebra. The action of the coordinate algebra on the basis in the Gauss decomposition can be calculated in a completely analogous way to how this action in the Iwasawa decomposition \eqref{eq:coordinateAlgebraActionIwasawa} was derived in Appendix~\ref{sec:actionsIwasawaDecomp}. In particular, after acting with $\alpha_{s,qt} = \alpha_{\infty,qt} - s^{-1}\gamma_{\infty, qt}$ on the basis elements $h^{\pm, (st)}_{n,y,\xi}$ as given in \eqref{eq:preGaussBasis}, one recognises the relations for the little $q^2$-Jacobi functions or the Al-Salam-Chihara polynomials as given in Lemma~\ref{lemma:relationsqJacobi} and Lemma~\ref{lemma:relationsAlSalamChihara}. Applying these relations to action on the basis elements $\delta^{\pm, (st)}_{\eta, y, \xi}$ then gives the result of \eqref{eq:Gaussquantumgroupaction}.\\

Next, we will calculate the action of the universal enveloping algebra in the Gauss decomposition. Note that $\delta^{\pm, (\infty, t)}_{n, x, \xi}.K^{-1} = q^{\frac{n}{2}}\delta^{\pm, (\infty, t)}_{n, x, \xi}$. If we now consider the right action of $K^{-1}$ on basis elements $h^{\pm, (st)}_{n,y,\xi}$ as given in \eqref{eq:preGaussBasis}, one recognises the $q$-difference relations for the $q^2$-Jacobi functions (if $e(x)=1$) and the Al-Salam-Chihara polynomials (if $e(x) = -1$) as given in Lemma~\ref{lemma:secondqdifferenceqJacobi} or Lemma~\ref{lemma:thirdqdifferenceAlSalamChihara}, respectively. For $\mu(y) \in \sigma(\tilde \rho_{st}) \setminus [-1,1]$ this therefore gives
\begin{equation}
    \begin{gathered}
        h_{n, y, \xi}^{\pm, (st)}.K^{-1} = A_s(y)h_{n, qy, \xi}^{\pm, (q^{-1}s,t)} + A_s(y^{-1})h_{n, q^{-1}y, \xi}^{\pm, (q^{-1}s, t)},\\
        A_s(y) = q^{\frac{n}{2}}\sqrt{-e(x)\kappa(y)qst^{\kappa(y)}y\frac{(1-qs^{-1}ty)(1-qs^{-1}t^{-1}y)}{(1-y^2)(1-q^2y^2)}}.
    \end{gathered}
\end{equation}
Since $K$ is self-adjoint, one can take the adjoint of the above action to obtain
\begin{equation}
    h_{n, y, \xi}^{\pm, (st)}.K^{-1} = A_{qs}(q^{-1}y)h_{n, q^{-1}y, \xi}^{\pm, (qs, t)} + A_{qs}(q^{-1}y^{-1})h_{n, qy, \xi}^{\pm, (qs, t)}.
\end{equation}
By combining these two actions, one now finds that the right action of $K^{-2}$ is given by the $q$-difference equation
\begin{equation}
    \label{eq:K2actionGauss}
    \begin{gathered}
        e(x)\kappa(y)q^{-n}h_{n,y, \xi}^{\epsilon, (st)}.K^{-2} = A(y)h_{n,q^{-2}y, \xi}^{\epsilon, (st)} + B(y)h_{n, y, \xi}^{\epsilon, (st)} + A(y^{-1})h_{n,q^{2}y, \xi}^{\epsilon, (st)},\\
        A(y) = t^{\kappa(y)}\sqrt{\frac{y^2(1 - q^{-1}s^{\pm}t^{\pm}y)}{(1-y^2)(1-q^{-2}y^2)^2(1-q^{-4}y^2)}}, B(y) =qt^{\kappa(y)}\frac{\mu(t)\mu(s) - \mu(s)\mu(y)}{\mu(y)^2 - \mu(q)^2},
    \end{gathered}
\end{equation}
where
\begin{equation}
    (1 - q^{-1}s^{\pm}t^{\pm}y) = (1 - q^{-1}sty)(1 - q^{-1}s^{-1}t y)(1 - q^{-1}st^{-1} y)(1 - q^{-1}s^{-1}t^{-1} y).
\end{equation}
Note here that $\pm$ does not refer to a variable in the basis element anymore. That is, we changed notations $\delta^{\pm} \to \delta^{\epsilon}$, such that the notation of the actions can be `compactified'. We can now derive the action of the Casimir element on the basis $\delta^{\epsilon, (st)}_{\eta, x, \xi}$. Firstly, we can split up the Casimir action in the Iwasawa decomposition \eqref{eq:CasimiractionIwasawa} as
\begin{equation*}
    \tilde \Omega. \delta^{\epsilon, (\infty, t)}_{n, x, \xi} = \tilde \Omega^L.\delta^{\epsilon, (\infty, t)}_{n,x,\xi} + \tilde \Omega^C.\delta^{\epsilon, (\infty, t)}_{n,x,\xi} + \tilde \Omega^R.\delta^{\epsilon, (\infty, t)}_{n,x,\xi}.
\end{equation*}
where
\begin{equation}
    \begin{split}
        &\tilde \Omega^L.\delta^{\epsilon, (\infty, t)}_{m, x, \xi}\\
        &= -\frac{1}{2}e(x)\sqrt{(1+x^{-1})(1+t^{-2}x^{-1})(1 +\epsilon\xi^{-1}(tx)^{-1} q^{-n})(1 +\epsilon \xi (tx)^{-1}q^{-n})} \delta^{\epsilon, (\infty, t)}_{n, q^{-2}x, \xi},\\
        &\tilde \Omega^C.\delta^{\epsilon, (\infty, t)}_{n, x, \xi} = -\frac{1}{2}(2t^{-1}\mu(\xi) +\epsilon t^{-2}x^{-1}(1 + q^{-2} + (1+t^2)x)q^{-n})q^{-1}x^{-1}\delta^{\epsilon, (\infty, t)}_{n, x, \xi}.
    \end{split}
\end{equation}
When acting on $h^{\epsilon, (st)}_{n,y,\xi}$, it can now be immediately seen that the action of $\tilde \Omega^{C}$ can be rewritten in terms of the action of $K^{-2}$. In particular, one finds
\begin{equation}
    \begin{split}
        \tilde \Omega^C.h^{\epsilon, (st)}_{n, y, \xi}&= \epsilon \frac{1}{2}q^{-1}(1+t^{-2})q^{-n}.h^{\pm, st}_{n, y, \xi} \\
        &\quad + (q^{-1}t^{-1}\mu(\xi) q^{-n}+\epsilon\frac{1}{2}q^{-1}t^{-2}(1+q^{-2})q^{-2n})h^{\epsilon, (st)}_{n,y,\xi}.K^{-2}.
    \end{split}
\end{equation}
For the action of $\tilde \Omega^L$, let us consider the case where $\mu(y) \in \sigma(\tilde \rho_{st}|_W) \setminus [-1,1]$. In this case, the action of $\tilde \Omega_L$ on $h^{\epsilon,(st)}_{n,y,\xi}$ is given by
\begin{equation}
    \begin{split}
        \tilde \Omega^L.h^{\epsilon, (st)}_{n, y, \xi}& = \sum_{k} j_k(-\mu(y); q^{2}s^{-2}, -qs^{-1}t^{-1}; q^2, -1)\tilde \Omega^{L}.\delta^{\epsilon, (\infty, t)}_{n+2k, q^{-2k}, \xi}\\
        &= -\frac{1}{2}\sqrt{(1+\epsilon \xi^{-1}t^{-1}q^{-n})(1+\epsilon \xi t^{-1}q^{-n})}\\
        &\quad \times \sum_k \sqrt{(1+q^{2k})(q+t^{-2}q^{2k})}j_k(-\mu(y); q^{2}s^{-2}, -qs^{-1}t^{-1}; q^2, -1)\\
        &\qquad\qquad\times\delta^{\epsilon, (st)}_{n-2+2(k+1), q^{-2(k+1)}, \xi}.
    \end{split}
\end{equation}
Here, one can now apply Lemma~\ref{lemma:firstqdifferenceqJacobi} and Corollary~\ref{cor:firstqdifferenceqJacobicor} for the little $q^2$-Jacobi functions to rewrite the term in the sum. One gets a similar expression when $\mu(y)\in \sigma(\tilde \rho_{st}|_{U,V})\setminus [-1,1]$, and one can use Lemma~\ref{lemma:firstqdifferenceAlSalamChihara} and  Lemma~\ref{lemma:secondqdifferenceAlSalamChihara} to simplify the expression. In particular, the action of $\tilde \Omega^L$ can be calculated to be equal to
\begin{equation}
    \begin{gathered}
        \tilde \Omega^L.h_{n, y, \xi}^{\epsilon. (st)} = A(y)h^{\epsilon, (st)}_{n-2, q^{-2}y, \xi} + B(y)h^{\epsilon, (st)}_{n-2, y, \xi} + A(y^{-1})h^{\epsilon, (st)}_{n-2, q^2y, \xi},\\
        A(y) = \frac{1}{2}q^{-2}y^2\sqrt{\frac{(1-q^{-1}s^{\pm}t^{\pm}y)(1+\epsilon \xi^{\pm}t^{-1}q^{-n})}{(1-y^2)(1-q^{-2}y^2)(1-q^{-4}y^2)}},\\
        B(y) = \frac{1}{2}\sqrt{(1+\epsilon \xi^{\pm}t^{-1}q^{-n})}\frac{\mu(s)\mu(q) + \mu(t)\mu(y)}{\mu(y)^2 - \mu(q)^2}.
    \end{gathered}
\end{equation}
One can now finally consider the action of the Casimir element on the basis $\delta^{\epsilon, (st)}_{\eta, y, \xi}$. In particular, when writing out the actions of $\tilde \Omega^C$ and $\tilde \Omega^L$ as given above on basis elements  $\delta^{\epsilon, (st)}_{\eta, y, \xi}$, one will recognise the right action of $\tilde Y_s^R$. Moreover, the $q$-difference relations as given in Lemma~\ref{lemma:relationsqJacobi} and Lemma~\ref{lemma:relationsAlSalamChihara} for the little $q^2$-Jacobi functions and Al-Salam-Chihara functions will appear. Analogous to the derivation of the action of the Iwasawa decomposition, the action of the Casimir element in the Gauss decomposition can then be derived to be given by \eqref{eq:CasimiractionGauss}.

\end{appendices}

\bibliographystyle{JHEP}
\bibliography{bibliography}

\end{document}